\pdfoutput=1
\documentclass[]{jkpaper}
\usepackage{tikz}
\usepackage{tikz-cd}
\usetikzlibrary{decorations.markings}
\usetikzlibrary{decorations.pathreplacing}
\usetikzlibrary{arrows.meta}
\usetikzlibrary{patterns}
\usetikzlibrary{decorations.pathmorphing}
\tikzset{snake it/.style={decorate, decoration={snake, segment length=1.5mm, amplitude=0.5pt}}}
\usepackage{bigints}
\usepackage[backend=bibtex]{biblatex}
\usepackage{aas_macros}
\usepackage[outline]{contour}
\usepackage{soul}

\theoremstyle{definition}
\newtheorem{eg}{Example}

\newcommand{\hodge}{\mathop{\ast}} 
\newcommand{\Hodge}{\mathop{\star}} 
\newcommand{\Hodgee}{\mathop{\star'}} 
\newcommand{\supp}{\operatorname{supp}} 
\newcommand{\relsupp}{\operatorname{relsupp}} 

\newcommand{\OIST}{\raisebox{-0.08em}{\includegraphics[height=0.8em]{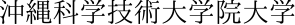}}}

\NewDocumentCommand{\newauthornote}{mmm}{
  \NewDocumentCommand{#1}{s m}{%
    \textcolor{#3}{\IfBooleanTF{##1}%
      {{##2}}
      {{\emph{\footnotesize \textbf{#2}:~##2}}} 
  }}%
}

\newauthornote{\jjvk}{JJVK}{blue}

\title{Fighting non-locality with non-locality: \texorpdfstring{\\}{}microcausality and boundary conditions in QED }

\abstr{

 In gauge theories, globally charged observables necessarily depend non-locally on the kinematical fields, with this dependence extending to the asymptotic boundary of spacetime. Despite this, we show that a subset of such observables can be consistently regarded as local to the bulk, in a manner that respects microcausality and leaves locality properties of uncharged observables untouched. A sufficient condition for this is to impose kinematically \emph{non-local} boundary conditions on the large gauge sector of the theory, and to invoke a \emph{relational} notion of localisation for observables. This reveals a relatively underappreciated link between boundary conditions, and different notions of microcausality and locality. We develop this point through a detailed case study in scalar QED, describing non-local boundary conditions that allow a large family of observables on a codimension-1 bulk surface to be viewed as local to that surface, despite being dressed by asymptotic Wilson lines. We show that this property continues to hold within a perturbative quantisation of the theory, and we argue that this leads to a consistent local net of algebras that includes these charged observables in bulk algebras. We explain how this setup may be understood in terms of a preferred dynamical reference frame for small gauge transformations appearing in the boundary conditions. Many features of the theory (such as microcausality, the vacuum state, and the net of algebras of observables) depend on the choice of this frame, and we briefly discuss some repercussions of this for algebraic formulations of QFT. While our analysis is performed in QED, we expect our results to carry over qualitatively to more complicated theories including gravity. 

}

\author{Philipp A.\ H\"ohn,\texorpdfstring{\textsuperscript{1\,$*$}}{} and Josh Kirklin\texorpdfstring{\textsuperscript{2\,$\dagger$}}{}}
\institution{\texorpdfstring{\textsuperscript{1}}{}Qubits and Spacetime Unit,\texorpdfstring{\\}{ } Okinawa Institute of Science and Technology \emph{(}\OIST\emph{)},\texorpdfstring{\\}{ } 1919-1 Tancha, Onna-son, Kunigami-gun, Okinawa, Japan 904-0495
\texorpdfstring{\\\vspace*{1.5em}}{ } \texorpdfstring{\textsuperscript{2}}{}Perimeter Institute for Theoretical Physics,\\ 31 Caroline Street North, Waterloo, ON, N2L 2Y5, Canada}

\email{\textsuperscript{$*$}\emaillink{philipp.hoehn@oist.jp}\\ \textsuperscript{$\dagger$}\emaillink{jkirklin@perimeterinstitute.ca}}

\bibliography{refs}
\begin{document}
\maketitleandtoc

\section{Introduction}

Theories of physics are usually supposed to govern the behaviour of the entire universe, but are only ever experimentally tested by `local' observations. On the other hand, diffeomorphism invariance in gravity requires all degrees of freedom to be non-local~\cite{Torre1993}, even in perturbative settings leading to challenges to locality and microcausality \cite{Donnelly:2015hta,Donnelly:2016rvo,Donnelly:2017jcd,Giddings:2018koz,Giddings:2025xym,Giddings:2025bkp}. This seems at first to be problematic. If there are no local degrees of freedom in the theory, how can we ever hope to test it with local experiments? A seemingly physical way out is to abandon the usual na\"ive notion of \emph{kinematical} locality, defined relative to a background spacetime, in favour of \emph{relational} locality, defined through certain \emph{relationships} between degrees of freedom. What one calls a `local' observation is then really a measurement of such relationships. Thus, the notion of locality is not altogether separate from the theory -- rather it is a part of it \cite{Rovelli:1990ph,Rovelli:1990pi,Rovelli:2013fga,Giddings:2005id,Giddings:2025xym,Giddings:2025bkp,Dittrich:2004cb,Dittrich:2005kc,Giesel:2007wi,Tambornino:2011vg,Carrozza:2022xut,ReconcilingBulkLocality,Freidel:2025ous,Donnelly:2016auv}. In~\cite{ReconcilingBulkLocality} we explained how this perspective reconciles locality -- including microcausality under certain conditions -- with diffeomorphism invariance in gravity.

In this paper, we step back from gravity to the simpler setting of scalar quantum electrodynamics (QED), which already displays many of the features responsible for the tension between gauge symmetry and locality: the absence of strictly (kinematically) local charged observables, the need for dressing, and the appearance of edge/Goldstone modes \cite{Donnelly:2014fua,Donnelly:2016auv,Geiller:2019bti,Carrozza:2021gju,Araujo-Regado:2024dpr,Ball:2024hqe,Fewster:2025ijg}. Our focus is a concrete case study that isolates an aspect of these issues that has received  apparently little attention: the observation that there exists a multitude of valid, yet \emph{inequivalent} definitions of microcausality (a key feature of locality defined below) in theories with gauge symmetry, and that their suitability
interplays intimately with the choice of boundary conditions.

We will see that, besides standard notions of microcausality defined with respect to ordinary \emph{kinematical} notions of observable support, there also exist \emph{relational} notions of microcausality defined in terms of a dressing (or gauge frame) dependent definition of relational support that we import from the gravity case \cite{ReconcilingBulkLocality}. Which notion of microcausality may be deemed most natural depends on the boundary conditions and arguably operational considerations. In particular, in some cases, relational notions will permit us to consistently associate certain charged observables with local bulk regions, in contrast to kinematical notions of locality. We will explore consequences of these observations for algebraic notions of locality in quantum field theories with gauge symmetry.
The relative simplicity of the QED setting means we can carry out an explicit analysis of many of its features, and its perturbative quantisation. We expect the lessons drawn from this analysis to carry over qualitatively to the gravitational case, refining our discussion in \cite{ReconcilingBulkLocality}.

In gauge theory, the bare value of a charged field $\psi$ at a point $x$ in spacetime cannot be observed because it is not gauge-invariant. To construct an observable out of it, a common technique involves ``dressing'' it with some other fields. For example, if $\psi$ is a complex scalar of charge $q$ in QED, one can attach a Wilson line to the bare quantity $\psi(x)$ to obtain the ``dressed observable''
\begin{equation}
  \Psi(x) = \exp(-iq\int_\gamma A)\,\psi(x),
  \label{Equation: dressed scalar}
\end{equation}
where $A$ is the gauge potential and $\gamma$ is some path extending between $x$ and a point on the boundary (asymptotic or finite, depending on the setting) of spacetime.

This observable $\Psi(x)$ is `boundary-dressed', because its dressing (the Wilson line) extends to the boundary. One could also construct observables which are `bulk-dressed'. For example, one could involve a second scalar field $\psi'$ of the opposite charge $-q$. Then $\psi(x)\psi'(x)$ is a bulk-dressed observable; it is the first scalar field dressed by the second. Beyond the fact that any boundary-dressed observable must have kinematical support extending all the way to the boundary, while bulk-dressed observables can have arbitrarily compact support, the key difference between the two is that the former are sensitive to large gauge transformations, while the latter are not; see Figure~\ref{Figure: bulk vs boundary-dressed}. In fact, any non-trivially globally charged observable must be boundary-dressed. We will mostly focus on boundary-dressed observables in this paper as these constitute the genuine challenges to standard notions of locality.

\begin{figure}[h]
  \centering
  \vspace{1em}
  \begin{tikzpicture}[remember picture,overlay,shift={(2,0)}]
    \draw[thick,red!70!black] (0,0) .. controls (1,0.2) and (4,-0.2) .. (10,0);
    \node[above] at (3,0) {$\gamma$};
    \fill[black!60] circle (0.07);
    \node[above] at (0,0.1) {$y$};
    \node[right] at (-0.6,-0.5) {$\Psi(y) = e^{-iq\int_\gamma A}\psi(y)$};

    \begin{scope}[shift={(-4,0)}]
      \fill[black!60] circle (0.07);
      \node[above] at (0,0.1) {$x$};
      \node[] at (-0.1,-0.5) {$\psi(x)\psi'(x)$};
    \end{scope}
  \end{tikzpicture}
  \vspace{2.5em}
  \caption{Bulk-dressed observables like $\psi(x)\psi'(x)$ can have arbitrarily compact support. Boundary-dressed observables like $\Psi(y)$ must extend to the boundary, and are sensitive to large gauge transformations (i.e.\ those which act non-trivially at the boundary).}
  \label{Figure: bulk vs boundary-dressed}
\end{figure}

The bulk-dressed observable $\psi(x)\psi'(x)$ is clearly local to $x$. On the other hand, the boundary-dressed observable $\Psi(x)$ does have some affiliation with the point $x$, and by measuring it one can obtain more information about the field configuration at $x$ than just what is contained in purely locally constructed gauge-invariant observables (such as $\psi(x)\psi'(x)$). So it is not unreasonable to hope that $\Psi(x)$ can be considered as ``part of the physics local to $x$''.

Of course, the full story is not quite so simple, because the value of $\Psi(x)$ also depends on the gauge potential at all points along the path $\gamma$. By doing an appropriate gauge transformation one could fix the value of the dressing to be $\exp(-iq\int_\gamma A) = 1$, which would appear to give the observable a fully local construction $\Psi(x)=\psi(x)$, and in some sense to move all the information measured by $\Psi(x)$ into $\psi(x)$.\footnote{This is an incarnation of Page--Wootters reduction in the study of quantum reference frames (QRFs)~\cite{PhysRevD.27.2885,1984IJTP...23..701W,Hoehn:2019fsy,delaHamette:2021oex}, where one gauge fixes the QRF into a specific configuration. Indeed, the Wilson line dressing may be understood as constituting a reference frame for small gauge transformations (see Section~\ref{Section: frame}).} But the required gauge transformation has a non-local effect on the fields. Moreover, different dressings will generally lead to different observables, and it would be overambitious to think of all such observables as local to $x$ simultaneously. For example, if $\gamma_1,\gamma_2$ are two choices of path leading to different dressed observables $\Psi_1(x),\Psi_2(x)$, both of which we attempt to consider as local to $x$, then clearly we would also have to think of their quotient
\begin{equation}
  \frac{\Psi_1(x)}{\Psi_2(x)} = \exp(-iq\int_{\gamma_1\cup \gamma_2}A)
\end{equation}
as local to $x$. But this is just a Wilson line along the path $\gamma_1\cup \gamma_2$, apparently no more local to $x$ than it is to any other point in $\gamma_1\cup \gamma_2$. This already hints at a dressing (or gauge frame) dependence of relational notions of locality.

Let us now focus on a more concrete way to define ``locality''. We will mainly focus on the principle of \emph{microcausality}, which states that any two spacelike separated local observables can be measured simultaneously and independently, meaning classically their Poisson bracket vanishes, and quantum mechanically they commute as operators (in both cases we will often just say that the observables commute). Essentially, microcausality forbids the faster-than-light propagation of information measured by local observables, and thus prevents causality paradoxes at arbitrary length scales.

More concretely, a reasonable formulation of locality involves what is known as a `causally local net of observables' obeying the Haag-Kastler axioms of algebraic quantum field theory (QFT) \cite{haag2012local,Fewster:2019ixc}. This is the assignment of an algebra of observables $\mathcal{A}_U$ (classically a Poisson algebra, quantum mechanically typically a $C^*$-algebra) to every subset $U$ of spacetime, such that $D(U_1) \subseteq D(U_2)$ implies $\mathcal{A}_{U_1}\subseteq \mathcal{A}_{U_2}$ (i.e.\ `isotony'), where $D(U)$ denotes the domain of dependence of $U$,\footnote{$D(U)$ is the set of points $x$ such that any causal curve through $x$ intersects $U$ at least once.} and such that $[\mathcal{A}_{U_1},\mathcal{A}_{U_2}]=0$ if $U_1$ and $U_2$ are spacelike separated (meaning there is no causal curve connecting $U_1$ with $U_2$). Each algebra $\mathcal{A}_U$ is supposed to consist of those observables that can be observed using only the degrees of freedom in $U$, and the latter property is microcausality and sometimes referred to as `Einstein causality'. There are additional conditions that are usually imposed on the local algebras \cite{haag2012local,Fewster:2019ixc}, but here we focus on those most pertinent to locality.

In practice, one typically has in mind that $\mathcal{A}_U$ is generated by observables that admit a local construction within $D(U)$. For example, in the case of scalar QED one could build a net of local algebras obeying isotony if one assigned any given  physical (i.e.\ small gauge-invariant) observable $O$ to any algebra $\mathcal{A}_U$ such that the kinematical support of $O$ resides entirely in $D(U)$. With standard local boundary conditions, it is also clear that such a net of algebras obeys Einstein causality. In this way, one could even sidestep the non-locality issues discussed in \cite{Donnelly:2015hta,Donnelly:2016rvo,Donnelly:2017jcd}. The price to pay would be that no non-trivially globally charged observable could ever be included in any local bulk region $U$; \emph{all} non-trivially charged observables would only reside in algebras associated with regions touching the asymptotic boundary.

From an algebraic point of view, however, this restriction of $\mathcal{A}_U$ to observables with kinematical support in $U$ is not strictly required.
In general, the assignment of $\mathcal{A}_U$ to $U$ can have a priori nothing to do with the way in which the observables in $\mathcal{A}_U$ depend on the fields (in particular, whether the observables are locally constructed from the fields in $U$). So, for example, there is nothing stopping $\Psi(x)$ from being a part of $\mathcal{A}_U$ for all $U$ such that $x\in D(U)$, provided that microcausality and isotony are respected. This reflects the fact that the notion of localisation relevant for microcausality need not coincide with the na\"ive kinematical support of an observable as a functional of the fields. On the other hand, it seems reasonable to insist that each $\mathcal{A}_U$ includes at a minimum all those observables which \emph{can} be locally constructed from the fields in $U$. The question then becomes: given this minimum, \emph{what else} can be included in each $\mathcal{A}_U$, and moreover, \emph{how much} else can be included, before the assignment $U\to\mathcal{A}_U$ becomes maximal? In particular, are there consistent ways to assign \emph{charged} observables to \emph{bulk} algebras associated with regions that do not touch the asymptotic boundary? The recent work \cite{Fewster:2025ijg} demonstrated that it \emph{is} possible to rigorously include charged observables and edge/Goldstone modes in algebras associated with finite regions. Importantly, however, these regions were considered in isolation and thus what was not addressed in this study is how these regions and algebras embed into a global spacetime and net of algebras, squaring up with the isotony, causality and other axioms of algebraic QFT.

The purpose of the present paper is to study this question as it pertains to dressed observables like $\Psi(x)$. We will show that there are cases when  such charged observables may be included in algebras associated with bulk regions not touching an asymptotic boundary, while consistently embedding into a net of local algebras obeying isotony and Einstein causality.

It has previously been argued that thinking of such observables as local to $x$ would violate microcausality, precluding their inclusion in $\mathcal{A}_U$ for all $U$ such that $x\in D(U)$, e.g.\ see~\cite{Donnelly:2015hta,Donnelly:2016rvo,Donnelly:2017jcd} for some discussion. The basic reason is that in many approaches to constructing the Poisson bracket in QED (such as via a Dirac bracket), the gauge field $A$ and the electric field are conjugate variables, so the bracket of $\Psi(x)$ and the electric field at any point along $\gamma$ would have to be non-vanishing. As a consequence, there would be non-vanishing brackets between $\Psi(x)$ and observables at points spacelike separated from $x$, within the domain of influence of $\gamma$.\footnote{The domain of influence is the set of all points reachable by causal curves.}

As we shall explain in this paper, the validity of the above argument depends on the class of boundary conditions one employs.  Consider the common approach to constructing the Dirac bracket in QED where one introduces a gauge-fixing term $\mathcal{L}_{\text{g.f.}}$ into the Lagrangian.  For example, $\mathcal{L}_{\text{g.f.}} = -\frac1{2\xi}A_t^2$ enforces temporal gauge $A_t = 0$ (where $t$ is a time coordinate), which can be satisfied by changing $A \to A - \dd{\Lambda}$ for some suitable $\Lambda$ (for a different choice, see \cite{Donnelly:2015hta}). For most $\Lambda$, this is just a gauge transformation, so nothing goes wrong. But if $\Lambda$ does not vanish on the boundary of spacetime, then this is a `large gauge transformation', which is not a true gauge transformation at all, but rather a physical change of the field configuration (for most kinds of boundary conditions). Thus, to enforce this gauge, in general one must impose some physical restriction on the fields at the boundary (in other words, a boundary condition). The same is true for most other kinds of gauge choices that are commonly employed -- but different such gauge choices typically will lead to different physical restrictions at the boundary.

\begin{figure}
  \centering
  \begin{tikzpicture}[scale=1.2]
    \begin{scope}
      \fill[blue!20] (0,0) rectangle (2,2);
      \draw[red,thick,snake it] (1.7,0.5) -- (2.0,0.8);
      \draw[red,thick,-stealth,snake it] (2.0,0.8) -- (1.7,1.1);
      \draw[red,thick,-stealth,snake it] (0.6,0.8) -- (1.2,1.4);
      \draw[blue!60!black,thick] (0,0) -- (0,2);
      \draw[blue!60!black,thick] (2,0) -- (2,2);
      \node[blue!70!black,left] at (0,1) {\large$\Gamma_1$};
      \node[blue!70!black,right] at (2,1) {\large$\Gamma_2$};
      \node at (1,-0.3) {\footnotesize Dirichlet};
    \end{scope}
    \begin{scope}[shift={(4,0)}]
      \fill[blue!20] (0,0) rectangle (2,2);
      \draw[red,thick,snake it] (1.7,0.5) -- (2.0,0.8);
      \draw[red,thick,-stealth,snake it] (0,0.8) -- (0.3,1.1);
      \draw[blue!60!black,thick] (0,0) -- (0,2);
      \draw[blue!60!black,thick] (2,0) -- (2,2);
      \node[blue!70!black,left] at (0,1) {\large$\Gamma_1$};
      \node[blue!70!black,right] at (2,1) {\large$\Gamma_2$};
      \node at (1,-0.3) {\footnotesize $\phi|_{\Gamma_1} = \phi|_{\Gamma_2}$};
    \end{scope}
  \end{tikzpicture}
  \caption{On the left, standard Dirichlet boundary conditions lead to a scalar field obeying microcausality; signals are reflected off the timelike boundaries $\Gamma_{1,2}$. On the right, periodic boundary conditions lead to a scalar field that violates microcausality (with respect to the underlying causal structure); signals instantaneously propagate between the timelike boundaries.}
  \label{Figure: periodic non-microcausality}
\end{figure}
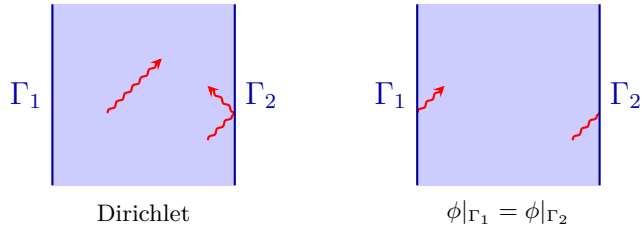

In field theory in general, the boundary conditions one imposes are intimately connected with the status of microcausality. As a simple demonstration of this fact, consider a massless scalar field propagating on a spacetime of the form $\mathcal{M}=I\times\mathcal{W}$, where $I$ is a spacelike interval and $\mathcal{W}$ is some spatially compact spacetime. The timelike part of the boundary then decomposes into two connected components $\Gamma_1\sqcup\Gamma_2 = \partial I\times\mathcal{W}$. It is well known that imposing Dirichlet boundary conditions on $\Gamma_{1,2}$ leads to a theory obeying microcausality. If we were to instead impose the condition that the value of the scalar on $\Gamma_1$ is equal to the value of the scalar on $\Gamma_2$, then the scalar field would evolve as if it was on $S^1\times \mathcal{W}$, so signals could propagate arbitrarily quickly from a neighbourhood of $\Gamma_1$ to a neighbourhood of $\Gamma_2$, and vice versa. Thus, the scalar field would violate microcausality (with respect to the causal structure of the underlying spacetime $I\times\mathcal{W}$). This is depicted in Figure~\ref{Figure: periodic non-microcausality}.

Since the scalar field equations of motion are local and causal, this example may lead one to the conclusion that violations of microcausality occurred only because of the non-locality of the boundary conditions --- and that perhaps choosing local boundary conditions would be sufficient to `maximise' microcausality. In a theory without gauge symmetry, this may be true. However, in gauge theories this is not necessarily the correct conclusion. In fact, as we will show, non-local boundary conditions in conjunction with relational notions of observable support in gauge theories can be used to \emph{enhance} microcausality for \emph{some} charged observables, while degrading it for other charged observables and leaving it unaffected for uncharged observables.  In particular, by picking such boundary conditions, we can allow certain families of dressed \emph{charged} observables into some bulk algebras $\mathcal{A}_U$, which wouldn't be allowed with ordinary local boundary conditions. Conversely, charged observables whose microcausality properties have degraded will now only fit into algebras associated with larger regions $U$ that still touch the asymptotic boundary. Hence, there is an interesting tradeoff: our construction allows charged observables to participate in bulk locality, at the cost of boundary non-locality.

For clarity, we emphasise that our observations are not in contradiction with those in \cite{Donnelly:2015hta,Donnelly:2016rvo,Donnelly:2017jcd}, where such non-local boundary conditions are not considered. Similarly, we note that the tension between microcausality results in perturbative gravity reported in \cite{Donnelly:2015hta,Donnelly:2016rvo} and \cite{Frob:2022ciq,Frob:2023gng} may also be related to boundary terms (and thus boundary conditions), e.g.\ see the discussion in \cite[footnote 14]{Giddings:2022hba}. Our discussion rather aims to highlight the importance of the type of boundary conditions and notions of observable support in explorations of microcausality. In particular, in our case study of non-local boundary conditions, the standard construction of a net of local algebras based on kinematical support, alluded to above, would in fact \emph{violate} Einstein causality.\footnote{Indeed, as explained in Section~\ref{ssec_other}, for some boundary-dressed observables the relational support is larger than the kinematical support. These observables would be responsible for violations of Einstein causality, if one were to form algebras based on kinematical support.} Thus, in this case, one is forced to resort to relational definitions.

Gauge theories by their nature contain fundamentally non-local degrees of freedom. So it is maybe not surprising that a more nuanced perspective on these theories can be obtained by considering also non-local boundary conditions. Indeed, it seems we can fight non-locality with non-locality, at least to some degree. A complementary systematic analysis of microcausality and the axioms of algebraic QFT in the presence of large gauge charged observables but \emph{without} non-local boundary conditions and relational notions of support will appear in \cite{AHLT}.

These results should be viewed as a step toward addressing analogous questions in gravity. The gravitational case is substantially more subtle, since localisation itself is field-dependent: spacetime points $x \in \mathcal{M}$ do not have an invariant meaning independent of the dynamical geometry. As a result, both the notion of dressing and the definition of microcausality must be formulated relationally from the outset \cite{ReconcilingBulkLocality}. An important open direction is to identify concrete gravitational analogues of the non-local boundary conditions introduced here, and to determine whether they similarly permit a relationally local, microcausal algebra of observables. We leave this question to future work.

Let us also note that the idea of multiple notions of microcausality being relevant for the same theory is a fundamental part of the holographic approach to quantum gravity. Indeed, boundary operators are guaranteed to commute if they are spacelike separated according to the boundary causal structure. But they also are guaranteed to commute if their bulk duals are spacelike separated according to the \emph{bulk} causal structure. These two criteria for commutativity can be quite distinct, because a boundary operator can have very non-local support, while the corresponding bulk dual has compact support. Reconciling this fact with  algebraic QFT ideas has led to the principle of subregion-subalgebra duality~\cite{Leutheusser:2021frk,Leutheusser:2021qhd,Leutheusser:2022bgi}, and shows promise for shedding light on the emergence of the bulk spacetime. 

The paper proceeds as follows. In Section~\ref{Section: boundary conditions}, we establish the family of theories we will use to demonstrate our proposal, and review how the boundary conditions affect the phase space structures of these theories. In Section~\ref{Section: non-local}, we will describe in detail the particular non-local boundary conditions that we will use, and construct the resulting phase space. Then, in Section~\ref{Section: microcausality} we analyse the status of microcausality under these boundary conditions, demonstrating that it is consistent with the bulk locality of a large class of dressed observables like~\eqref{Equation: dressed scalar}; in particular, we develop a general relational criterion for microcausality. The choice of boundary conditions is related to the existence of a preferred dynamical reference frame, which we refer to as the `microcausal frame', and describe in Section~\ref{Section: frame}. We carry out a canonical quantisation of the theory in Section~\ref{Section: quantisation}, showing that our results extend to the quantum domain, and comment on the implications of our results for algebraic QFT. In particular, we explain how the vacuum state and net of algebras depend on the choice of microcausal frame. Finally, we conclude in Section~\ref{Section: conclusion} with a recounting of the general lessons to be learned from our example, as well as a brief outlook on how our results may extend to non-Abelian gauge theories and gravity.

\section{Boundary conditions in scalar QED}
\label{Section: boundary conditions}

Consider a theory consisting of a Maxwell gauge potential $A$ coupled to a charged complex scalar $\psi$ of charge $q$. The action is
\begin{equation}
  S = \frac12\int_{\mathcal{M}} \qty(-F\wedge \hodge F + \Dd{\psi}\wedge\hodge\overline{\Dd{\psi}} + \hodge V(\abs{\psi}^2)) + S_{\partial}\,,
\end{equation}
where $\hodge$ is the Hodge star operator associated with spacetime $\mathcal{M}$, $V$ is a potential for the scalar, and as usual $F=\dd{A}$ and $\Dd{\psi} = \dd{\psi} - iq A\psi$. $S_{\partial}$ is a boundary term which we keep general for now. Maxwell gauge transformations are of the form
\begin{equation}
  A \to A - \dd{\Lambda}, \qquad \psi \to e^{-iq\Lambda}\psi,
  \label{Equation: gauge transformation}
\end{equation}
where $\Lambda$ is any real function on $\mathcal{M}$. The variation of the action may then be written
\begin{equation}
  \delta S = -\int_{\mathcal{M}}\qty[\delta A\wedge\qty(\dd{\hodge F}+\hodge j)
  +\Re\qty(\delta\overline{\psi}\,\qty(\Dd{\hodge\Dd{\psi}} -\hodge\psi V'(\abs{\psi}^2)))]
  + \int_{\partial\mathcal{M}} \theta + \delta S_{\partial},
\end{equation}
where
\begin{equation}\label{eq:theta}
  j = q\Im\big(\overline{\psi}\Dd{\psi}\big),\qquad
  \theta=-\delta A \wedge \hodge F + \Re\qty(\delta\psi \hodge \overline{\Dd{\psi}}).
\end{equation}

In what follows, we will assume that as $\abs{\psi}^2\to 0$ the potential $V(\abs{\psi}^2)$ diverges sufficiently quickly that we always have $\psi\ne0$ when the fields are on-shell or approximately on-shell. The reason for making this assumption is that it leads to some technical simplifications related to defining dressed observables.\footnote{A similar assumption was made in \cite{Vanrietvelde:2018dit} in a mechanical context to rule out certain pathological configurations that complicated the construction of quantum reference frames and relational observables. }

\begin{figure}
  \centering
  \begin{tikzpicture}[scale=0.8]
    \fill[blue!20] (0,2.2) -- (0,6) arc (180:0:2 and 1) -- (4,2.2) arc (360:180:2 and 1);
    \draw[thick,blue!60!black] (0,2.2) -- (0,6) arc (180:0:2 and 1) -- (4,2.2) arc (360:180:2 and 1);
    \draw[thick,blue!60!black] (0,6) arc (-180:0:2 and 1);
    \draw[thick,dashed,blue!60!black] (0,2.2) arc (180:0:2 and 1);

    \node at (2,4.1) {\Large$\mathcal{M}$};
    \node[left] at (0,4.1) {\Large$\Gamma$\,\,};
    \node at (2,2.2) {\Large$\Sigma^-$};
    \node at (2,6) {\Large$\Sigma^+$};
  \end{tikzpicture}
  \caption{We consider spacetimes whose boundaries decompose into a timelike component $\Gamma$ and two spacelike components $\Sigma^\pm$.}
  \label{Figure: spacetime}
\end{figure}
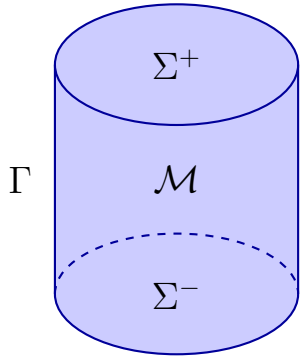

We will consider spacetimes whose boundaries decompose as $\partial\mathcal{M}=\Gamma\cup\Sigma^+\cup\Sigma^-$, where $\Sigma^-,\Sigma^+$ are spacelike components in the past/future respectively, and $\Gamma$ is a timelike component. This is depicted in Figure~\ref{Figure: spacetime}. For example, $\Gamma$ could be $i_0$ in asymptotically flat spacetimes, the global boundary in asymptotically anti-de Sitter spacetime, or simply a finite boundary delimiting a subregion \cite{Harlow:2019yfa,Geiller:2019bti,Carrozza:2021gju,Araujo-Regado:2024dpr}. In the conclusion, we will comment more on the case where $\Gamma$ is an asymptotic boundary.

\subsection{Large gauge transformations and boundary conditions}

The variational principle involves fixing initial/final data and/or conjugate momenta for the fields on $\Sigma^\pm$, and solving $\delta S=0$ for the configuration of the fields in the bulk $\mathcal{M}$. Here to simplify matters we fix $A|_{\Sigma^\pm},\psi|_{\Sigma^\pm}$, which means $\delta A|_{\Sigma^\pm}=0$ and $\delta\psi|_{\Sigma^\pm}=0$. One finds the equations of motion
\begin{equation}
  \dd{\hodge F}+\hodge j = 0, \qquad \Dd{\hodge\Dd{\psi}} -\hodge\psi V'(\abs{\psi}^2) = 0.
  \label{Equation: bulk equations of motion}
\end{equation}
For this procedure to be well-defined, we require two things. First, we have to impose boundary conditions such that $\int_{\partial\mathcal{M}} \theta + \delta S_{\partial}=0$ when the initial/final data are fixed and the bulk equations of motion are obeyed, and second, $\delta S=0$ should lead to a `physically unique' bulk field configuration, meaning that the values of any physical observables depending on $\phi$ are fully determined. In gauge theories, physical observables must take the same values on any two bulk field configurations related by a \emph{small} gauge transformation (meaning one whose action is trivial at the boundary $\partial\mathcal{M}$). Thus, the variational principle is well-defined only if it determines the field configuration up to small gauge transformations.

On the other hand, large gauge transformations, which are ones acting non-trivially at $\partial\mathcal{M}$, may lead to physically \emph{distinct} field configurations.\footnote{We refer to transformations \eqref{Equation: gauge transformation} with non-trivial support on $\partial\mathcal{M}$ as large gauge  also when $\partial\mathcal{M}$ is the boundary of a bulk region $\mathcal{M}$ in a global spacetime. However, in that case they do not correspond to gauge transformations of the global theory. Rather, they are \emph{physical} reorientations of an edge mode reference frame (for the small gauge group) with support in the complement of $\mathcal{M}$   \cite{Carrozza:2021gju}. Upon gauge fixing the edge mode, these reorientations take the form~\eqref{Equation: gauge transformation} on $\partial\mathcal{M}$ and thus appear as though they are gauge transformations \cite{Araujo-Regado:2024dpr}.} Since the equations of motion~\eqref{Equation: bulk equations of motion} are invariant under all gauge transformations, including large ones,  to preserve the theory, large gauge transformations must preserve the chosen boundary conditions. Whenever such transformations can be chosen to act `locally' on $\Gamma$ in-between $\Sigma^+$ and $\Sigma^-$ in the sense that they do not change initial/final data, they must be considered as redundancies. For example, this is the case with traditional Neumann boundary conditions fixing a profile of $*F|_\Gamma$, which are invariant under \emph{all} gauge transformations. In this case, all large gauge transformations are physical redundancies too \cite{Carrozza:2021gju}. By contrast, whenever the boundary-condition-preserving large gauge transformations do not act locally in the above sense, they change the physical field configuration. Clearly, this is only possible when the boundary conditions break large gauge invariance at least partially. In what follows, we will be interested in boundary conditions that are not preserved under all large gauge transformations, yielding an ensuing set of physical symmetries. This can be understood in terms of Goldstone modes associated with these symmetries, and their equations of motion on the timelike boundary $\Gamma$  \cite{Araujo-Regado:2024dpr}, as we shall now explain.

It is convenient to decompose the boundary conditions into two separate pieces: first, we impose boundary conditions which are preserved by the action of all large gauge transformations. Thus, unless we impose additional boundary conditions, all of these transformations are redundancies. Then, we impose the rest of the boundary conditions; the surviving large gauge transformations become physical when their time dependence no longer is arbitrary. In this paper, for the first piece, we will use ``Dirichlet up to large gauge transformations'' boundary conditions, meaning that
\begin{equation}
  A|_{\Gamma} = {A}_\bullet-\dd{\varphi},\qquad \psi|_\Gamma = \psi_\bullet e^{-iq\varphi},
  \label{Equation: Dirichlet up to gauge}
\end{equation}
where ${A}_\bullet,\psi_\bullet$ are some fixed boundary configurations, and $\varphi$ is a dynamical object which we call the Goldstone mode. It is small gauge-invariant and will thus be a \emph{physical} boundary mode in what follows. For simplicity, we will take ${A}_\bullet=0$ (so $F|_\Gamma=0$) and $\psi_\bullet\ne 0$. In terms of the field variations we have
\begin{equation}
  \delta A|_{\Gamma} = -\dd(\delta\varphi),\qquad \delta\psi|_\Gamma = -iq\,\delta\varphi\,\psi|_{\Gamma}.
  \label{Equation: variation of Dirichlet up to gauge}
\end{equation}
In other words, the field variation on $\Gamma$ is indistinguishable from that of some infinitesimal large gauge transformation. As such a variation leaves $F|_\Gamma=0$ invariant, and the configuration of $\varphi$ parametrises the ``classical vacuum degeneracy'' of the theory on $\Gamma$ -- this is the reason for the name Goldstone mode \cite{Araujo-Regado:2024dpr}.\footnote{To be clear, we could use other types of boundary condition here, which would lead to different kinds of Goldstone modes -- we only use Dirichlet for concreteness of the example.}

The next step is to give additional boundary conditions that determine $\varphi|_{\Gamma}$ when initial/final data is fixed. We will refer to these as `Goldstone boundary conditions'. Really, they should be viewed as the equations of motion for the Goldstone mode, which are not determined by the bulk theory; indeed, as the exact piece of $A|_\Gamma$, this mode is not radiative. There are infinitely many possible Goldstone boundary conditions, each leading to \emph{physically distinct} theories, because large-gauge-dependent physical observables (such as those dressed by Wilson lines extending to $\Gamma$) take on different characteristics in each theory. On the other hand, this distinction has no consequences for observables with support away from the boundary, because they must be large-gauge-invariant. So from the point of view of the bulk physics the theories are equivalent (indeed, the bulk equations of motion are always the same). Only large-gauge-dependent physical observables are sensitive to the differences.

Let us emphasise that the choice of a Goldstone boundary condition is distinct from a choice of gauge. Gauge fixing selects representatives within a single phase space, leaving the symplectic structure, Hamiltonian, and algebra of observables unchanged. By contrast, the Goldstone boundary conditions imposed here modify the phase space itself: they alter the symplectic form, change the Hamiltonian, and render large gauge transformations dynamical rather than redundant. Consequently, theories defined by different Goldstone boundary conditions are not related by gauge fixing, even though they share the same bulk equations of motion and agree on strictly local, completely gauge-invariant observables.

One particularly obvious thing to try would be to just set $\varphi=0$ on all of $\Gamma$, breaking \emph{all} large gauge transformations. Together with \eqref{Equation: Dirichlet up to gauge} this amounts to a traditional Dirichlet boundary condition \cite{Harlow:2019yfa,Carrozza:2021gju,Araujo-Regado:2024dpr}. However, this condition would in general be somewhat stronger than what we may want, as it would restrict the freedom in setting initial/final data \cite{Araujo-Regado:2024dpr}. In particular, this would end up putting unnecessary restrictions on the allowed configurations of $A|_{\Sigma^\pm}$ (assuming the fields are continuous at $\partial\Sigma^\pm$).\footnote{A general class of less restrictive boundary conditions, encompassing ``Dirichlet up to gauge'' and certain Goldstone boundary conditions --- together termed \emph{soft boundary conditions} --- was recently introduced in \cite{Araujo-Regado:2024dpr}. These admit an infinite-dimensional subgroup of (time-independent) large gauge transformations  and also encompass the earlier proposed ``dynamical edge mode conditions'' of \cite{Ball:2024hqe}.}

Let us now give two examples of Goldstone boundary conditions which fix $\varphi|_{\Gamma}$ when initial/final data are fixed, but which do not restrict the possible configurations of the initial/final data.
We use these examples as a warmup. From Section~\ref{Section: non-local} onward, we will focus on a qualitatively different kind of boundary condition for the second step -- one that is non-local.
\begin{eg}
  Consider temporal gauge\footnote{Recall this is a ``large gauge'' and so a physical boundary condition; it can also be phrased in terms of dressed observables \cite{Araujo-Regado:2024dpr}.} for the boundary gauge potential: $A_t|_{\Gamma}=0$, where $t$ is a timelike coordinate along $\Gamma$.\footnote{Together with \eqref{Equation: Dirichlet up to gauge} this yields  an example of \emph{soft Dirichlet boundary conditions} \cite{Araujo-Regado:2024dpr}.} In this case, fixing initial/final data (which may now be chosen arbitrarily, up to conservation to be mentioned shortly sets $\delta\varphi|_{\partial\Sigma^\pm}=0$, and integrating $\delta A_t|_\Gamma=\partial_t\delta\varphi=0$ yields ${\delta\varphi}|_\Gamma=0$, so $\varphi|_\Gamma$ is completely determined. Note that, for this choice of boundary condition, any large gauge transformations must act on the initial data on $\Sigma^-$ in the same way that they act on the final data on $\Sigma^+$ (if we map between points in $\Sigma^-$ and $\Sigma^+$ using the integral curves of $\partial_t$); i.e.\ they are \emph{time-independent}. The Goldstone mode is a function on $\Gamma$ that is conserved, i.e. remains constant in time.
  \end{eg}
  \begin{eg}
    Suppose we impose that the pullback of the gauge potential $A$ to $\Gamma$ is divergence-free with respect to the metric on $\Gamma$. This amounts to setting $\Box_\Gamma \varphi = 0$, where $\Box_\Gamma$ is the d'Alembertian of $\Gamma$. Fixing initial/final data (which again may be chosen arbitrarily) and solving $\Box_\Gamma\varphi=0$ yields  $\varphi|_\Gamma$ as required. Unlike the previous example, the remaining large gauge transformations can act independently on the initial/final data, since the full action of the large gauge transformation on $\Gamma$ is determined by a wave equation, which is second order.   Thus, with their time dependence now fixed, large gauge transformations are physical and there is no additional conservation arising from this boundary condition. The Goldstone mode may be thought of as a massless scalar field on $\Gamma$.
  \end{eg}

  These kinds of boundary conditions can be imposed in two slightly different ways. The first way is more direct: we only extremise the action over field configurations obeying the boundary conditions. The second way is to vary over a larger class of field configurations which do not necessarily obey the full boundary conditions, but choose the boundary action $S_{\partial{\mathcal{M}}}$ such that $\delta S=0$ leads to the rest of the boundary conditions \cite{Carrozza:2021gju}. From now on we will consider the second way for the partially large gauge symmetry breaking part of the boundary conditions.

  Here is how this can work in the above examples. In all the following cases (including the rest of this paper), we extremise the action over field configurations obeying Dirichlet up to gauge boundary conditions.
  \setcounter{eg}{0}
  \begin{eg}
    Let us introduce a new boundary field $\pi$ which couples to the gauge potential in the boundary action via
    \begin{equation}
      S_\partial = \int_{\Gamma} \varepsilon\, \pi A_t,
    \end{equation}
    where $\varepsilon$ is the induced volume form on $\Gamma$. Clearly, $\pi$ acts as a Lagrange multiplier enforcing the appropriate gauge boundary condition $A_t|_\Gamma=0$. The equation of motion associated with $\varphi$ is $\partial_t\pi=0$ --- so on-shell $\pi$ is constant in time.
  \end{eg}
  \begin{eg}
    To get the Goldstone mode to behave like a massless scalar field on the boundary, we can simply introduce a Klein-Gordon action for $\varphi$. In terms of the gauge potential, this is
    \begin{equation}
      S_\partial = -\frac12\int_\Gamma\varepsilon A^{||}\cdot A^{||},
    \end{equation}
    where $A^{||}=A - (n\cdot A)n$ is the tangential component of $A$ to $\Gamma$.
  \end{eg}

  \subsection{Phase space and symplectic form}

  Generally speaking, distinct choices of boundary conditions for a given set of fields result in theories with different phase spaces. Let us now see how to construct these phase spaces, and how they relate to the choice of Goldstone boundary condition. We will use the covariant phase space formalism \cite{Crnkovic:1987tz,Lee:1990nz,Iyer:1994ys,Khavkine:2014kya,Harlow:2019yfa,Gieres:2021ekc}.

  First, we consider the set $\mathcal{S}$ of field configurations on $\mathcal{M}$ which solve the equations of motion, and which obey the boundary conditions. This is the so-called `space of solutions'. We then define the phase space as $\mathcal{P}=\mathcal{S}/\!\sim$, where two field configurations in $\mathcal{S}$ are considered equivalent under $\sim$ if they are related by a small gauge transformation.

  From this description it is clear that $\mathcal{P}$ depends on the Goldstone boundary condition that constrains $\varphi$, as this is part of the definition of $\mathcal{S}$. However, since we are interested in Poisson brackets (and microcausality specifically), we need to go further than just constructing the space $\mathcal{P}$ --- we also need to construct the symplectic form $\Omega$ on $\mathcal{P}$. It too will depend on the boundary conditions.

  In general, the way to construct $\Omega$ is as follows. First, we take the field space exterior derivative (denoted $\delta$) of the on-shell action $S$ (which we view as a function on $\mathcal{S}$). We write this as a sum of two terms $\delta S = \Theta^+ - \Theta^-$, where $\Theta^-,\Theta^+$ only depend on variations of initial/final data on the surfaces $\Sigma^-,\Sigma^+$ respectively, but are otherwise schematically similar. Then, we define the `presymplectic form' $\Omega_*$ on $\mathcal{S}$ as the exterior derivative of either $\Theta^-$ or $\Theta^+$:
  \begin{equation}
    \Omega_* = \delta\Theta^- = \delta\Theta^+
  \end{equation}
  (it does not matter which because $\delta\Theta^+-\delta\Theta^- = \delta^2 S = 0$). If we have done this correctly, genuine (i.e.\ small) gauge transformations will correspond exactly with the degenerate directions of $\Omega_*$, which allows us to define $\Omega$ as the unique 2-form on $\mathcal{P}$ whose pullback to $\mathcal{S}$ is $\Omega_*$.

  Let us see how this works for the case at hand. The variation of the action, on-shell of the bulk equations of motion, is
  \begin{equation}
    \delta S = \int_{\Sigma^+}\theta - \int_{\Sigma^-}\theta + \int_{\Gamma}\theta + \delta S_\partial
    \label{Equation: action variation J 0}\,.
  \end{equation}
  On the timelike boundary, we can use~\eqref{eq:theta},~\eqref{Equation: Dirichlet up to gauge} and the equations of motion to write
  \begin{equation}
    \theta|_{\Gamma} = \qty\big[\dd{\delta\varphi} \wedge \hodge F -\delta\varphi \hodge j]_{\Gamma} = \dd(\delta \varphi\hodge F)|_{\Gamma}.
  \end{equation}
  As explained in the previous subsection, the boundary action is chosen to have its own equations of motion which enforce extra boundary conditions. On-shell of these boundary equations of motion, the variation of the boundary action reduces to some boundary terms:
  \begin{equation}
    \delta S_\partial = C_{\partial\Sigma^+} - C_{\partial\Sigma^+}.
  \end{equation}
  Here $C_{\partial\Sigma^\pm}$ are 1-forms in field space that depend on the fields at $\partial\Sigma^\pm$ respectively. Collecting all the terms, we obtain $\delta S = \Theta^+-\Theta^-$ where
  \begin{equation}
    \Theta^\pm = \int_{\Sigma^\pm}\theta - \int_{\partial\Sigma^\pm}\delta \varphi\hodge F + C_{\partial\Sigma^\pm}.
  \end{equation}
  As a result, the presymplectic form is
  \begin{equation}
    \Omega_* = \delta\Theta^\pm = \int_{\Sigma^\pm}\qty( \delta A \wedge \hodge \delta F - \Re\qty(\delta\psi \hodge \delta\qty(\overline{\mathrm{D}\psi}))) + \int_{\partial\Sigma^\pm}\delta \varphi \hodge  \delta F + \delta C_{\partial\Sigma^\pm}.
    \label{Equation: naive presymplectic form}
  \end{equation}
  (We use here the convention in which the field space wedge product is implicit, so e.g.\ $\delta f\, \delta g$ is the wedge product of $\delta f$ and $\delta g$.)

  Using the ``Dirichlet up to large gauge transformations'' boundary conditions \eqref{Equation: Dirichlet up to gauge}, the equations of motion, and Stokes' theorem, the presymplectic form may further be rewritten as
  \begin{equation}
    \Omega_* =  \int_{\Sigma^\pm}\qty( \delta \tilde{A} \wedge \hodge \delta F - \Re\qty(\delta\psi \hodge \delta\qty(\overline{\mathrm{D}\psi}))) + \delta C_{\partial\Sigma^\pm},
    \label{Equation: naive presymplectic form 2}
  \end{equation}
  where $\tilde{A}:=A+\dd{\varphi}$, so that $\tilde{A}|_\Gamma=0$ (where we have extended $\varphi$ into the bulk in such a manner that $\tilde{A}$ is large gauge-invariant). It may be checked that the scalar field contribution vanishes on the corner $\partial\Sigma^\pm$ thanks to the boundary conditions. The advantage of this form of $\Omega_*$ is that it makes manifest that the only non-trivial corner term is $\delta C_{\partial\Sigma^\pm}$. Thus, the $\Sigma^\pm$-integral is large gauge-invariant and only $C_{\partial\Sigma^\pm}$ contributes to the charges.

  It is helpful to see how this works in the previously considered examples.
  \setcounter{eg}{0}
  \begin{eg}
    After introducing the boundary Lagrange multiplier $\pi$, the resulting presymplectic form is
    \begin{equation}
      \Omega_* = \delta\Theta^\pm = \int_{\Sigma^\pm} \qty\Big(\delta A \wedge \delta(\hodge F) - \Re\qty\big(\delta\psi \hodge \delta(\overline{\mathrm{D}\psi}))) + \int_{\partial\Sigma^\pm}\delta \varphi\qty\Big(\hodge  \delta F+\epsilon\delta \pi),
    \end{equation}
    where $\epsilon$ is the induced volume form on $\partial\Sigma^\pm$. With this we may roughly speaking identify $\pi$ as the conjugate momentum of the Goldstone mode $\varphi$, and the generator of large gauge transformations.
  \end{eg}
  \begin{eg}
    $\varphi$ behaves as a scalar field on the boundary, and one obtains the appropriate resulting presymplectic form
    \begin{equation}
      \Omega_* = \delta\Theta^\pm = \int_{\Sigma^\pm} \qty\Big(\delta A \wedge \delta(\hodge F) - \Re\qty\big(\delta\psi *\delta(\overline{\mathrm{D}\psi}))) + \int_{\partial\Sigma^\pm}\delta\varphi\qty\Big(\hodge \delta F+\epsilon\delta (m\cdot A)),
    \end{equation}
    where $m$ is the future-directed unit normal to $\partial\Sigma^\pm$ as a submanifold of $\Gamma$. The contribution from the boundary action is the symplectic form of a scalar field.
  \end{eg}

  In both of these examples, the boundary conditions were enforced by some \emph{local} differential equations involving the fields on the boundary. The same is almost always true of boundary conditions that are used across the literature --- but, in fact, this locality is unnecessary. Indeed, starting in the next section we will instead consider boundary conditions which break large gauge invariance \emph{non-locally}. These boundary conditions will essentially be those of Example 2, but ``pulled into'' into the bulk using dressing Wilson lines.

  \section{Non-local boundary condition}
  \label{Section: non-local}

  Let us now describe the non-local Goldstone boundary conditions that we will study for most of the remainder of the paper. Specifically, our focus will lie on how they impact on microcausality of small gauge-invariant observables. We will henceforth also assume that the ``Dirichlet up to large gauge'' boundary conditions are imposed, so that we have a complete set of boundary conditions.

\begin{figure}
    \centering
    \begin{tikzpicture}[scale=1.4]
      \fill[blue!20] (0,2.2) -- (0,6) arc (180:0:2 and 0.6) -- (4,2.2) arc (360:180:2 and 0.6);

      \fill[red!40!blue,opacity=0.15] (1.3,2.2) -- (1.3,6) arc (180:0:1 and 0.3) -- (3.3,2.2) arc (0:180:1 and 0.3);
      \fill[red!40!blue,opacity=0.15] (1.3,2.2) -- (1.3,6) arc (180:360:1 and 0.3) -- (3.3,2.2) arc (360:180:1 and 0.3);

      \draw[thick,blue!60!black] (0,2.2) -- (0,6) arc (180:0:2 and 0.6) -- (4,2.2) arc (360:180:2 and 0.6);
      \draw[thick,blue!60!black] (0,6) arc (-180:0:2 and 0.6);
      \draw[thick,dashed,blue!60!black] (0,2.2) arc (180:0:2 and 0.6);

      \draw[red!40!blue] (2.3,6) ellipse (1 and 0.3);
      \draw[red!40!blue] (3.3,6) -- (3.3,2.2) arc (360:180:1 and 0.3) -- (1.3,2.2) -- (1.3,6);
      \draw[dashed,red!40!blue] (1.3,2.2) arc (180:0:1 and 0.3);

      \node[red!20!black] at (2.6,4.1) {\large$\mathcal{N}$};
      \node[red!20!black] at (1,2.1) {$K^-$};
      \node[red!20!black] at (1,6.1) {$K^+$};

      \begin{scope}[decoration={
            markings,
            mark=at position 0.5 with {\arrow{<}}
        }]
        \draw[red!80!black,thick,postaction={decorate}] (0.3,4.6) .. controls (0.8,4.6) and (1.2,4.3) .. (1.6,4.3);
        \draw[red!80!black,thick,postaction={decorate}] (0.4,3.2) .. controls (0.8,3.2) and (1.2,3.6) .. (1.7,3.6);
      \end{scope}
      \fill[red!40!black] (0.3,4.6) circle (0.05);
      \fill[red!40!black] (0.4,3.2) circle (0.05);
      \fill[red!40!black] (1.6,4.3) circle (0.05);
      \fill[red!40!black] (1.7,3.6) circle (0.05);
      \node[above] at (0.4,4.6) {\footnotesize$f(y')$};
      \node[above] at (0.3,3.25) {\footnotesize$f(y)$};
      \node[above] at (1.6,4.3) {\footnotesize$y'$};
      \node[above] at (1.7,3.6) {\footnotesize$y$};
      \node[red!50!black,below] at (0.85,4.45) {\footnotesize$\gamma(y')$};
      \node[red!50!black,below] at (1,3.35) {\footnotesize$\gamma(y)$};

      \node[left] at (0,4.1) {\Large$\Gamma$\,\,};
    \end{tikzpicture}
    \caption{The non-local boundary conditions studied in this paper involve the choice of a timelike codimension 1 surface $\mathcal{N}\subset\mathcal{M}$, a bijection $f:\mathcal{N}\to\Gamma$ that preserves causal ordering, and for each $y\in\mathcal{N}$ a curve $\gamma(y)$ connecting $y$ to $f(y)$.}
    \label{Figure: wilson lines}
  \end{figure}
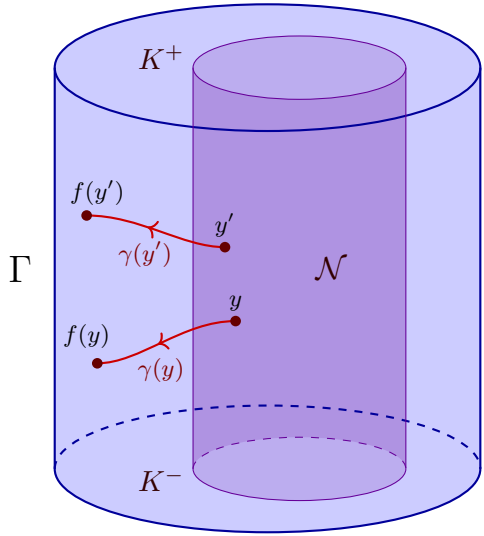

  The first thing these boundary conditions involve is the choice of a timelike codimension 1 surface $\mathcal{N}$ in the bulk whose boundary decomposes as $\partial\mathcal{N} = K^-\cup K^+$ with $K^\pm\subset\Sigma^\pm$, such that we can pick a diffeomorphism $f:\mathcal{N}\to \Gamma$ that preserves causal ordering. In other words, $f(y)$ is in the causal future of $f(y')$ iff $y$ is in the causal future of $y'$. For each $y\in\mathcal{N}$, let $\gamma(y)$ be a curve from $y$ to $f(y)$. This is shown in Figure~\ref{Figure: wilson lines}.

  By integrating $A$ along the curves $\gamma(y)$ we get a set of Wilson lines which can be used to dress observables on $\mathcal{N}$. For our construction let us single out a particular family of such observables:
  \begin{equation}
    \alpha(y) = \frac1q\arg(\Psi(y)),
    \label{Equation: dressed phase}
  \end{equation}
  where $\Psi(y) = \psi(y) \exp\big(iq\int_{\gamma(y)}A)$ is the scalar field on $\mathcal{N}$ dressed by the Wilson line. This observable is invariant under small gauge transformations. Here we are just picking some fixed branch of the argument function. For a complete definition of $\alpha(y)$, we set $\arg(0)=0$ -- but note that when the field configuration is on-shell or almost on-shell we will not need this, because then $\Psi(y)\ne 0$ (since we are assuming that the potential $V$ diverges at $\psi = 0$ sufficiently quickly that we always have $\psi\ne 0$ on-shell).

  We may view $\alpha$ as a function on $\mathcal{N}$. The Goldstone boundary condition we will use is $\Box_{\mathcal{N}}\alpha=0$, where $\Box_{\mathcal{N}}$ is the d'Alembertian of $\mathcal{N}$. In other words, $\alpha$ is required to obey the massless wave equation on $\mathcal{N}$, i.e.\ it behaves like a massless scalar. Even though $\alpha$ is non-locally constructed from the fields in the bulk, this really is a genuine \emph{boundary} condition, because $\Box_{\mathcal{N}}\alpha=0$ can be satisfied by constraining the $\varphi$ that appears in~\eqref{Equation: Dirichlet up to gauge}.

  In particular, under a large gauge transformation~\eqref{Equation: gauge transformation} we have $\alpha\to \alpha - \Lambda_f$, where $\Lambda_f=f^*\Lambda=\Lambda\circ f$. If we start off in a field configuration where $\Box_{\mathcal{N}}\alpha \ne 0$, we can just solve $\Box_{\mathcal{N}}\Lambda_f=\Box_{\mathcal{N}}\alpha$ to find a (field-dependent) gauge transformation that gets us to the right large gauge (this gauge transformation is unique up to a harmonic function).\footnote{Discontinuities in the argument function imply that the solution for $\Lambda_f$ will generically also be discontinuous. These discontinuities involve jumps of $2\pi/q$, so $e^{iq\Lambda_f}$ will be continuous.} As is well known, if $\Lambda_f$ satisfies this equation, then $\dd{\Lambda_f}$ is entirely determined (on all of $\mathcal{N}$) by initial/final data for $\dd{\Lambda_f}$ on any two Cauchy slices of $\mathcal{N}$. Using $\dd{\Lambda}|_\Gamma=f_*(\dd{\Lambda_f}|_{\mathcal{N}})$, and the fact that $f$ preserves causal ordering, we may conclude that the action of a large gauge transformation on $\Gamma$ is determined by its action on initial/final data at $\partial\Sigma^\pm$. As explained in the previous section, this property ensures that $\Box_{\mathcal{N}}\alpha=0$ is a well-defined Goldstone boundary condition. We may view these non-local boundary conditions as those of Example 2 ``pulled onto $\mathcal{N}$'' using the Wilson lines $\gamma$.

  For later purpose, we note that we can regard the family of Wilson lines associated with $\{\gamma(y)\}_{y\in\mathcal{N}}$ as a dynamical reference frame for \emph{small} gauge transformations on $\mathcal{N}$ \cite{Carrozza:2021gju,Araujo-Regado:2024dpr}. Indeed, these Wilson lines transform gauge covariantly on $\mathcal{N}$, take value in $\rm{U}(1)$, and, as we will see more in later sections, they can be used to dress non-invariant degrees of freedom on $\mathcal{N}$, describing them relative to the frame. From the point of view of the subregion delimited by $\mathcal{N}$, the small gauge frame constituted by the Wilson lines amounts to an \emph{extrinsic} edge mode frame, being built by composite fields in the complement of this subregion \cite{Araujo-Regado:2024dpr,Araujo-Regado:2025ejs}. By contrast, one may also consider an \emph{intrinsic} edge mode frame for small gauge transformations on $\mathcal{N}$, e.g.\ given by the scalar field $\psi(y)$, $y\in\mathcal{N}$, which too transforms gauge covariantly (though does not take value in the gauge group) and is built entirely locally from within the subregion. Now the small gauge-invariant edge field $\alpha(y)$ really is the (phase of the) relational observable describing the intrinsic relative to the extrinsic edge mode frame. In \cite{Araujo-Regado:2024dpr,Araujo-Regado:2025ejs}, this object was called a subregional Goldstone mode (here distinguishing it from our global Goldstone mode $\varphi$ on $\Gamma$). As we have just seen, $\alpha$ transforms covariantly under \emph{large} gauge transformations and as such it constitutes (the phase of) a dynamical reference frame field for those. In particular, its configurations parametrise large gauge symmetry breaking. We will repeatedly come back to this throughout our discussion.

  \subsection{Action and variational principle}

  Since $\alpha$ behaves as a massless scalar field on $\mathcal{N}$, it is clear what we should use as an action:
  \begin{equation}
    S = \frac12\int_{\mathcal{M}} \qty(-F\wedge \hodge F + \Dd{\psi}\wedge\hodge\overline{\Dd{\psi}} + \hodge V(\abs{\psi}^2)) + \frac12\int_{\mathcal{N}}\dd{\alpha}\wedge\Hodge\dd{\alpha}\,,\label{nonlocalaction}
  \end{equation}
  where $\star$ is the Hodge star operator associated with $\mathcal{N}$. This is the ordinary bulk action of scalar QED, augmented by the Klein-Gordon action for $\alpha$ on $\mathcal{N}$, which constitutes the ``boundary term''. This term is not an integral over $\Gamma$, unlike the local boundary terms given in the examples in the previous section. However, as will be shown below, its only effect is to implement the Goldstone boundary condition we are considering here, and for this reason one may still call it a boundary term. It is really quite a non-local expression, involving three integrations (one over $\mathcal{N}$, and two over the curves $\gamma(y)$). It provides the equation of motion for a composite field and only has a local appearance due to the fact that we have written it in terms of $\alpha$. This also means that if we write it in the gauge in which all the Wilson lines become fixed to the identity it appears local. Later we will see that this defines a relational notion of locality relative to a gauge frame constituted by the system of Wilson lines.
  Despite this non-locality, the action is as good as any other that yields the Maxwell and Klein-Gordon equations in the bulk in the sense that it provides a consistent set of equations -- including the boundary conditions -- that can be solved for arbitrary on-shell configurations of the large-gauge-invariant degrees of freedom.

  Let us verify that this boundary term enforces the right boundary condition. To this end, we can at first consider the extremisation of the action under the restricted class of field variations given by large gauge transformations (under which the standard bulk piece is invariant). Let us from now on use $\delta_\Lambda\phi=(\delta_\Lambda A,\delta_\Lambda\psi)$ to denote a gauge transformation with parameter $\Lambda$, as in~\eqref{Equation: gauge transformation}. Then, using $\delta_\Lambda\alpha=-\Lambda_f$, we have
  \begin{equation}
    \delta_\Lambda S = \int_{\mathcal{N}} \Lambda_f\dd{\Hodge\dd{\alpha}} - \int_{K^+}\Lambda_f\Hodge\dd{\alpha} + \int_{K^+}\Lambda_f\Hodge\dd{\alpha}.
  \end{equation}
  When initial/final data are fixed, the latter two terms vanish\footnote{Note that the causal order preserving property of $f$ implies $f(K^\pm)=\partial\Sigma^\pm$, so $\Lambda|_{\partial\Sigma^\pm}=0$ implies $\Lambda_f|_{K^\pm}=0$.}, and we are left with the first term. For $\delta_\Lambda S$ to vanish for arbitrary $\Lambda$ that vanishes on $\partial\Sigma^\pm$, we then must have $\dd{\Hodge\dd{\alpha}}=0$, which is equivalent to the required condition $\Box_\mathcal{N}\alpha=0$. We will see shortly that extremisation under arbitrary field variations will not introduce additional constraints on $\alpha$. As argued previously, $\Box_\mathcal{N}\alpha=0$ ensures a complete breaking of large gauge symmetry by any solution. Indeed, any permissible large gauge transformation that does not modify initial/final data must now obey $\Box_\mathcal{N}\Lambda_f=0$ and $\Lambda_f|_{K^\pm}=0$, leaving $\Lambda|_\Gamma=0$ as the only solution.

  Our next task is to check whether the Klein-Gordon action for $\alpha$ interferes in any undesired way with the bulk equations of motion~\eqref{Equation: bulk equations of motion}. Given the non-local nature of this boundary term, it may not be particularly obvious that this interference is absent. To show that everything is fine, let us now consider how the action changes under a general field variation. We have:
  \begin{multline}
    \delta S = -\int_{\Gamma} \Lambda(\dd{\hodge F}+\hodge j) - \int_{\mathcal{M}}\bigg(\delta A\wedge\qty(\dd{\hodge F}+\hodge j +\hodge  \int_{y\in\mathcal{N}}\fdv{\alpha(y)}{A}\dd{\Hodge\dd{\alpha}}|_y) \\
    + \Re\qty(\overline{\delta\psi}\,\qty( \Dd{\hodge\Dd{\psi}}-*\psi V'(\abs{\psi}^2)+\hodge  \int_{y\in\mathcal{N}}\fdv{\alpha(y)}{\overline\psi}\dd{\Hodge\dd{\alpha}}|_y))\bigg)\\
    + \int_{\Sigma^+} \qty(-\delta A \wedge \hodge F + \Re\qty(\delta\psi \hodge \overline{\mathrm{D}\psi}))
    - \int_{\Sigma^-} \qty(-\delta A \wedge \hodge F + \Re\qty(\delta\psi \hodge \overline{\mathrm{D}\psi})) \\
    - \int_{\partial\Sigma^+} \Lambda \hodge F + \int_{\partial\Sigma^-} \Lambda \hodge F + \int_{K^+}\delta\alpha \Hodge\dd{\alpha} - \int_{K^-}\delta\alpha \Hodge\dd{\alpha}\,,
    \label{Equation: action variation}
  \end{multline}
  where we have used the ``Dirichlet up to large gauge transformations'' boundary condition~\eqref{Equation: Dirichlet up to gauge}, and
  \begin{align}
    \fdv{\alpha(y)}{A_\mu(x)} &= \int_{x'\in\gamma(y)}\dd{x'^\mu}\delta^{(D)}(x-x'),\\
    \fdv{\alpha(y)}{\overline\psi(x)} &= \frac{i}{2q\overline\psi(y)}\delta^{(D)}(x-y)\,,
  \end{align}
  are the functional derivatives of $\alpha$.

  The latter two lines of~\eqref{Equation: action variation} are past/future boundary terms, so we would like to set them to zero by fixing initial/final data. A minor technicality presents itself here. So far, initial/final data has for us meant $A|_{\Sigma^\pm},\psi|_{\Sigma^\pm}$. However, fixing these quantities does not in general imply $\delta\alpha|_{K^\pm}=0$, because for $y\in K^\pm$, $\alpha(y)$ depends on $A$ at all points along the curve $\gamma(y)$, and we have not demanded $\gamma(y)\subset\Sigma^\pm$. However, this is nothing to be worried about. There are two ways to address it:
  \begin{itemize}
    \item The most straightforward thing to do would be to simply pick the curves $\gamma(y)$ such that $\gamma(y)\subset\Sigma^\pm$ whenever $y\in K^\pm$. Then fixing $A|_{\Sigma^\pm},\psi|_{\Sigma^\pm}$ would indeed set $\delta\alpha|_{K^\pm}=0$.
    \item More generally, we can redefine what we mean by initial/final data. In particular, we can define initial/final data as: $(A|_{\Sigma^\pm},\psi|_{\Sigma^\pm})$ up to gauge transformations (including large ones), \emph{and} $\alpha|_{K^\pm}$. Fixing this data is enough to make the latter two terms in~\eqref{Equation: action variation} vanish. Also, this is a complete set of initial/final data, in the sense that fixing it determines a single solution of the equations of motion (up to small gauge transformations), and thus a single physical state, and that any physical state can be determined in this way.
  \end{itemize}
  In either case, fixing initial/final data causes the latter two lines of~\eqref{Equation: action variation} to vanish.

  We are left with the first two lines of~\eqref{Equation: action variation}. Ignoring the integral over $\Gamma$ for now, we have the equations of motion
  \begin{align}
    \dd{\hodge F}+\hodge j&=-\hodge  \int_{y\in\mathcal{N}}\fdv{\alpha(y)}{A}\dd{\Hodge\dd{\alpha}}|_y, \label{Equation: A eom}\\
    \Dd{\hodge\Dd{\psi}}-\hodge \psi V'(\abs{\psi}^2)&=-\hodge  \int_{y\in\mathcal{N}}\fdv{\alpha(y)}{\overline\psi}\dd{\Hodge\dd{\alpha}}|_y. \label{Equation: psi eom}
  \end{align}
  These equations may at first sight seem complicated. But since we know that on-shell field configurations must obey $\Box_{\mathcal{N}}\alpha=0$, there is a clear strategy for solving them. We just solve the ordinary equations of motion for a Maxwell field coupled to a charged scalar~\eqref{Equation: bulk equations of motion}. Those equations of motion are invariant under all gauge transformations, including large ones, so we can solve them while simultaneously ensuring that $\Box_{\mathcal{N}}\alpha=0$ holds, using the large gauge transformation described above. Since $\Box_{\mathcal{N}}\alpha=0$ and~\eqref{Equation: bulk equations of motion} imply~(\ref{Equation: A eom}--\ref{Equation: psi eom}), we have thus succeeded in extremising the action.  Actually, there are no solutions to~(\ref{Equation: A eom}--\ref{Equation: psi eom}) other than those which can be found in this way. This is because~(\ref{Equation: A eom}--\ref{Equation: psi eom}) imply $\Box_{\mathcal{N}}\alpha=0$, which can be confirmed by (for example) taking an exterior derivative of~\eqref{Equation: A eom} and substituting in~\eqref{Equation: psi eom} and its complex conjugate.

  Finally, if the field configuration obeys the field equations~\eqref{Equation: bulk equations of motion}, then it is clear that the integral over $\Gamma$ in~\eqref{Equation: action variation} vanishes. Thus, we have succeeded in finding the set of field configurations which extremise the action: they are the solutions to the ordinary equations of motion~\eqref{Equation: bulk equations of motion} in a large gauge satisfying $\Box_{\mathcal{N}}\alpha=0$. As claimed, the Klein-Gordon action for $\alpha$ does not interfere with the bulk equations of motion.

  \subsection{Phase space structure}

  Let us now extract the presymplectic form of the theory we are describing. When the equations of motion are obeyed, the variation of the action~\eqref{Equation: action variation} may be written in the form $\delta S = \Theta^+-\Theta^-$, where
  \begin{equation}
    \Theta^\pm = \int_{\Sigma^\pm} \qty(-\delta A \wedge \hodge F + \Re\qty(\delta\psi \hodge \overline{\mathrm{D}\psi}))
    - \int_{\partial\Sigma^\pm} \delta\varphi \hodge F + \int_{K^\pm}\delta\alpha\Hodge\dd{\alpha}.
  \end{equation}
  The presymplectic form is then given by the field space exterior derivative of either of these quantities:
  \begin{equation}
    \Omega_* = \delta\Theta^\pm = \int_{\Sigma} \qty(\delta A \wedge \hodge \delta F - \Re\qty(\delta\psi \hodge \delta(\overline{\mathrm{D}\psi})))
    + \int_{\partial\Sigma} \delta\varphi \hodge \delta F - \int_{K}\delta\alpha\Hodge\dd{\delta\alpha}.
  \end{equation}
  We have dropped the ${}^\pm$ superscripts because now it does not matter which initial data surface $\Sigma$ we evaluate this on (we generally set $K=\Sigma\cap\mathcal{N}$), since the presymplectic form is conserved by $\delta^2 S=0$. Repeating the arguments leading to \eqref{Equation: naive presymplectic form 2}, $\Omega_*$ may be rewritten in the form
  \begin{equation}
    \Omega_*  = \int_{\Sigma} \qty(\delta \tilde{A} \wedge \hodge \delta F - \Re\qty(\delta\psi \hodge \delta(\overline{\mathrm{D}\psi})))
    - \int_{K}\delta\alpha\Hodge\dd{\delta\alpha},\label{omegastar2}
  \end{equation}
  underscoring that our non-local Goldstone boundary conditions have eliminated the asymptotic corner term on $\partial\Sigma$, replacing it with a bulk corner term on $K$.

  What does this presymplectic form tell us? First, it is easy to verify that small gauge transformations are degenerate directions of $\Omega_*$. Thus, one can consistently gauge reduce and get a symplectic form $\Omega$ on the space of solutions modulo small gauge transformations. On the other hand, large gauge transformations are not degenerate directions. Indeed, if we contract a large gauge transformation into the presymplectic form, we find that the integrals over $\Sigma$ in $\Omega_*$ vanish as we discussed below Eq.~\eqref{Equation: naive presymplectic form 2}, so that
  \begin{equation}
    \Omega_*(\delta_\Lambda\phi) = \int_K\epsilon\qty(\Lambda_f\delta\dot\alpha + \dot\Lambda_f \delta \alpha),
  \end{equation}
  where $\cdot$ denotes a derivative along the unit future-directed normal to $K$ as a submanifold of $\mathcal{N}$, and $\epsilon$ is the volume form on $K$. Thus, the charges associated with large gauge transformations take a form we are perhaps not used to, and moreover there are two distinct families of them:
  \begin{equation}
    Q[F] = \int_K \epsilon F \alpha, \qquad P[F] = \int_K \epsilon F \dot \alpha.
  \end{equation}
  Here $F$ is any real function on $K$, and $P[F]$ generates a large gauge transformation obeying $\Lambda_f|_K= F$, $\dot\Lambda_f|_K=0$, while $Q[F]$ generates a large gauge transformation obeying $\Lambda_f|_K=0$, $\dot\Lambda_f|_K=F$ (the action of the large gauge transformation on all of $\Gamma$ is then determined by $\Box_{\mathcal{N}}\alpha=\Box_{\mathcal{N}}\Lambda_f=0$). By computing how $Q[F'],P[F']$ change under the action of these gauge transformations for some other function $F'$, we can deduce that the algebra of these charges is
  \begin{equation}
    \pb{Q[F]}{Q[F']} = \pb{P[F]}{P[F']} = 0, \quad \pb{Q[F]}{P[F']} = -\int_K \epsilon F F'.
    \label{Equation: QP}
  \end{equation}
  It is worth noting that these charges are just smeared versions of $\alpha$ and its conjugate momentum. They do not form a representation of the Lie algebra of large gauge transformations, since that Lie algebra is Abelian. However, they do form a representation of a central extension of that algebra (in other words, they form a projective representation of the algebra), which is all that is required of these charges. Later, we will see that these large gauge transformations amount to reorientations of the large gauge frame field $\alpha$.

  Let us now obtain the Hamiltonian of the theory. In general, a time-independent Hamiltonian only exists if time evolution is a symmetry. For simplicity, we will restrict to this time-independent case, and in the present situation it suffices to assume that there is a unit norm timelike Killing vector field $\xi$ which is tangential to $\Gamma$ and $\mathcal{N}$, and such that if $y\in\mathcal{N}$ maps to $y'\in\mathcal{N}$ under the flow generated by $\xi$, then $\gamma(y)$ maps to $\gamma(y')$. It can then be shown that the field variation generated by $\xi$:
  \begin{equation}
    \delta_\xi A = \lie_\xi A = \iota_\xi F + \dd(\iota_\xi A), \qquad \delta_\xi \psi = \lie_\xi \psi = \xi(\psi),
    \label{Equation: time evolution}
  \end{equation}
  leads to
  \begin{equation}
    \delta_\xi\alpha=\xi(\alpha),
  \end{equation}
  and preserves the bulk equations of motion~\eqref{Equation: bulk equations of motion} and $\Box_{\mathcal{N}}\alpha=0$ (since time evolution preserves the metrics appearing in these equations), and the ``Dirichlet up to large gauge transformations'' boundary condition~\eqref{Equation: Dirichlet up to gauge} (since $\xi$ is tangent to $\Gamma$, we have $(\iota_\xi F)|_{\Gamma} = \iota_\xi(F|_{\Gamma})$, and $F|_{\Gamma}=0$ by the boundary conditions). Thus, time evolution is a genuine symmetry of the theory. Furthermore, contracting~\eqref{Equation: time evolution} into the presymplectic form yields
  \begin{equation}
    \Omega_*(\delta_\xi\phi) = \delta H,
  \end{equation}
  where the Hamiltonian is
  \begin{equation}
    H = \frac12\int_\Sigma \operatorname{vol}_\Sigma \qty(\abs{\vb{E}}^2 + \abs{\vb{B}}^2 + \abs*{\dot\psi}^2 + \abs{\mathrm{D}_\Sigma\psi}^2 + 2V(\abs{\psi}^2)) + \frac12\int_K\epsilon \qty(\dot\alpha^2 + \abs{\partial_K\alpha}^2)\,,
    \label{Equation: Hamiltonian}
  \end{equation}
  where $\operatorname{vol}_\Sigma$ is the volume form of $\Sigma$, we have chosen a $\Sigma$ which is normal to $\xi$, $\abs{E}$ and $\abs{B}$ are the electric and magnetic fields on $\Sigma$, $\cdot$ denotes a time derivative, $\mathrm{D}_\Sigma$ denotes the components of the covariant derivative along $\Sigma$, and $\partial_K$ is the gradient on $K$. We can see that $H$ is just the ordinary Hamiltonian of Maxwell-scalar theory, plus the Klein-Gordon energy of $\alpha$.

  The Hamiltonian is invariant under small gauge transformations, but notably not invariant under large gauge transformations. Indeed, we have the algebraic relations
  \begin{equation}
    \pb{Q[F]}{H} = P[F], \qquad \pb{P[F]}{H} = Q[\Delta_K F],
    \label{Equation: Hamiltonian QP}
  \end{equation}
  where $\Delta_K$ is the Laplacian of $K$. Therefore, in this theory, \emph{carrying out large gauge transformations usually takes energy}.

  In fact, as we shall see shortly, the only large gauge transformation which doesn't take energy is the one for which $\Lambda$ is constant, i.e.\ the global $U(1)$ symmetry.\footnote{This statement holds if $K$ only has one connected component. Let us for simplicity assume this is true.} Setting $\Lambda=1$, this symmetry is generated by $Q[1] = \int_K \epsilon \alpha$. Usually, the generator of the global $U(1)$ symmetry has the interpretation of measuring the total charge in spacetime. Although it may not be obvious from its form, $Q[1]$ has a similar interpretation.

  To see this, let us consider what happens when we deform the action by a background current, $S\to S[J] = S - \int_{\mathcal{M}}A\wedge\hodge J$, with $\dd{\hodge J}=0$. Then, following the same steps as above but with this deformed action, we find that the  boundary condition $\dd{\Hodge\dd{\alpha}}=0$ is deformed to
  \begin{equation}
    \dd{\Hodge\dd{\alpha}} + f^*(\hodge J|_\Gamma) = 0,
  \end{equation}
  while the bulk equations of motion take the expected form
  \begin{equation}
    \dd{\hodge F}+\hodge j +\hodge J = 0, \qquad \Dd{\hodge\Dd{\psi}} -\hodge\psi V'(\abs{\psi}^2) = 0.
  \end{equation}
  The difference in total charge due to the background current in between the initial/final surfaces $\Sigma^\pm$ may then be written
  \begin{equation}
    \int_{\Sigma^+}\hodge J-\int_{\Sigma}\hodge J = -\int_\Gamma \hodge J = -\int_{\mathcal{N}}f^*(\hodge J|_\Gamma) = \int_{\mathcal{N}}\dd{\Hodge\dd{\alpha}} = \underbrace{\int_{K^+}\Hodge\dd{\alpha}}_{=Q[1]|_{K^+}} - \underbrace{\int_{K^-}\Hodge\dd{\alpha}}_{ = Q[1]|_{K^-}}.
  \end{equation}
  Thus, the change in $Q[1]$ is exactly equal to the change in total charge due to the background current. When there is no background current we see that $Q[1]$ is a conserved charge, which is consistent with the global $U(1)$ symmetry having zero energy --- both statements are equivalent to $\pb{Q[1]}{H}=0$.

  \subsection{Motivations of some details}

  Let us now provide a few motivations for some choices we made when constructing the boundary conditions described above.

  First, the boundary conditions are based upon the field $\alpha$ on $\mathcal{N}$, which was defined~\eqref{Equation: dressed phase} as the argument of the dressed scalar field $\Psi$ divided by $q$. It may be tempting to remove the scalar field $\psi$ from this definition, and to just set $\alpha$ equal to the integral of $A$ along the curve $\gamma(y)$:
  \begin{equation}
    \alpha(y) \stackrel{?}{=} \int_{\gamma(y)}A.
  \end{equation}
  This appears to be a simpler construction, but it is unsatisfactory for our purposes. With such an $\alpha$ it would be possible to satisfy $\Box_{\mathcal{N}}\alpha=0$ with a \emph{small} gauge transformation, since such an $\alpha$ would transform as
  \begin{equation}
    \alpha(y) \stackrel{?}{\to} \alpha(y) - \Lambda(f(y)) + \Lambda(y).
  \end{equation}
  Indeed, by choosing $\Lambda$ such that it vanishes on $\Gamma$ and satisfies $\Box_{\mathcal{N}}\Lambda=-\alpha$ on $\mathcal{N}$ (this is a small gauge transformation), one would set $\alpha\to 0$. Then the condition $\Box_{\mathcal{N}}\alpha=0$ would not introduce deterministic dynamics for the Goldstone modes -- it would just amount to a small-gauge-fixing at $\mathcal{N}$. Thus, it is crucial to include $\psi$ in the definition of $\alpha$ as in~\eqref{Equation: dressed phase}, so that it is small-gauge-invariant. More generally, it is desirable for the Goldstone boundary condition to be small-gauge-invariant.

  In the next section, we will see that the boundary conditions described here will allow us to treat observables such as $\Psi(y)$, dressed by Wilson lines from $\Gamma$ to $y\in\mathcal{N}$, as local to $y$, in a way that is consistent with microcausality. Note that we required $\mathcal{N}$ to have codimension 1. This means we will get a ``codimension 1's worth'' of such dressed observables, because such observables have to live on $\mathcal{N}$. It is reasonable to ask whether one can go beyond this, and use a codimension 0 $\mathcal{N}$ such as the full spacetime; then one could microcausally dress observables throughout spacetime. The reason this does not work as desired is that a codimension 0 $\mathcal{N}$ would have one extra dimension compared to $\Gamma$, which means that there would not exist a diffeomorphism $f$ between them, since it could not be one-to-one. We could try to relax the requirement that $f$ is one-to-one, but this would lead to other problems. In particular, it would not be possible to solve the equation $\Box_{\mathcal{N}}\alpha=0$ by performing a large gauge transformation -- since the large gauge transformation is parametrised on the codimension 1 surface $\Gamma$, and so does not contain enough degrees of freedom to satisfy this equation everywhere in the codimension 0 region $\mathcal{N}$. For these reasons, it is essential for $\mathcal{N}$ to be codimension 1, at least for the scope of this paper.

  Of course, it is possible to construct \emph{other} kinds of dressed observables in a codimension 0 region compatible with microcausality -- for example, the phase of a bulk complex scalar field provides a good dressing, as described in the Introduction. However, in this paper we are primarily interested in \emph{boundary}-dressed observables. It seems plausible that microcausal boundary-dressed observables should always have some kind of codimension 1 restriction in this way. This seems to hint at a holographic description -- we leave full exploration of this to future work.

  Finally, recall that the function $f:\mathcal{N}\to\Gamma$ was chosen to preserve causal ordering of events. We could relax this condition. However, as we shall see in the next section, it is required to restore microcausality to observables dressed by the large gauge frame $\alpha$, because it guarantees that the dynamics of $\alpha$ respects the causal structure of spacetime.

  \section{Which observables are microcausal?}
  \label{Section: microcausality}

  Having carefully described the theory we are considering, including the construction of its variational principle and its phase space structure, we now come to the main point of the paper: the status of microcausality.

  The result will be as follows: we will define a spacetime subset for any observable, called its relational support. The relational support is the region to which the observable should be considered local, in the sense that if two observables' relational supports are spacelike separated, then they commute. We adopt this relational notion of support from our previous discussion of microcausality in gravity \cite{ReconcilingBulkLocality}. In the case of completely gauge-invariant observables, the relational support reduces to the usual support. For observables that transform non-trivially under large gauge transformations, the relational support can be much smaller than the usual support (in the case of observables on $\mathcal{N}$ dressed by Wilson lines along the curves $\gamma(y)$, $y\in\mathcal{N}$). For example, the relational support of $\alpha(y) $ is $y$. But the relational support can also be larger than the usual support (in the case of observables dressed in some other way), as we will illustrate  with an example in section~\ref{ssec_other}.

  We will consider both local (supported at a point) and quasilocal (supported over a larger region) observables. We emphasize that our discussion depends on our choice of boundary conditions, though we suspect qualitative features to carry over to other types of Goldstone boundary conditions also.

  \subsection{Completely gauge-invariant observables}

  Let us first consider observables $O$ which are completely gauge-invariant, including under large gauge transformations. An example of such an observable is $\abs{\psi(x)}$, i.e.\ the absolute value of the scalar $\psi$ at $x$. Other examples involving the connection are $\int_C A$ for some contractible curve $C$, i.e.\ the Wilson loop around $C$, or $\psi(x)\exp(-iq\int_{\gamma(x,y)} A)\overline{\psi}(y)$ for some curve $\gamma$ connecting event $y$ with $x$. The supports of these observables are $x$, $C$, and $\gamma$ respectively.
  We will now show that such observables are consistent with  microcausality, meaning they commute if their supports are spacelike separated. This result is essentially folklore (e.g.\ see \cite{Donnelly:2015hta,Donnelly:2016rvo,Donnelly:2017jcd} for some  discussion of examples in scalar QED), and not affected by the choice of Goldstone boundary conditions on $\varphi$ --- but for completeness we will give a full argument below. In fact, this microcausality property is even promoted to an axiom in the algebraic approach to quantum field theories and applies in particular to the  local (completely gauge-invariant) algebras in electromagnetic theories \cite{Buchholz:1981fj,haag2012local,Fewster:2019ixc}; e.g., realizations via a universal $C^*$-algebra are discussed in \cite{Buchholz:2015epa,Buchholz:2021tpv} (see also \cite{Ciolli:2013pta}).

  The intuition is as follows: the completely gauge-invariant degrees of freedom obey causal wave equations with local boundary conditions, so their propagators have support only within the lightcone. The result then follows from the fact that the Poisson bracket is essentially just a combination of propagators evaluated between the relevant observables. It is perhaps not obvious that the non-local boundary conditions we have employed do not invalidate this intuition. The reason they do not is that the non-local part of the boundary conditions are more or less `decoupled' from the completely gauge-invariant degrees of freedom. To make sure this is the case, we will now give a much more detailed version of events, which will in any case be useful for the generalisation to boundary-dressed observables in the following sections. This argument has appeared in a similar, but general form in \cite[Sec.~3.4.3]{ReconcilingBulkLocality}; here we adapt it to our setting of scalar electrodynamics.

  We want to understand the properties of the transformation generated by such an observable, i.e.\ the on-shell field variation $\delta_O\phi$ satisfying
  \begin{equation}
    \delta O = \Omega_*(\delta_O\phi).
    \label{Equation: O generates}
  \end{equation}
  This equation uniquely determines $\delta_O\phi$, up to small gauge transformations. Actually, rather than directly using~\eqref{Equation: O generates} to compute $\delta_O\phi$, it turns out to be simpler to first turn to an alternative (but equivalent) means of computing $\delta_O\phi$ due to Peierls \cite{Peierls:1952cb,DeWitt:1984sjp,Gieres:2021ekc,Khavkine:2014kya,Kirklin:2019xug}.

  The procedure starts by deforming the action $S\to S-\lambda O$, where $0<\lambda\ll1$. This results in a deformed set of equations of motion. For simplicity, we shall assume the support of $O$ lies in between $\Sigma^-$ and $\Sigma^+$. We then compute two solutions to the deformed equations of motion:
  \begin{equation}
    \phi_O^\pm = \phi + \lambda\delta^\pm_O\phi+\order{\lambda^2},
    \label{Equation: retarded advanced}
  \end{equation}
  where $\phi$ is some fixed solution to the \emph{un}deformed equations of motion, and $\delta_O^\pm\phi$ are independent of $\lambda$. These field variations $\delta_O^+\phi,\delta_O^-\phi$, known as the retarded and advanced variations respectively, are required to vanish in an open neighbourhood of $\Sigma^-,\Sigma^+$ respectively, which (assuming the deformed equations of motion are deterministic) uniquely determines $\delta_O^\pm\phi$ (up to small gauge transformations). It then turns out that the on-shell field variation generated by $O$ is given by the difference of the retarded and advanced variations:
  \begin{equation}
    \delta_O\phi = \delta_O^+\phi - \delta_O^-\phi,
    \label{Equation: peierls difference}
  \end{equation}
  where both sides are evaluated on the field configuration $\phi$ appearing on the right-hand side of~\eqref{Equation: retarded advanced}. In particular, $\phi+\lambda\,\delta_O\phi$ solves the \emph{un}deformed equations of motion (to linear order in $\lambda$) \cite{ReconcilingBulkLocality}, and therefore represents an on-shell field variation.

  The change in $O = O[\phi]$ under a (not necessarily on-shell) field variation $\delta\phi$ may (after integrating by parts when $O$ depends on derivatives of the fields) be written in the form
  \begin{equation}
    \delta O = \int_{\mathcal{M}} \qty(\delta A \wedge \hodge  \fdv{O}{A} + \delta\psi \hodge \fdv{O}{\psi} + \delta\overline{\psi}\hodge \fdv{O}{\overline{\psi}}) + \int_{\partial\mathcal{M}} \mathscr{F}[\delta A, \delta\psi, \delta\overline\psi],
  \end{equation}
  where $\fdv{O}{A},\fdv{O}{\psi},\fdv{O}{\overline{\psi}}$ are the functional derivatives of $O$, and $\mathscr{F}[\delta A, \delta\psi, \delta\overline\psi]$ is some quantity that depends linearly and locally on $\delta A,\delta\psi,\delta\overline\psi$ at $\partial\mathcal{M}$.

  The invariance of $O$ under gauge transformations implies
  \begin{nalign}
    0 = \delta_\Lambda O &= \int_{\mathcal{M}} \qty(-\dd{\Lambda} \wedge \hodge  \fdv{O}{A} -iq\Lambda\psi \hodge \fdv{O}{\psi} + iq\Lambda\overline{\psi}\hodge \fdv{O}{\overline{\psi}}) + \int_{\partial\mathcal{M}} \mathscr{F}[\delta_\Lambda A, \delta_\Lambda\psi, \delta_\Lambda\overline\psi] \\
    &= \int_{\mathcal{M}}\Lambda\qty(\dd(\hodge  \fdv{O}{A}) -iq\Lambda\psi \hodge \fdv{O}{\psi} + iq\Lambda\overline{\psi}\hodge \fdv{O}{\overline{\psi}}) + \int_{\partial\mathcal{M}} \qty(\mathscr{F}[\delta_\Lambda A, \delta_\Lambda\psi, \delta_\Lambda\overline\psi] - \Lambda \hodge \fdv{O}{A}).
  \end{nalign}
  For this to be true for arbitrary $\Lambda$, we must have
  \begin{equation}
    \dd(\hodge  \fdv{O}{A}) -iq\Lambda\psi \hodge \fdv{O}{\psi} + iq\Lambda\overline{\psi}\hodge \fdv{O}{\overline{\psi}} = 0,
    \label{Equation: functional derivative conservation}
  \end{equation}
  and it must also be the case that we can choose $\mathscr{F}[\delta A, \delta\psi, \delta\overline\psi]$ such that
  \begin{equation}
    \mathscr{F}[\delta_\Lambda A, \delta_\Lambda\psi, \delta_\Lambda\overline\psi] = \Lambda \hodge \fdv{O}{A}
    \label{Equation: scr F gauge}
  \end{equation}
  for all $\Lambda$. Such a $\mathscr{F}[\delta A, \delta\psi, \delta\overline\psi]$ cannot depend on derivatives of $\delta A,\delta\psi,\delta\overline\psi$ normal to $\partial\mathcal{M}$, or on normal components of $\delta A$ (since then normal derivatives of $\Lambda$ would have to appear in the right-hand side of~\eqref{Equation: scr F gauge}). Thus, $\mathscr{F}[\delta A,\delta\psi,\delta\overline\psi]$ can be chosen to depend only on the field variations via their pullback to $\partial\mathcal{M}$. But note that the pullback of $\delta A,\delta\psi,\delta\overline\psi$ to $\Gamma$ is indistinguishable from a large gauge transformation, according to the ``Dirichlet up to large gauge transformations'' boundary conditions~\eqref{Equation: Dirichlet up to gauge}. This means that for a \emph{general} field variation we can write
  \begin{equation}
    \mathscr{F}[\delta A, \delta\psi, \delta\overline\psi]\Big|_\Gamma = \left.\Lambda \hodge \fdv{O}{A}\right|_\Gamma,
  \end{equation}
  where now $\Lambda$ is the same one that appears in~\eqref{Equation: Dirichlet up to gauge}.

  In summary, we may write
  \begin{multline}
    \delta O = \int_{\mathcal{M}} \qty(\delta A \wedge \hodge  \fdv{O}{A} + \delta\psi \hodge \fdv{O}{\psi} + \delta\overline{\psi}\hodge \fdv{O}{\overline{\psi}}) + \int_\Gamma \Lambda\hodge \fdv{O}{A}\\
    + \int_{\Sigma_+} \mathscr{F}[\delta A|_{\Sigma^+}, \delta\psi|_{\Sigma^+}, \delta\overline\psi|_{\Sigma^+}] - \int_{\Sigma_-} \mathscr{F}[\delta A|_{\Sigma^-}, \delta\psi|_{\Sigma^-}, \delta\overline\psi|_{\Sigma^-}]
    \label{Equation: delta O}
  \end{multline}
  for some $\mathscr{F}$, and the functional derivatives of $O$ satisfy~\eqref{Equation: functional derivative conservation}.

  We may use this to obtain the deformed equations of motion from the deformed action $S - \lambda O$. First, since $\delta_\Lambda O=0$, the equation of motion resulting from extremising the deformed action with respect to large gauge transformations is unchanged: it is still $\Box_{\mathcal{N}}\alpha=0$, i.e.\ the Goldstone boundary condition. More generally, the variation of the deformed action may be obtained by combining~\eqref{Equation: delta O} with~\eqref{Equation: action variation}; when initial/final data are fixed we find
  \begin{multline}
    \delta (S-\lambda O) = -\int_{\Gamma} \Lambda\qty(\dd{\hodge F}+\hodge j-\lambda\hodge \fdv{O}{A}) \\
    - \int_{\mathcal{M}}\bigg(\delta A\wedge\qty(\dd{\hodge F}+\hodge j -\lambda\hodge \fdv{O}{A}+\hodge  \int_{y\in\mathcal{N}}\fdv{\alpha(y)}{A}\dd{\Hodge\dd{\alpha}}|_y) \\
    + \Re\qty(\overline{\delta\psi}\,\qty( \Dd{\hodge\Dd{\psi}}-*\psi V'(\abs{\psi}^2)+2\lambda*\fdv{O}{\overline{\psi}}+\hodge  \int_{y\in\mathcal{N}}\fdv{\alpha(y)}{\overline\psi}\dd{\Hodge\dd{\alpha}}|_y))\bigg).
    \label{Equation: deformed action variation}
  \end{multline}
  As before, we can extremise the action by solving
  \begin{nalign}
    \dd{\hodge F}+\hodge j = \lambda\hodge \fdv{O}{A}, \qquad \Dd{\hodge \Dd{\psi}} - \hodge \psi V'(\abs{\psi}^2) = -2\lambda\hodge \fdv{O}{\overline\psi},
    \label{Equation: O deformed eom}
  \end{nalign}
  and then doing the large gauge transformation required to set $\Box_{\mathcal{N}}\alpha=0$.\footnote{Doing this large gauge transformation will not lead to violations of~\eqref{Equation: O deformed eom}, because the equations~\eqref{Equation: O deformed eom} are invariant under all gauge-transformations. To see this, note that~\eqref{Equation: O deformed eom} are the equations of motion that would directly result from the deformed action if we were to remove the Klein-Gordon action for $\alpha$. But then the action would be invariant under \emph{all} gauge transformations, so if one field configuration were to extremise it, any gauge transformation of that field configuration would also extremise it.} These may therefore be taken as the deformed equations of motion.

  Note that~\eqref{Equation: functional derivative conservation} is consistent with the deformed equations of motion~\eqref{Equation: O deformed eom}. Indeed, taking the exterior derivative of the first equation and substituting in the second multiplied by $\bar\psi$ yields back~\eqref{Equation: functional derivative conservation}.

  The support of $O$ is defined as
  \begin{equation}
    \supp(O) = \supp\qty(\fdv{O}{A})\cup\supp\qty(\fdv{O}{\psi})\cup\supp\qty(\fdv{O}{\overline\psi}).
  \end{equation}
  In other words, it is the region of spacetime in which $O$ has a dependence on the fields. We can observe that the deformed equations of motion~\eqref{Equation: O deformed eom} only differ from the undeformed ones~\eqref{Equation: bulk equations of motion} on $\supp(O)$.

  At this point we can use an important property of the Maxwell-scalar field equations: they are \emph{causal}. What this means is that, whenever these equations hold, the field configuration in any open subset $U\subset\mathcal{M}$ fully determines the field configuration in the domain of dependence $D(U)$ of that open subset (up to gauge transformations). The equation $\Box_{\mathcal{N}}\alpha=0$ is also causal.

  Consider now the solutions $\phi^\pm_O$ to the deformed equations of motion defined in~\eqref{Equation: retarded advanced}. These field configurations obey the undeformed equations of motion in $\mathcal{M}\setminus\supp(O)$. Since $\phi^\pm_O=\phi$ in an open neighbourhood of $\Sigma^\mp$ respectively, we have that $\phi^\pm_O=\phi$ (up to gauge transformations) in $D_{\mathcal{M}\setminus\supp(O)}(\Sigma^\mp)$, which is the domain of dependence of $\Sigma^\mp$ when considered as a subset of $\mathcal{M}\setminus\supp(O)$, respectively. Since the causal future/past of $\supp(O)$ are respectively given by
  \begin{equation}
    J^\pm(\supp(O)) = \mathcal{M}\setminus D_{\mathcal{M}\setminus\supp(O)}(\Sigma^\mp),
  \end{equation}
  we may conclude that (up to gauge transformations)
  \begin{equation}
    \supp(\delta^\pm_O\phi) \subset J^\pm(\supp(O)),
  \end{equation}
  and hence, using~\eqref{Equation: peierls difference}, that (up to gauge transformations)
  \begin{equation}
    \supp(\delta_O\phi) \subset J^+(\supp(O))\cup J^-(\supp(O)).
    \label{Equation: lightcone support completely gauge invariant}
  \end{equation}
  The region on the right-hand side is the domain of influence of $\supp(O)$, and consists of all points which are timelike or null separated from $\supp(O)$.

  This result is sufficient to show that completely gauge-invariant observables obey microcausality. To be completely clear, the statement of microcausality for such observables is:
  \begin{quote}
    If $O_1,O_2$ are completely gauge-invariant, and $\supp(O_1)$ is spacelike separated from $\supp(O_2)$, then $\pb{O_1}{O_2}=0$.
  \end{quote}
  To show this holds, suppose $O_1,O_2$ are two such observables, and consider their Poisson bracket
  \begin{equation}
    \pb{O_1}{O_2} = \delta_{O_2} O_1 = \Omega_*(\delta_{O_1}\phi,\delta_{O_2}\phi).
    \label{Equation: Poisson Peierls bracket}
  \end{equation}
  Since $\delta_{O_2}\phi$ is indistinguishable from a gauge transformation at spacelike separation from $\supp(O_2)$, and since $O_1$ is gauge-invariant, we must have $\delta_{O_2}O_1=0$, i.e.\ the bracket vanishes.

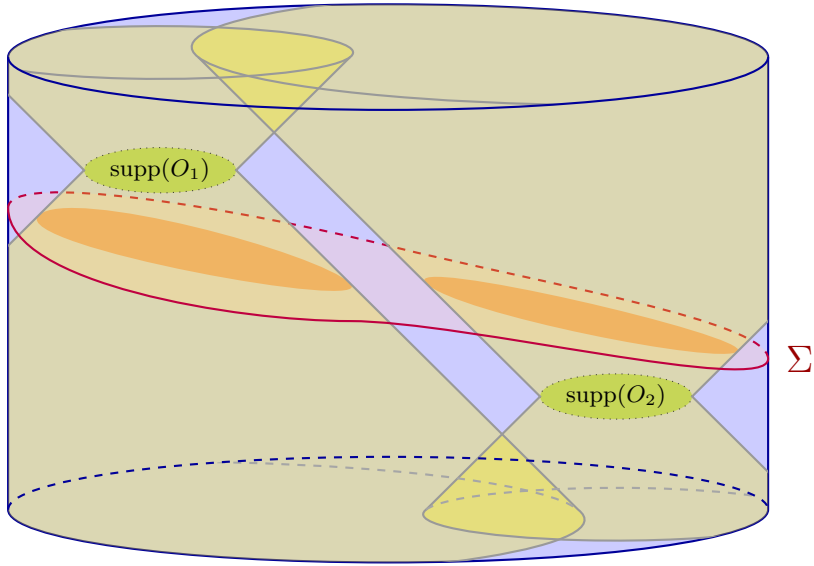
\begin{figure}
    \centering
    \begin{tikzpicture}[scale=1]
      \fill[blue!20] (0,0) -- (0,6) arc (180:0:5 and 0.7) -- (10,0) arc (360:180:5 and 0.7);
      \draw[thick,blue!60!black] (0,0) -- (0,6) arc (180:0:5 and 0.7) -- (10,0) arc (360:180:5 and 0.7);

      \fill[purple!20,opacity=0.5] (0,4) .. controls (0,3) and (3,2.5) .. (4.5,2.5) .. controls (6,2.5) and (10,1.5) .. (10,2)
      .. controls (10,2.5) and (7,3) .. (5,3.5) .. controls (3,4) and (0,4.5) .. (0,4);

      \draw[purple,dashed,thick] (10,2) .. controls (10,2.5) and (7,3) .. (5,3.5) .. controls (3,4) and (0,4.5) .. (0,4);

      \begin{scope}
        \clip (0,0) -- (0,6) arc (180:0:5 and 0.7) -- (10,0) arc (360:180:5 and 0.7);

        \fill[yellow,opacity=0.3] (7.5,0) -- (3,4.5) -- (4.5,6) arc (-10:190:{2.5/cos(10)} and 0.35) -- (1,4.5) -- (-3.5,0) arc (170:370:{5.5/cos(10)} and 0.77);
        \fill[yellow,opacity=0.3] (10.5,0) -- (9,1.5) -- (13.5,6) arc (-10:190:{5.5/cos(10)} and 0.77) -- (7,1.5) -- (5.5,0) arc (170:370:{2.5/cos(10)} and 0.35);

        \fill[red!50!yellow, opacity=0.4] (0.4,3.9) .. controls (0,3.5) and (4.9,2.6) .. (4.5,3) .. controls (4.1,3.4) and (0.8,4.3) .. (0.4,3.9);
        \fill[red!50!yellow, opacity=0.4] (5.5,3) .. controls (5.1,3.4) and (9.8,2.3) .. (9.6,2.1) .. controls (9.4,1.9) and (5.9,2.6) .. (5.5,3);

        \draw[dotted,fill=yellow!70!green,opacity=0.7] (2,4.5) ellipse (1 and 0.3);
        \node at (2,4.5) {\footnotesize$\operatorname{supp}(O_1)$};
        \draw[dotted,fill=yellow!70!green,opacity=0.7] (8,1.5) ellipse (1 and 0.3);
        \node at (8,1.5) {\footnotesize$\operatorname{supp}(O_2)$};
        \draw[gray!80, thick] (7.5,0) -- (3,4.5) -- (4.5,6) arc (-10:190:{2.5/cos(10)} and 0.35) -- (1,4.5) -- (-3.5,0) arc (170:370:{5.5/cos(10)} and 0.77);
        \draw[gray!80, thick] (10.5,0) -- (9,1.5) -- (13.5,6) arc (-10:190:{5.5/cos(10)} and 0.77) -- (7,1.5) -- (5.5,0) arc (170:370:{2.5/cos(10)} and 0.35);
      \end{scope}
      \begin{scope}
        \clip (5,6) ellipse (5 and 0.7);
        \draw[gray!80,thick] (4.5,6) arc (350:190:{2.5/cos(10)} and 0.35);
        \draw[gray!80,thick] (13.5,6) arc (350:190:{5.5/cos(10)} and 0.77);
      \end{scope}
      \begin{scope}
        \clip (5,0) ellipse (5 and 0.7);
        \draw[gray!70, thick, dashed] (-3.5,0) arc (170:10:{5.5/cos(10)} and 0.77);
        \draw[gray!70, thick, dashed] (5.5,0) arc (170:10:{2.5/cos(10)} and 0.35);
      \end{scope}

      \draw[purple,thick] (0,4) .. controls (0,3) and (3,2.5) .. (4.5,2.5) .. controls (6,2.5) and (10,1.5) .. (10,2);

      \draw[thick,blue!60!black] (0,6) arc (-180:0:5 and 0.7);
      \draw[thick,dashed,blue!60!black] (0,0) arc (180:0:5 and 0.7);

      \node[right, red!60!black] at (10.1,2) {\Large$\Sigma$};
    \end{tikzpicture}
    \caption{For $O_1$, $O_2$ with spacelike separated supports, we can always choose a Cauchy surface $\Sigma$ with an open neighbourhood that only ever intersects one of the domains of influence of $\operatorname{supp}(O_1)$, $\operatorname{supp}(O_2)$.}
    \label{Figure: microcausal}
  \end{figure}

  For a more symmetric argument, one may also use the two field variations $\delta_{O_1}\phi,\delta_{O_2}\phi$ and the structure of $\Omega_*$ to show that the bracket vanishes. Indeed, the large gauge invariance of $O_1,O_2$ implies $\delta_{O_1}\alpha=\delta_{O_2}\alpha=0$ (since $\alpha$ is the generator of large gauge transformations), so we may ignore the integral over $K$ appearing in $\Omega_*$. What is left is a pair of integrals over $\Sigma,\partial\Sigma$ whose integrands depend linearly and locally on the field variations $\delta_{O_1}\phi,\delta_{O_2}\phi$. Also, these remaining integrals are not affected if we change the choice of gauge (large or small) for either $\delta_{O_1}\phi$ or $\delta_{O_2}\phi$, so without loss of generality let us change the gauge so that these field variations vanish outside $J^+(\supp(O_1))\cup J^-(\supp(O_1))$, $J^+(\supp(O_2))\cup J^-(\supp(O_2))$ respectively. Since the supports of the two observables are spacelike separated, it is possible for us to pick the Cauchy surface $\Sigma$ such that
  \begin{equation}
    U_\Sigma \cap \qty(J^+(\supp(O_1))\cup J^-(\supp(O_1))) \cap \qty(J^+(\supp(O_2))\cup J^-(\supp(O_2)))=\emptyset,
  \end{equation}
  where $U_\Sigma$ is a sufficiently small open neighbourhood of $\Sigma$ (see Figure~\ref{Figure: microcausal}). In other words, $U_\Sigma$ only ever intersects one of the domains of influence of the two observables. Therefore, by~\eqref{Equation: lightcone support completely gauge invariant}, in an open neighbourhood of every point in $\Sigma$ at least one of $\delta_{O_1}\phi,\delta_{O_2}\phi$ vanishes. But this implies that the remaining integrands in $\Omega_*(\delta_{O_1}\phi,\delta_{O_2}\phi)$ vanish. Therefore, $\pb{O_1}{O_2}=0$, as required. This argument was presented in a similar form for gravity in \cite{ReconcilingBulkLocality} (and in a related, but somewhat abbreviated manner also in \cite{Marolf:2015jha}).

  In terms of the causally local net of observables described in the introduction, we can include $O\in\mathcal{A}_U$ whenever $\supp(O)\subset D(U)$, in line with the axioms of algebraic quantum field theory \cite{Buchholz:1981fj,haag2012local,Fewster:2019ixc} and as realized by universal local $C^*$-algebras in electromagnetic theories \cite{Buchholz:2015epa,Buchholz:2021tpv,Ciolli:2013pta}.\footnote{These do not yet encompass finite mass charges \cite{Buchholz:2021tpv}, as necessary in the present context.}

  We should note that $\supp(O)$ is the smallest region to which $O$ may be considered local. This is because the field variation $\delta_O\phi$ must always be non-trivial in $\supp(O)$, so for any subset $V\subset\supp(O)$ there is always some other observable $O'$ local to $V$ such that $\pb{O'}{O} = \delta_O O' \ne 0$.

  \subsection{Observables dressed by the Wilson lines}

  It is now that we really come to the point of the paper. We will now show that any small-gauge-invariant observable formed by taking some bare quantity on $\mathcal{N}$, and dressing it with Wilson lines along the curves $\gamma(y),y\in\mathcal{N}$, may be considered as local to the support of the bare quantity, in a way that is consistent with microcausality. Thus, our Goldstone boundary conditions ensure that microcausality is obeyed also for large-gauge-charged observables on a codimension-one surface in the bulk. This microcausality is essentially ``pulled in'' from $\Gamma$.

  For our purposes, the simplest such observables are $\alpha(y)$, where $y\in\mathcal{N}$. Recall that $\alpha(y)$ is the phase of $\Psi(y)$, i.e.\ of $\psi(y)$ (which is the bare quantity in this case) dressed by the Wilson line along $\gamma(y)$. Since $\alpha(y)$ is the generator of a large gauge transformation, we automatically have $\pb{O}{\alpha(y)}=0$ for any completely gauge-invariant observable. Moreover, the Poisson bracket $\pb{\alpha(y)}{\alpha(y')}$ for any two points $y,y'\in\mathcal{N}$ reduces to the Poisson bracket of a scalar field on $\mathcal{N}$ (indeed, this may in principle be derived from~\eqref{Equation: QP} and~\eqref{Equation: Hamiltonian QP}). Therefore, we can immediately conclude that
  \begin{equation}
    \pb{\alpha(y)}{\alpha(y')}=0 \text{ if $y,y'$ are spacelike separated.}
    \label{Equation: alpha microcausal}
  \end{equation}
Note that by differentiating~\eqref{Equation: alpha microcausal}, one finds that derivatives of $\alpha$ also commute at spacelike separation.
  It is perhaps worthwhile to emphasise what this means in terms of the standard kinematical notion of support of $\alpha(y),\alpha(y')$ (and their derivatives). These composite observables thus commute if their \emph{endpoints} on $\mathcal{N}$ are spacelike separated, which does \emph{not} imply that the Wilson lines built into them must be spacelike separated too. With standard boundary conditions, we would not have been able to arrive at this conclusion.

  So the set of observables $\{\alpha(y),y\in\mathcal{N}\}$ obeys microcausality relative to all completely gauge invariant observables, and it obeys microcausality relative to itself, if we consider each $\alpha(y)$ to be local to $y$. It is therefore consistent with microcausality to include $\alpha(y)\in\mathcal{A}_U$ whenever $y\in D(U)$.  By linearity, we can also include derivatives of $\alpha(y)$. Given that large gauge generators are included in this discussion, our observation also applies to regions $U$ that touch the asymptotic boundary $\Gamma$. Indeed, the form of the presymplectic form in \eqref{omegastar2} clarifies that it is now  $\mathcal{N}$ rather than $\Gamma$ that assumes a special role in the theory.

  Furthermore, by the linearity of the Poisson bracket, and the Leibniz/chain rules, it is entirely consistent with microcausality to `complete' $\mathcal{A}_U$ by including arbitrary sums, products, and functions of the observables that we have already included in it. This is enough to show that any small-gauge-invariant observable formed by dressing bare quantities on $\mathcal{N}$ with Wilson lines along $\gamma(y), y\in\mathcal{N}$ is consistent with microcausality.

  For example, consider the observable
  \begin{equation}
    \Psi(y) = \psi(y) e^{iq\int_{\gamma(y)}A},
  \end{equation}
  i.e.\ a dressed scalar of the kind noted at the beginning of the paper. We may equivalently write this observable as
  \begin{equation}
    \Psi(y) = \abs{\psi(y)} e^{iq\alpha(y)}.
  \end{equation}
  Now $\abs{\psi(y)}$ is a completely gauge-invariant observable whose support is $y$. Thus, $\Psi(y)$ is formed by combining quantities that are microcausally local to $y\in\mathcal{N}$. By the properties of the Poisson bracket, $\Psi(y)$ is also microcausally local to $y$, i.e.\ we should include $\Psi(y)\in \mathcal{A}_U$ whenever $y\in D(U)$.

  More generally, suppose $O$ is any observable formed by dressing bare quantities with Wilson lines. Then we can decompose $O$ as a function of bare quantities $o\in\mathscr{B}$ in some set $\mathscr{B}$, where each $o$ depends on the fields $A,\psi$ and their derivatives only in some fixed subset $V\subset\mathcal{N}$, along with the Wilson lines along the curves $\gamma(y)$ for $y\in V$:
  \begin{equation}
    O = O\qty[\Big\{o\bigm\vert o\in\mathscr{B}\Big\};\qty{\int_{\gamma(y)}A\Bigm\vert y\in V}].
    \label{Equation: dressed observable decomposition}
  \end{equation}
  Now let us do a small gauge transformation~\eqref{Equation: gauge transformation} with $\Lambda(y) = \frac1q \arg(\psi(y))$. Under this gauge transformation $O$ does not change. However, the quantities it depends on do change. In particular, each $o\in\mathscr{B}$ changes in some way $o\to o'$. The resulting $o'$ is a so-called `relational observable' that measures $o$ relative to the phase of $\psi$ \cite{Carrozza:2021gju}, is completely gauge-invariant,\footnote{This follows from the fact that doing the small gauge transformation with $\Lambda(y) = \frac1q \arg(\psi(y))$ completely fixes the gauge on $\mathcal{N}$. See \cite{Carrozza:2021gju} for a more in-depth discussion of how to build such relational observables in gauge theories using field-dependent gauge transformations.} and obeys $\supp(o')\subset V$. Let us use $\mathscr{B}'$ to denote the set of all such $o'$. On the other hand, under this small gauge transformation the Wilson lines change to $\int_{\gamma(y)}A \to \alpha(y)$. Since $O$ is small-gauge-invariant, we can now write
  \begin{equation}
    O = O\qty[\Big\{o'\bigm\vert o'\in\mathscr{B}'\Big\};\Big\{\alpha(y)\bigm\vert y\in V\Big\}].
  \end{equation}
  We have therefore written $O$ as a combination of observables all of which we already know are in $\mathcal{A}_U$ whenever $V\subset D(U)$. Therefore, by the properties of the Poisson bracket, we should also include $O\in \mathcal{A}_U$ whenever $V\subset D(U)$.

  Another notable special case of the above result is as follows. Let
  \begin{equation}
    \mathbf{A} = A|_{\mathcal{N}} + \dd{W}, \qq{where} W(y) = \int_{\gamma(y)} A.
  \end{equation}
  This is the pullback of the gauge potential to $\mathcal{N}$, dressed by Wilson lines along $\gamma(y)$. From the above, we have that $\mathbf{A}(y)\subset \mathcal{A}_U$ whenever $y\in D(U)$.

  Let us now give a more precise statement of the kind of microcausality that is obeyed by these observables, in terms of a kind of generalised support. The most general kind of observable that we have considered so far can be formed by combining completely gauge-invariant observables with observables of the form~\eqref{Equation: dressed observable decomposition}, and may therefore be decomposed as
  \begin{equation}
    O = O\qty[\Big\{o'\bigm\vert o'\in\mathscr{C}\Big\};\Big\{\alpha(y)\bigm\vert y\in V\Big\}],
    \label{Equation: O decomposition}
  \end{equation}
  where $\mathscr{C}$ is some collection of completely gauge-invariant observables, and $V\subset\mathcal{N}$ (we no longer require $\supp(o')\subset \mathcal{N}$). Note that there are different ways to do this decomposition.

  The observables $o'\in\mathscr{C}$ fail to commute with other observables in $\supp(o')$, while the observables $\alpha(y),y\in V$ fail to commute with other $\alpha(y'),y'\in V$ if $y,y'$ are not spacelike separated. This motivates the construction of the spacetime region
  \begin{equation}
    V\cup \bigcup_{o'\in\mathscr{C}} \supp(o').
  \end{equation}
  We define the `relational support' of $O$ (see \cite{ReconcilingBulkLocality} for the gravitational case), denoted $\relsupp(O)$, as the smallest region that can be constructed in this way, over all possible decompositions of $O$. For example, we have
  \begin{equation}
    \relsupp\qty(\abs{j(x)}^2 + \Psi(y) + \partial\cdot \mathbf{A}(y') ) = \{x,y,y'\}.
  \end{equation}
  For completely gauge-invariant observables, the relational support reduces to the ordinary support.

  One way to find the relational support of $O$ is to consider $\delta O$ for all field variations $\delta\phi$ which leave all the Wilson lines invariant, i.e.\ $\delta\qty(\int_{\gamma(y)}A)=0$, for all $y\in\mathcal{N}$. Then we have $x\in\relsupp(O)$ iff $\delta O = 0$ implies that $\delta\phi$ is indistinguishable from a small gauge transformation in an open neighbourhood of $x$. In this way, we can think of the relational support of an observable as similar to the support of that observable, but neglecting any dependence of the observable on the fields via the Wilson lines along $\gamma(y)$.

  The precise statement of microcausality is then as follows:
  \begin{quote}
    If $O_1,O_2$ are such that $\relsupp(O_1),\relsupp(O_2)$ are spacelike separated, then $\pb{O_1}{O_2}=0$.
  \end{quote}
  For the reasons explained above, this statement holds in the theory we have constructed. Therefore, we include $O\in\mathcal{A}_U$ whenever $\relsupp(O)\subset D(U)$.

  \subsection{Other microcausal observables?}\label{ssec_other}

  We have so far succeeded in showing that completely gauge invariant observables, and observables on $\mathcal{N}$ dressed by Wilson lines along the curves $\gamma(y)$, and any combinations of these observables, are consistent with microcausality, if one views them as local to their relational support. Let us now ask the important question: are there any other observables which are consistent with this notion of microcausality, but which we have not already considered?

  The answer to this question is no. It turns out that \emph{any} (small-gauge-invariant) observable can be written as a combination of the ones considered above. The relational support thus extends to all physical observables, and defines the regions to which such observables should be considered local, in terms of the statement of microcausality given above.

  For example, suppose $x$ is a spacetime point not in $\mathcal{N}$, and that $\gamma'$ is some curve extending from $x$ to a point $x'\in\Gamma$. We can choose to do this in such a way that $\gamma'$ is spacelike separated from $f^{-1}(x')$. Consider the observable
  \begin{equation}
    \tilde\Psi(x) = \psi(x) e^{iq\int_{\gamma'}A},
  \end{equation}
  i.e.\ the scalar dressed by a Wilson line along $\gamma'$. We might hope that $\tilde\Psi(x)$ could be considered as local to $x$, but this is not true. In fact, $\tilde\Psi(x)$ is not even local to $\gamma'$, which is its support, because the bracket of $\tilde\Psi(x)$ with $\dot\alpha(f^{-1}(x'))$ does not vanish (since $\dot\alpha(f^{-1}(x'))$ generates a large gauge transformation which acts on $\tilde\Psi(x)$ non-trivially).

  To find the region to which $\tilde\Psi(x)$ is actually local, we can write it as
  \begin{equation}
    \tilde\Psi(x) = e^{iq\alpha(f^{-1}(x'))} \times \psi(x)  \frac{\overline{\psi}(f^{-1}(x'))}{\abs{\psi(f^{-1}(x'))}}e^{iq\int_{\gamma'}A-iq\int_{\gamma(f^{-1}(x'))}A}.
    \label{Equation: tilde Psi decomposition}
  \end{equation}
  The second factor on the right-hand side is a completely gauge-invariant observable whose support is $\gamma'\cup \gamma(f^{-1}(x'))$, while the first factor is an observable dressed by the Wilson line along $\gamma(f^{-1}(x'))$. From this expression we may deduce that the relational support of $\tilde\Psi(x)$ is
  \begin{equation}\label{relsuppex}
    \relsupp\qty(\tilde\Psi(x)) = \gamma'\cup \gamma(f^{-1}(x')),
  \end{equation}
  and this is the true region to which $\tilde\Psi(x)$ should be considered local. Thus, the relational support of an observable may be larger than its usual support.

  Let us now argue more generally. Suppose $O$ is any physical observable. Under a large gauge transformation parametrised by $\Lambda$, let us denote the transformation of this observable as $O\to O_\Lambda$. We may view this transformation in two ways: as evaluating the same observable $O$ at a large-gauge-transformed field configuration (`active picture'), or, equivalently, as a change of observable at the same field configuration (`passive picture'). Given that we are interested in decomposing $O$ into a combination of completely gauge-invariant observables and $\alpha$, it will be more natural to adopt the passive picture in what follows. Under a second large gauge transformation parametrised by $\Lambda'$, the observable transforms again, $O_\Lambda\to (O_\Lambda)_{\Lambda'}$. In particular, setting $\Lambda'=-\Lambda$ we have $(O_\Lambda)_{-\Lambda} = O$. So far for field-independent $\Lambda$.

  Now consider the observable $O_{f_*\alpha}:=O_\Lambda\big|_{\Lambda=f_*\alpha}$. This is what $O$ would become were we to evaluate the large gauge transformation at the \emph{field-dependent} parameter $\Lambda=f_*\alpha$. This large gauge transformation actually sets $\alpha \to 0$, so the observable $O_{f_*\alpha}$ may be thought of as ``the value of $O$ when $\alpha=0$'' (this is a relational observable). In the language of dynamical reference frames \cite{Carrozza:2021gju,Araujo-Regado:2024dpr} that we turn to shortly, $\alpha$ is (the phase of) a \emph{gauge-invariant} frame for \emph{large} gauge transformations. Similarly to the discussion below \eqref{Equation: dressed observable decomposition} for small gauge transformations, this field-dependent large gauge transformation is (in the passive picture) equivalent to dressing $O$ with $\alpha$, which is why $O_{f_*\alpha}$ is completely gauge-invariant. Indeed, any large gauge transformations we may do to it will be cancelled out by the ${}_{f_*\alpha}$. In the example of $O=\tilde\Psi(x)$, the second factor in~\eqref{Equation: tilde Psi decomposition} is $O_{f_*\alpha}$, i.e.\ $O_{f_*\alpha}=\tilde\Psi(x)e^{-iq\alpha(f^{-1}(x'))}$. Complete gauge invariance means that now $(O_{f_*\alpha})_{\Lambda'}=O_{f_*\alpha}$ for all $\Lambda'$ and so also $(O_{f_*\alpha})_{\Lambda'}\big|_{\Lambda'=-f_*\alpha}=O_{f_*\alpha}$.

  To transform $O_{f_*\alpha}$ back into $O$ and thereby obtain a suitable observable decomposition, we may proceed differently. Rather than evaluating the first large gauge transformation at a field-dependent parameter \emph{before} the second one, we first perform two independent such transformations and \emph{then} evaluate them at suitable field-dependent parameters. This avoids also transforming the dressing ${}_{f_*\alpha}$ in $O_{f_*\alpha}$, which is responsible for the latter's complete gauge invariance. Clearly, we have
  \begin{equation}
    O_{f_
    *\alpha}(\Lambda')\Big|_{\Lambda'=-f_*\alpha}=O,
  \end{equation}
  where
  \begin{equation}
    O_{f_
    *\alpha}(\Lambda'):=(O_\Lambda)_{\Lambda'}\big|_{\Lambda=f_*\alpha}\,,
  \end{equation}
  is a $\Lambda'$-parameter family of completely gauge-invariant observables. Hence, we may write  $O_{f_*\alpha}(\Lambda')=O(\{O_i\},\Lambda')$ as a functional of some basis of completely gauge-invariant observables $O_i$ and $\Lambda'$. As a consequence, $O=O(\{O_i\},-f_*\alpha$), and so we have succeeded in decomposing $O$ in terms of a combination of completely gauge-invariant observables $\{O_i\}$, and the observable $-f_*\alpha$, which is a bare quantity on $\mathcal{N}$ dressed by Wilson lines along $\gamma(y), y\in\mathcal{N}$, and which absorbs $O$'s large gauge dependence. In the example of $O=\tilde\Psi(x)$, the first factor in~\eqref{Equation: tilde Psi decomposition} corresponds to the second transformation by $\Lambda'=-f_*\alpha$. Using this decomposition, we can in principle compute the relational support as above, and thus determine the spacetime region to which $O$ should be considered local.

  Let us note one further way in which the relational support may be computed:
  \begin{equation}
    \relsupp(O) = \supp(O_{f_*\alpha}).
  \end{equation}
  In other words, the relational support of an observable is the support of the corresponding completely gauge-invariant relational observable dressed with the large gauge frame $\alpha$.\footnote{An already completely gauge-invariant $O$ is left invariant by this dressing.}
  Thus, all the differences between  the relational and standard support of a physical observable can be attributed to $\alpha$.

  \section{Dynamical reference frames and microcausality}
  \label{Section: frame}

  In physics in general, small-gauge-dependent `bare' quantities are non-physical and so cannot be observed. In order to extract physical information from them, we need to pair them with some other small-gauge-dependent quantities, sometimes known as `dynamical reference frames' (or just `frames') \cite{Carrozza:2021gju,Carrozza:2022xut,ReconcilingBulkLocality,Araujo-Regado:2024dpr}, to form a single small-gauge-invariant quantity, i.e.\ a physical observable. Observables formed in this way measure the bare quantity with respect to the frame \cite{ReconcilingBulkLocality,Carrozza:2021gju,Rovelli:1990ph,Rovelli:1990pi,Rovelli:2013fga,Giddings:2005id,Giddings:2025xym,Giddings:2025bkp,Dittrich:2004cb,Dittrich:2005kc,Giesel:2007wi}.

  A frame is associated with -- and transforms under --  the pertinent gauge symmetry, and is thus not necessarily a spatiotemporal frame. For example, in our case of scalar QED, a system of Wilson lines, such as those along the particular curves $\gamma(y),y\in\mathcal{N}$, is a small gauge frame \cite{Carrozza:2021gju,Araujo-Regado:2024dpr}. We can use this frame to construct physical observables from bare quantities on $\mathcal{N}$. But there are other frames, for example involving different systems of Wilson lines, the phase of the scalar field $\psi$, or different kinds of dressings altogether. We could use any of these frames to construct small gauge-invariant observables.

  However, the results of the previous section suggest that, in the particular theory we are studying, there appears to be a preferred frame, among all the possible choices: the Wilson lines along $\gamma(y),y\in\mathcal{N}$. This frame is preferred because observables that we construct with it enjoy an enhanced microcausality -- for other frames, this is not generically the case. It will be convenient to give the privileged frame a name; let us call it the \emph{microcausal frame} (of the given theory).

  The reason for the privileged role of the microcausal frame is of course the particular non-local boundary condition $\Box_{\mathcal{N}}\alpha=0$ that went into the definition of the theory. The quantity $\alpha(y)$ is itself defined in terms of the Wilson line along $\gamma(y)$ --- so it is not surprising that such Wilson lines have a special status. In this way, a choice of a microcausal frame may be viewed as essentially equivalent to a choice of boundary conditions on the gauge sector of the theory. More general choices of such boundary conditions than those described in this paper will have their own particular associated microcausal frames. As such, the choice of microcausal frame is part of the \emph{definition} of the theory. Operationally speaking, the microcausal frame specifies the relational structure required for a distributed apparatus within $\mathcal{N}$ to perform measurements that are jointly compatible with microcausality.

  On the other hand, given two theories defined with different microcausal frames, it is sometimes possible to map states and observables between them in a canonical way, such that brackets are preserved. We will describe an example of this below. As a consequence, one can have two (and usually more) distinct notions of microcausality existing within the \emph{same} theory: one associated with the original microcausal frame of the original theory, and the other associated with the microcausal frame of the second theory, applied to observables pulled back through the canonical mapping between the theories. These `notions of microcausality' can be alternatively understood as `criteria for commutation' of observables. The criterion is whether the relational supports of the observables are spacelike separated, and each frame has a \emph{different} associated definition of relational support. Note that one only needs one such frame's criterion to be satisfied, in order to have commuting observables. Checking this property may involve mapping the respective support into another theory.

  A very similar observation of coexisting but distinct criteria for spacelike commutation has previously been made in the holographic gravitational context and goes under the name `subregion-subalgebra duality'~\cite{Leutheusser:2021frk,Leutheusser:2021qhd,Leutheusser:2022bgi}. There, the duality (in the strict $G_N\to0$, or equivalently $N\to\infty$ limit) between the bulk and boundary theories makes the picture particularly direct, and seeks to identify which boundary operator algebra is dual to the operator algebra associated with a given causally complete bulk subregion. One can either check for spacelike separation of bulk observables, or spacelike separation of their boundary duals, and in general the two criteria do not agree. For example, the boundary dual algebras of bulk causal diamonds may be delocalized and not admit a geometric interpretation. If one took two spacelike separated bulk diamonds, their bulk algebras will commute, as will their corresponding boundary avatars, even though they may not be localizable to spacelike separate subregions of the boundary theory. The results of this paper may be viewed as steps towards a generalisation of this phenomenon. In particular, mapping localised spacelike separated relational supports from one theory to another may not necessarily result in the same localisation relative to the new microcausal frame.

  Before continuing, we note that, besides frames for small gauge transformations, we may also consider small gauge-\emph{invariant} frames that instead are associated with large gauge transformations \cite{Araujo-Regado:2024dpr,Araujo-Regado:2025ejs}. For instance, we saw in the previous section that $\alpha$ is an example of a large gauge frame, and we will shortly encounter it again.

  Let us now make these considerations more precise in this section.
  We will begin by confirming that, within the same theory, other frames than the microcausal one do not give rise to a valid relational criterion for microcausality. More precisely, spacelike separated relational support relative to other frames than the preferred one does not imply commutation of the corresponding observables.
  Thus, \emph{microcausality is frame-dependent}. We will then consider how the relational support of large gauge dependent observables transforms under reorientations of the $\alpha$-frame, before exploring at a general level what it may mean to switch between theories with different microcausal frames. Finally, we briefly discuss cases when these transformations preserve commutation relations, so that we can trade microcausality criteria between theories.

  \subsection{Microcausality is frame-dependent}

  As explained above, a `frame' is a very general kind of object -- it is just something we can use to dress non-invariant bare observables. But, to be explicit, we will focus on frames of the kind constructed in the previous sections. In our context, we can view such frames as being parametrised by the collection of objects $\mathcal{F} = (\mathcal{N},f,\gamma)$. By changing these parameters, we get a different frame, and each such frame has its own notion of relational support.

  Suppose
  \begin{equation}
    \mathcal{F}=(\mathcal{N},f,\gamma), \qquad \mathcal{F}'=(\mathcal{N}',f',\gamma')
  \end{equation}
  are two such frames, and let us use
  \begin{equation}
    \relsupp, \qquad \relsupp'
  \end{equation}
  to denote the relational supports associated with $\mathcal{F},\mathcal{F}'$ respectively. We shall assume that the first frame $\mathcal{F}$ is the microcausal frame, i.e.\ the one that went into defining the theory through its boundary conditions. Thus, spacelike separation of $\relsupp$ is a good criterion for commutativity. The second frame $\mathcal{F}'$ a priori has nothing to do with the boundary conditions, but we can still ask: is spacelike separation of $\relsupp'$ also a good criterion for commutativity? The answer is: in general, no. Let us demonstrate this now.

  For simplicity, let us take $\mathcal{N}=\mathcal{N}'$ and $f=f'$, so that the endpoints of $\gamma(y)$ and $\gamma'(y)$ coincide for all $y\in\mathcal{N}$. Let us consider the two observables
  \begin{equation}
    {O}_1 = \psi(y) \exp\qty(-iq\int_{\gamma'(y)}A),\qquad
    {O}_2 = F(x),
  \end{equation}
  where $x$ is some point outside $\mathcal{N}$, spacelike separated from $y$. Both of these are small-gauge-invariant; the first is the scalar dressed along the curve $\gamma'(y)$, while the second is the field strength tensor at $x$.

\begin{figure}
    \centering
    \begin{tikzpicture}[scale=2.6]
      \fill[blue!20] (0,2.2) -- (0,4) arc (180:0:2 and 0.3) -- (4,2.2) arc (360:180:2 and 0.3);

      \fill[red!40!blue,opacity=0.15,shift={(0.4,0)}] (1.3,2.2) -- (1.3,4) arc (180:0:1 and 0.15) -- (3.3,2.2) arc (0:180:1 and 0.15);
      \fill[red!40!blue,opacity=0.15,shift={(0.4,0)}] (1.3,2.2) -- (1.3,4) arc (180:360:1 and 0.15) -- (3.3,2.2) arc (360:180:1 and 0.15);

      \draw[thick,blue!60!black] (0,2.2) -- (0,4) arc (180:0:2 and 0.3) -- (4,2.2) arc (360:180:2 and 0.3);
      \draw[thick,blue!60!black] (0,4) arc (-180:0:2 and 0.3);
      \draw[thick,dashed,blue!60!black] (0,2.2) arc (180:0:2 and 0.3);

      \draw[red!40!blue,shift={(0.4,0)}] (2.3,4) ellipse (1 and 0.15);
      \draw[red!40!blue,shift={(0.4,0)}] (3.3,4) -- (3.3,2.2) arc (360:180:1 and 0.15) -- (1.3,2.2) -- (1.3,4);
      \draw[dashed,red!40!blue,shift={(0.4,0)}] (1.3,2.2) arc (180:0:1 and 0.15);

      \fill[yellow,opacity=0.3] (1.4,2.6) -- (2,3.6) .. controls (1.6,3.7) and (1.2,3.7) .. (0.8,3.6);
      \fill[yellow,opacity=0.3] (1.4,2.6) -- (1.7,2.1) .. controls (1.5,2) and (1.3,2) .. (1.1,2.1);

      \begin{scope}[decoration={
            markings,
            mark=at position 0.5 with {\arrow[scale=0.8]{<}}
        }]
        \draw[red!80!black,thick,postaction={decorate}] (0.4,3.2) .. controls (0.9,3.4) and (1.6,3.4) .. (1.9,3.1);
        \draw[red!80!black,thick,postaction={decorate}] (0.4,3.2) .. controls (0.9,2.8) and (1.6,2.8) .. (1.9,3.1);
      \end{scope}
      \begin{scope}[decoration={
            markings,
            mark=at position 0.3 with {\arrow[scale=0.8]{<}}
        }]
        \draw[red!80!black,thick,postaction={decorate}] (0.5,3.18) .. controls (0.9,2.9) and (1.6,2.87) .. (1.8,3.1);
        \draw[red!80!black,thick,postaction={decorate}] (1.8,3.1) .. controls (1.6,3.3) and (0.9,3.35).. (0.5,3.18);
      \end{scope}
      \begin{scope}
        \clip(1.4,2.6) -- (2,3.6) .. controls (1.6,3.7) and (1.2,3.7) .. (0.8,3.6);
        \draw[red!60!yellow,line width=1.4pt] (0.5,3.18) .. controls (0.9,2.9) and (1.6,2.87) .. (1.8,3.1);
        \draw[red!60!yellow,line width=1.4pt] (1.8,3.1) .. controls (1.6,3.3) and (0.9,3.35).. (0.5,3.18);
      \end{scope}

      \fill[red!40!black] (0.4,3.2) circle (0.02);
      \fill[red!40!black] (1.9,3.1) circle (0.02);
      \fill[red!40!black] (1.4,2.6) circle (0.02);
      \node[right] at (1.4,2.6) {$x$};
      \node[above] at (0.3,3.23) {$f(y)$};
      \node[above] at (1.93,3.1) {$y$};
      \node[red!50!black,below] at (1,3.6) {$\gamma(y)$};
      \node[red!50!black,below] at (0.8,2.95) {$\gamma'(y)$};
      \node[red!50!black] at (1,3.1) {$C$};
      \node[fill=blue!10!white, fill opacity=0.8,text opacity=1, rounded corners] at (1.4,2.3) {${O}_2=F(x)$};
      \node[fill=red!40!blue!10!white, fill opacity=0.8,text opacity=1,rounded corners] at (2.86,3) {$
        \begin{aligned}
          {O}_1&=\psi(y)e^{-iq\int_{\gamma'(y)}A}\\
          &= \Psi(y) e^{-iq\int_C A}
      \end{aligned}$};
    \end{tikzpicture}
    \caption{$O_1$ and $O_2$ are spacelike separated according to the relational support $\relsupp'$ of frame $\mathcal{F}'$, but not according to the relational support $\relsupp$ of the \emph{microcausal} frame $\mathcal{F}$. Therefore, they do not necessarily commute; indeed they will generically not commute if $C=\gamma(y)\sqcup\gamma(y')$ intersects the lightcone of $x$.}
    \label{Figure: failed primed microcausality}
  \end{figure}
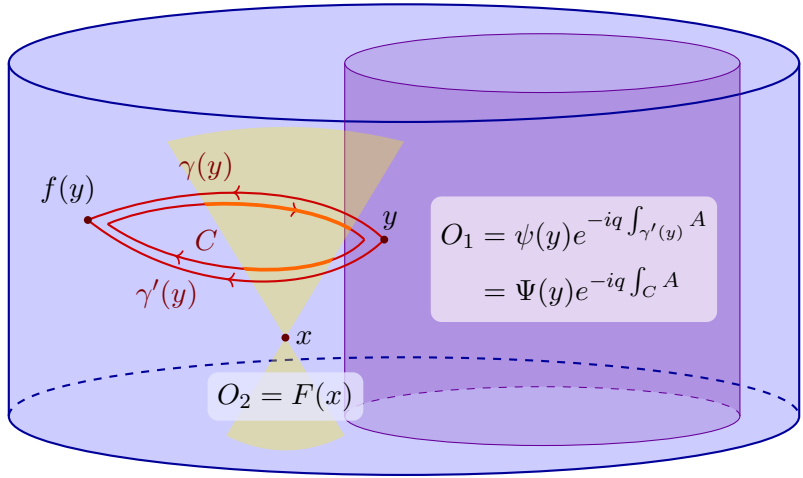

  We have
  \begin{equation}
    \relsupp'({O}_1) = y, \qquad \relsupp'({O}_2) = x,
  \end{equation}
  so the $\relsupp'$ of these observables are spacelike separated. A simple test of the criterion would therefore be to ask whether these two observables commute -- but this test fails. Indeed, let us decompose the first observable as ${O}_1 = \Psi(y)W_C$, where
  \begin{equation}
    \Psi(y) = \psi(y) \exp\qty\Big(-iq\int_{\gamma(y)}A), \qquad W_C = \exp\qty\Big(-iq\int_CA),
  \end{equation}
  and $C=\gamma'(y)\sqcup \gamma(y)^{\rm T}$, with $\gamma(y)^{\rm T}$ denoting the orientation reversal of $\gamma(y)$. Now
  \begin{equation}
    \relsupp(\Psi(y)) = y, \qquad \relsupp({O}_2) = x,
  \end{equation}
  so we know that $\pb{\Psi(y)}{{O}_2}=0$, and therefore
  \begin{equation}
    \pb{{O}_1}{{O}_2} = \Psi(y)\pb{W_C}{F(x)}
  \end{equation}
  so ${O}_1$ and ${O}_2$ commute only if $W_C$ and $F(x)$ commute. But both of these are now completely gauge-invariant observables,\footnote{One can optionally assume $\gamma(y)$ and $\gamma'(y)$ coincide in a neighbourhood of $\Gamma$, so that both $W_C$ and $F(x)$ have compact bulk support. But this is not strictly necessary for this argument.} so we have
  \begin{equation}
    \relsupp({O}_1) = C = \gamma(y)\cup \gamma'(y), \qquad \relsupp({O}_2) = x,
  \end{equation}
  The brackets of completely gauge-invariant observables do not depend on the Goldstone boundary conditions. All that matters is the Dirichlet (up to large gauge) part of the boundary conditions, for which it is well-known that the field strength $F(x)$ does not commute with Wilson loops from which $x$ is \emph{not} spacelike separated (and here there is no requirement for $C$ to be spacelike separated from $x$). So $\pb{{O}_1}{{O}_2}\ne 0$. This is depicted in Figure~\ref{Figure: failed primed microcausality}.

  So microcausality is frame-dependent. To be precise, for an arbitrary frame, spacelike separation of relational support will not suffice for commutativity -- so microcausality for that frame will generically fail. Only certain special frames (such as $\mathcal{F}$) will have an associated notion of relational support whose spacelike separation guarantees commutativity -- so microcausality for those frames holds.

  \subsection{Orientations and reorientations}

  The internal state of a frame is known as its `orientation'. For example, the orientation of an inertial reference frame in a Poincar\'e-invariant theory includes data such as its position, and the orientation of a Lorentz tetrad. Notably, these orientations take value in the group under which they transform -- in this case the Poincar\'e group.  A frame `reorientation' is a change in state which modifies the orientation of the frame, but which does not modify any of the bare quantities the frame can be used to dress \cite{Carrozza:2021gju,Carrozza:2022xut,Araujo-Regado:2024dpr,Hoehn:2023ehz,delaHamette:2021oex,DeVuyst:2024uvd,Araujo-Regado:2025ejs,Freidel:2025ous}. For an inertial reference frame, reorientations include translations and boosts of that frame.

  To emphasise further that $\mathcal{F}=(\mathcal{N},f,\gamma)$ should be understood as a frame, let us briefly explain how these concepts apply to it. The orientations of $\mathcal{F}$ may be understood as the collection of values taken by the Wilson lines $\gamma(y)$, $y\in\mathcal{N}$, since these are what the frame equips us with to dress bulk observables. Similarly to the inertial frame case, these Wilson lines take value in the local structure group $\rm{U}(1)$; a given $\mathcal{F}$ thus takes value in the small gauge transformations on $\mathcal{N}$. Large gauge transformations are one way to realize reorientations of $\mathcal{F}$ \cite{Araujo-Regado:2024dpr}; these leave the bare quantities on $\mathcal{N}$ unchanged.  As we have already explained, such transformations are generated by $Q[F],P[F]$, and the algebra of these charges is given by~\eqref{Equation: QP}.

  In what follows, we may view large gauge transformations as reorientations of both the small gauge frame $\mathcal{F}$ given by the Wilson lines, and the large gauge frame $\alpha$. Indeed, it was already observed that $\alpha$ is the difference between a \emph{charged} small gauge frame, namely the Wilson lines of $\mathcal{F}$, and an \emph{uncharged} small gauge frame, namely the phase of the scalar field on $\mathcal{N}$. We also saw in the previous section that, in line with reorientations of $\alpha$, all small gauge-invariant observables can be decomposed into completely gauge-invariant observables and $\alpha$. Thus, there exist decompositions such that large gauge transformations only change $\alpha$ and leave all other small gauge-invariant data invariant.

  The charges $Q[F],P[F]$ only generate \emph{orientation-independent} reorientations, since they each yield a large gauge transformation whose parameter is determined by the fixed function $F$. More generally, \emph{orientation-dependent} reorientations are generated by any functionals $R$ of $\alpha|_{\mathcal{N}}$.

  It is worth noting that a general orientation-dependent frame reorientation will change the relational support of whatever it acts on. For example, if $\mathcal{M}$ is a stationary spacetime, and $\mathcal{N}$ is preserved by the corresponding timelike Killing vector field $\xi$ then an example of an orientation-dependent reorientation is the transformation under which $\alpha$ undergoes time evolution, but completely gauge-invariant observables are unaffected. This is generated by the Hamiltonian of $\alpha$:
  \begin{equation}
    h = \frac12\int_K \epsilon\qty(\dot\alpha^2 + \abs{\partial_K\alpha}^2).
  \end{equation}
  Here we are taking a $K$ which is normal to $\xi$. We have
  \begin{equation}
    \pb{\alpha(y)}{h} = \dot\alpha(y),
  \end{equation}
  so the action of $h$ moves along $\xi$ the relational support of observables formed only out of $\alpha$. On the other hand, $h$ commutes with any completely gauge-invariant observable, so the relational support of such observables is unaffected.

  More generally, if $\zeta$ is a Killing vector field of the induced metric on $\mathcal{N}$, then the evolution of $\alpha$ along $\zeta$ is an orientation-dependent reorientation, and it is generated by
  \begin{equation}
    R_\zeta = \int_K \epsilon m^i\zeta^j t_{ij},
  \end{equation}
  where $m$ is the unit normal to $K$ as a submanifold of $\mathcal{N}$, $i,j,\dots$ are abstract indices of tensors on $\mathcal{N}$, and
  \begin{equation}
    t_{ij} = \partial_i\alpha \partial_j\alpha - \frac12 \eta_{ij}\abs{\partial\alpha}^2
  \end{equation}
  is the `stress tensor' of $\alpha$, where $\eta$ is the induced metric on $\mathcal{N}$. We have
  \begin{equation}
    \pb{\alpha(y)}{R_\zeta} = \zeta^i\partial_i\alpha(y),
  \end{equation}
  so the action of $R_\zeta$ moves along $\zeta$ the relational support of observables formed out of $\alpha$, while leaving invariant the relational support of completely gauge-invariant observables.

  We can call a frame reorientation `local' if it does not change the relational support of whatever it acts on. Clearly, $Q[F],P[F]$ generate local frame reorientations, but there are also local orientation-dependent reorientations, for example the rescaling $\alpha\to c\alpha$, for some constant $c\in\RR$. This is generated by $Z=\int_K\epsilon \alpha\dot\alpha$.

  \subsection{Switching between theories with different microcausal frames}

  As we have already indicated, the privileged status of the frame given by the Wilson lines along $\gamma(y),y\in\mathcal{N}$ is due to the choice of Goldstone boundary condition $\Box_{\mathcal{N}}\alpha=0$. Therefore, if we want to \emph{change} the privileged microcausal frame, we have to change this boundary condition, which consequently changes the theory. For brevity, we shall henceforth simply refer to this as a `change of microcausal frame', even though it entails a change of theory also. For concreteness, let us consider what happens when we change from $\mathcal{F}=(\mathcal{N},f,\gamma)$ to $\mathcal{F}'=(\mathcal{N}',f',\gamma')$, which correspond to two distinct theories, with phase spaces $(\mathcal{P},\Omega)$ and $(\mathcal{P}',\Omega')$ respectively. Let $\alpha,\alpha'$ denote the appropriate dressed phases of $\psi$ on $\mathcal{N},\mathcal{N}'$ along $\gamma,\gamma'$ in each case respectively.

  Given a state of the first theory, a field configuration in $\mathcal{P}$, how should we map it to a state in the second theory, a field configuration in $\mathcal{P}'$? Here is one way. We keep the initial data on the initial and final Cauchy surfaces the same, but change the boundary conditions on $\Gamma$ from those associated with $\mathcal{F}$ to those associated with $\mathcal{F}'$. This defines a map $m:\mathcal{P}\to\mathcal{P}'$ from the first theory to the second theory.

  Since the boundary conditions of the two theories are determined by boundary action, the map $m$ can be thought of as being implemented by a change in action. In the quantum theory, this would be done via the insertion of operators in the path integral, but we reserve this discussion for the next section. The analogue of this procedure in the classical theory (within the Hamilton-Jacobi formalism) is a deformation of the action by some observable.

  Before doing the change of microcausal frame, the action is just that of the original theory. Including the change of microcausal frame deforms the action to that of the second theory, with a different Goldstone boundary condition. The actions in the two theories may be written as
  \begin{align}
    S &= S_0 + \frac12\int_{\mathcal{N}}\dd{\alpha}\wedge\Hodge\dd{\alpha}, \label{Equation: action first frame}\\
    S' &= S_0 + \frac12\int_{\mathcal{N}'}\dd{\alpha'}\wedge\Hodgee\dd{\alpha'}, \label{Equation: action second frame}
  \end{align}
  where $\star'$ is the Hodge star of $\mathcal{N}'$, and
  \begin{equation}
    S_0 = \frac12\int_{\mathcal{M}} \qty(-F\wedge \hodge F + \Dd{\psi}\wedge\hodge\overline{\Dd{\psi}} + \hodge V(\abs{\psi}^2)),
  \end{equation}
  is the standard bulk contribution.

  Due to the different Goldstone boundary conditions, observables respect different kinds of microcausality after deforming the action. Indeed, after deforming the action we should use $\relsupp'$ to test for microcausality, while before deforming the action we should use $\relsupp$.

  One can use this setup to `pullback' observables between the two theories: we can change from the first theory to the second theory, measure an observable $O'$  from that second theory, and then change back to the first theory. The end result is a procedure for measuring an observable $m^*O'$ in the first theory. But note that, in general, one should not expect this map to be a canonical transformation, meaning that the Poisson brackets of pulled back observables will not agree with the pullback of the Poisson brackets of observables:
  \begin{equation}
    \pb{m^*O_1'}{m^*O_2'}\ne m^*\pb{O_1'}{O_2'} \qquad \text{in general.}
  \end{equation}
  This means that even if $O_1'$ and $O_2'$ commute in the second theory, their pullbacks do not necessarily commute in the first theory.

  Let us be clear that the setups associated with the two sides of the above inequality are subtly distinct. On the left, we do the change of frames procedure each time we measure an observable. Then we compute the \emph{overall} bracket between two such pullback observables. On the right, we only do the change of frames procedure \emph{once}, computing the bracket entirely with the new boundary conditions.

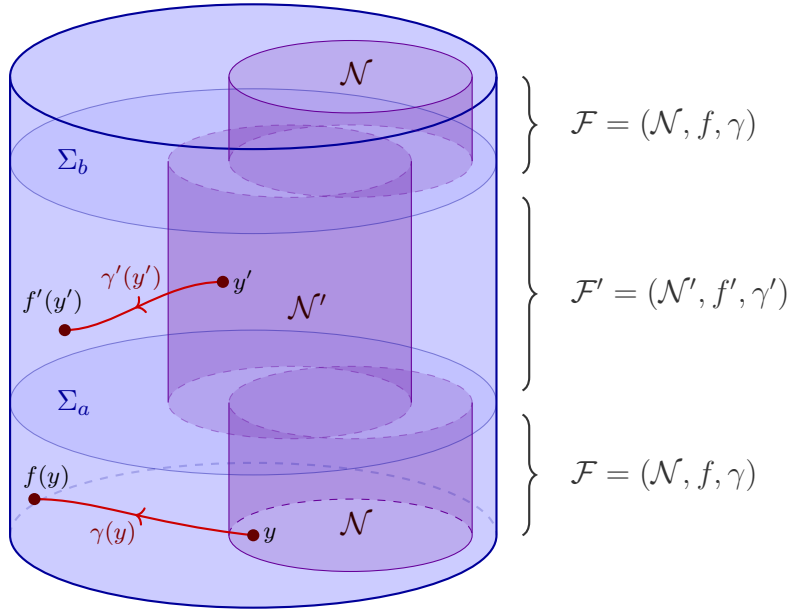
\begin{figure}
    \centering
    \hspace{8.5em}
    \begin{tikzpicture}[scale=1.6]
      \fill[blue!20] (0,2.2) -- (0,6) arc (180:0:2 and 0.6) -- (4,2.2) arc (360:180:2 and 0.6);

      \begin{scope}[shift={(0.5,0)}]
        \fill[red!40!blue,opacity=0.15] (1.3,2.2) -- (1.3,3.3) arc (180:0:1 and 0.3) -- (3.3,2.2) arc (0:180:1 and 0.3);
        \fill[red!40!blue,opacity=0.15] (1.3,2.2) -- (1.3,3.3) arc (180:360:1 and 0.3) -- (3.3,2.2) arc (360:180:1 and 0.3);
      \end{scope}

      \fill[blue,opacity=0.05] (2,3.3) ellipse (2 and 0.6);

      \fill[red!40!blue,opacity=0.15] (1.3,3.3) -- (1.3,5.3) arc (180:0:1 and 0.3) -- (3.3,3.3) arc (0:180:1 and 0.3);
      \fill[red!40!blue,opacity=0.15] (1.3,3.3) -- (1.3,5.3) arc (180:360:1 and 0.3) -- (3.3,3.3) arc (360:180:1 and 0.3);

      \fill[blue,opacity=0.05] (2,5.3) ellipse (2 and 0.6);

      \begin{scope}[shift={(0.5,0)}]
        \fill[red!40!blue,opacity=0.15] (1.3,5.3) -- (1.3,6) arc (180:0:1 and 0.3) -- (3.3,5.3) arc (0:180:1 and 0.3);
        \fill[red!40!blue,opacity=0.15] (1.3,5.3) -- (1.3,6) arc (180:360:1 and 0.3) -- (3.3,5.3) arc (360:180:1 and 0.3);
      \end{scope}

      \draw[thick,blue!60!black] (0,2.2) -- (0,6) arc (180:0:2 and 0.6) -- (4,2.2) arc (360:180:2 and 0.6);
      \draw[thick,blue!60!black] (0,6) arc (-180:0:2 and 0.6);
      \draw[thick,dashed,blue!60!black,opacity=0.2] (0,2.2) arc (180:0:2 and 0.6);

      \draw[blue!60!black,opacity=0.4](2,3.3) ellipse (2 and 0.6);
      \draw[blue!60!black,opacity=0.4] (2,5.3) ellipse (2 and 0.6);

      \node[right,blue!60!black] at (0.3,3.3) {$\Sigma_a$};
      \node[right,blue!60!black] at (0.3,5.3) {$\Sigma_b$};

      \draw[red!40!blue,opacity=0.4,dashed] (2.8,3.3) ellipse (1 and 0.3);
      \draw[red!40!blue] (3.8,3.3) -- (3.8,2.2) arc (360:180:1 and 0.3) -- (1.8,2.2) -- (1.8,3.3);
      \draw[dashed,red!40!blue] (1.8,2.2) arc (180:0:1 and 0.3);

      \draw[red!40!blue,opacity=0.4,dashed] (2.3,5.3) ellipse (1 and 0.3);
      \draw[red!40!blue] (3.3,5.3) -- (3.3,3.3);
      \draw[red!40!blue] (1.3,3.3) -- (1.3,5.3);
      \draw[red!40!blue,opacity=0.4,dashed] (2.3,3.3) ellipse (1 and 0.3);

      \draw[red!40!blue] (2.8,6) ellipse (1 and 0.3);
      \draw[red!40!blue] (3.8,6) -- (3.8,5.3);
      \draw[red!40!blue] (1.8,5.3) -- (1.8,6);
      \draw[red!40!blue,opacity=0.4,dashed] (2.8,5.3) ellipse (1 and 0.3);

      \node[red!20!black] at (2.85,2.3) {\large$\mathcal{N}$};
      \node[red!20!black] at (2.45,4.1) {\large$\mathcal{N}'$};
      \node[red!20!black] at (2.85,6.03) {\large$\mathcal{N}$};

      \begin{scope}[shift={(-0.1,-2)}]
        \begin{scope}[decoration={
              markings,
              mark=at position 0.5 with {\arrow{<}}
          }]
          \draw[red!80!black,thick,postaction={decorate}] (0.3,4.5) .. controls (0.8,4.5) and (1.2,4.3) .. (2.1,4.2);
        \end{scope}
        \fill[red!40!black] (0.3,4.5) circle (0.05);
        \fill[red!40!black] (2.1,4.2) circle (0.05);
        \node[above] at (0.4,4.5) {\footnotesize$f(y)$};
        \node[right] at (2.1,4.2) {\footnotesize$y$};
        \node[red!50!black,below] at (0.95,4.4) {\footnotesize$\gamma(y)$};
      \end{scope}

      \begin{scope}[shift={(0.05,0.7)}]
        \begin{scope}[decoration={
              markings,
              mark=at position 0.5 with {\arrow{<}}
          }]
          \draw[red!80!black,thick,postaction={decorate}] (0.4,3.2) .. controls (0.8,3.2) and (1.2,3.6) .. (1.7,3.6);
        \end{scope}
        \fill[red!40!black] (0.4,3.2) circle (0.05);
        \fill[red!40!black] (1.7,3.6) circle (0.05);
        \node[above] at (0.3,3.25) {\footnotesize$f'(y')$};
        \node[right] at (1.7,3.6) {\footnotesize$y'$};
        \node[red!50!black,above] at (0.95,3.45) {\footnotesize$\gamma'(y')$};
      \end{scope}

      \draw [black!80,thick,decorate,decoration={brace,amplitude=5pt,mirror,raise=4ex}]
      (3.8,2.2) -- (3.8,3.2) node[midway,xshift=3em,right]{\large $\mathcal{F}=(\mathcal{N},f,\gamma)$};
      \draw [black!80,thick,decorate,decoration={brace,amplitude=5pt,mirror,raise=4ex}]
      (3.8,3.4) -- (3.8,5) node[midway,xshift=3em,right]{\large $\mathcal{F}'=(\mathcal{N}',f',\gamma')$};
      \draw [black!80,thick,decorate,decoration={brace,amplitude=5pt,mirror,raise=4ex}]
      (3.8,5.2) -- (3.8,6) node[midway,xshift=3em,right]{\large $\mathcal{F}=(\mathcal{N},f,\gamma)$};
    \end{tikzpicture}
    \caption{At $\Sigma_a$ we change the microcausal frame associated with the boundary conditions from $\mathcal{F}=(\mathcal{N},f,\gamma)$ to $\mathcal{F}'=(\mathcal{N}',f',\gamma')$. At $\Sigma_b$ we change back to $\mathcal{F}$.}
    \label{Figure: change frame}
  \end{figure}

  We should note also that one may generalise the above procedure such that the change of boundary conditions only takes place between two particular Cauchy surfaces $\Sigma_a$ and $\Sigma_b$. The general situation is depicted in Figure~\ref{Figure: change frame}. This would define a map $m_{\Sigma_a,\Sigma_b}$, which in general depends on the choice of $\Sigma_a,\Sigma_b$. Physically speaking, one can imagine that one is switching frame $\mathcal{F}\to\mathcal{F}'$ at $\Sigma_a$, and then switching back $\mathcal{F}'\to\mathcal{F}$ at $\Sigma_b$.

  Before considering microcausal frame changes more explicitly, let us pause and inquire about covariance. By frame covariance of a given theory we usually mean that the descriptions of the \emph{totality} of possible physical states and observables relative to different frames from the pertinent covariance class are exactly identical. The paradigmatic example are Lorentz frames in special relativistic theories. While a fixed physical situation looks different to different frames, this covariance means that we cannot operationally distinguish, in an absolute sense, one frame from the covariance class from another.

  It is already clear that any of our theories, with fixed Goldstone boundary conditions, is not covariant in this sense. After all, we have \emph{preferred} frames $\mathcal{F}$ -- those for which microcausality holds. These could be operationally distinguished from non-microcausal frames. We may thus inquire about a more general notion of frame covariance, one \emph{across} different theories. More precisely, if $\mathcal{F}$ is the microcausal frame of theory $\mathcal{T}$ and $\mathcal{F}'$ the microcausal frame of theory $\mathcal{T}'$, is there a sense in which the respective spaces of solutions look `the same' when described with the respective frame dressings? $\mathcal{T},\mathcal{T}'$ may differ, for instance, by the Goldstone boundary conditions on $\mathcal{N}$ and $\mathcal{N}'$.

  Also the answer to this question is negative. First, if such a covariance across theories existed, it would be impossible to operationally distinguish $\mathcal{T}$ from $\mathcal{T}'$. But due to the different boundary conditions, the same initial data sets will generically yield solutions of $\mathcal{T}$ and $\mathcal{T}'$ that look different relative to the spacetime background. Second, even in a weaker relational sense these solutions will generically appear distinct.   For example, we may have that $\mathcal{F}'$ is simply the image of $\mathcal{F}$ under some (possibly infinitesimal) bulk diffeomorphism, as described below. The new action may then still be given by~\eqref{Equation: action second frame}  and corresponds to a valid choice within the class of frames and theories under consideration. Since our theories are not diffeomorphism-invariant, however, we cannot simply also transform field configurations by the same diffeomorphism. While this would lead to the same spacetime coincidences of the fields -- in this relational sense looking `the same' -- it would in general map solutions of $\mathcal{T}$ to non-solutions of $\mathcal{T}'$. Thus, we cannot expect a microcausal frame covariance across the class of theories under consideration. In this sense, our construction with frame-dependent boundary conditions manifestly breaks dynamical frame covariance both within and across theories.

  \subsection{Generators of microcausal frame changes}
  \label{Section: generators}

  Let us now derive the generators of changes of the microcausal frame. We can do so by considering the difference between the two actions~\eqref{Equation: action first frame} and~\eqref{Equation: action second frame}:
  \begin{equation}
    \Delta S = S-S' = \frac12\int_{\mathcal{N}}\dd{\alpha}\wedge\Hodge\dd{\alpha} - \frac12\int_{\mathcal{N}'}\dd{\alpha'}\wedge\Hodgee\dd{\alpha'}.
    \label{Equation: action difference}
  \end{equation}
  Clearly, if we deform the action $S$ of the first theory via $S\to S-\Delta S$, we get $S'$. Therefore, we can think of this difference $\Delta S$ as leading to the change of microcausal frames $\mathcal{F}\to\mathcal{F}'$. This deformation can be compared to the deformation $S\to S-\lambda O$ that one considers when computing the Peierls bracket.

  For concreteness, let us consider an infinitesimal change of frames, meaning $\mathcal{N}'$ and the curves $\gamma'$ can be reached from $\mathcal{N}$ and the curves $\gamma$ by flowing an amount $\lambda$ along a spacetime vector field $\upsilon$, where $0<\lambda\ll1$. We can recover finite changes of frame by exponentiating such infinitesimal changes. Without loss of generality, we can take $\upsilon$ to vanish at $\Gamma$, so that $\gamma(y)$ and $\gamma'(y')$ are related in this way if $f(y)=f'(y')$.

  We can then write
  \begin{equation}
    \alpha'(y') = \alpha(y) +\lambda B_\upsilon(y) + \order{\lambda^2}, \qq{where} B_\upsilon(y) = - \frac{\iota_\upsilon j (y)}{q^2\abs{\psi(y)}^2}+ \int_{\gamma(y)}\iota_\upsilon F
  \end{equation}
  and $f(y)=f'(y')$. Using this in~\eqref{Equation: action difference} yields
  \begin{equation}
    \Delta S = -\lambda\mathcal{C}_\upsilon+\order{\lambda^2},\qq{where} \mathcal{C}_\upsilon=\int_{\mathcal{N}}\qty(\dd{B_\upsilon}+Y_\upsilon\dd{\alpha})\wedge\Hodge\dd{\alpha},
  \end{equation}
  with $Y_\upsilon$ the conformal factor of the diffeomorphism $\mathcal{N}\to\mathcal{N}'$ generated by $\upsilon$ (it appears here due to the change in Hodge star operator $\star$). From the above discussion, $\mathcal{C}_\upsilon$ can be understood as the generator of the change of microcausal frame. An interesting special case is changing $f\to f'$ and $\gamma\to\gamma'$, but leaving $\mathcal{N}$ invariant. This corresponds to a change of dressing on the same surface.

  Here, by $\mathcal{C}_\upsilon$ being a `generator', we mean that the change of frames is carried out by inserting $\mathcal{C}_\upsilon$ in the action. This is not necessarily the same as $\mathcal{C}_\upsilon$ performing the change of frames via Poisson brackets, which will only be true if the change of frames is a canonical transformation. We will explore the conditions under which this is true in the following subsections.

  The change of frames generator $\mathcal{C}_\upsilon$ is not a large-gauge-invariant quantity; indeed, one has that it changes by
  \begin{equation}
    \delta_\Lambda \mathcal{C}_\upsilon = -\int_\mathcal{N} (\dd{B_\upsilon} +2Y_\upsilon\dd{\alpha})\wedge\Hodge\dd{\Lambda_f}
  \end{equation}
  under a large gauge transformation parametrised by $\Lambda$.
  Using the equations of motion $\dd{\Hodge\dd{\alpha}}=0$ and the condition that large gauge transformations must preserve the Goldstone boundary conditions, i.e.\  $\dd{\Hodge\dd{\Lambda_f}}=0$, one can write $\delta_\Lambda\mathcal{C}_\upsilon$ in the form
  \begin{equation}
    \delta_\Lambda\mathcal{C}_\upsilon = \int_{\mathcal{N}}\Lambda_f\dd(2Y_\upsilon\Hodge\dd{\alpha}) - \int_{\partial\mathcal{N}}\qty(B_\upsilon\Hodge\dd{\Lambda_f}+2Y_\upsilon\Lambda_f\Hodge\dd{\alpha})
    \label{Equation: large gauge variation of C_upsilon}
  \end{equation}
  We can interpret the first term on the right-hand side of~\eqref{Equation: large gauge variation of C_upsilon}, which may also be written
  \begin{equation}
    \int_{\mathcal{N}}\Lambda_f\dd(2Y_\upsilon\Hodge\dd{\alpha}) = \int_{\Gamma}\Lambda f^*\qty(\dd(2Y_\upsilon\Hodge\dd{\alpha})),
  \end{equation}
  as accounting for a `flux' through $\mathcal{N}$ (or $\Gamma$), while the second term on the right-hand side of~\eqref{Equation: large gauge variation of C_upsilon} accounts for `corner charges' at $K_\pm$. Indeed, using the Goldstone boundary condition, we may explicitly decompose $\mathcal{C}_\upsilon$ into flux and corner terms:
  \begin{equation}
    \mathcal{C}_\upsilon = \mathscr{F}^\upsilon - X^\upsilon_-+X^\upsilon_+,
  \end{equation}
  where
  \begin{equation}
    \mathscr{F}^\upsilon = -\int_{\mathcal{N}}\alpha\dd{Y_\upsilon}\wedge\Hodge\dd{\alpha}, \qquad X^\upsilon_\pm = -\int_{K_\pm} (B_\upsilon+Y_\upsilon\alpha)\star\dd{\alpha}.
  \end{equation}
  Note that the flux term $\mathscr{F}^\upsilon$ only depends on $\alpha$.

  One can similarly compute the generator $\mathcal{C}^\upsilon_{a,b}$ of the change of frames between any two Cauchy surfaces $\Sigma_{a,b}$, as depicted in Figure~\ref{Figure: change frame}, finding again a decomposition into a flux term and corner charges:
  \begin{equation}
    \mathcal{C}_{a,b}^\upsilon = \mathscr{F}_{a,b}^\upsilon - X^\upsilon_a+X^\upsilon_b,
  \end{equation}
  where
  \begin{equation}
    \mathscr{F}_{a,b}^\upsilon = -\int_{\mathcal{N}_{a,b}}\alpha\dd{Y_\upsilon}\wedge\Hodge\dd{\alpha}, \qquad X^\upsilon_a = -\int_{K_a} (B_\upsilon+Y_\upsilon\alpha)\Hodge\dd{\alpha}
  \end{equation}
  with $\mathcal{N}_{a,b}$ the portion of $\mathcal{N}$ in between $K_a$ and $K_b$, and $K_{a,b}=\Sigma_{a,b}\cap\mathcal{N}$.

  \subsection{Canonical changes of microcausal frame}
  \label{Section: canonical change}

  It is worth asking: when is the change of microcausal frame described above canonical, i.e.\ a symplectomorphism? In general, this will not be the case. But, as we will argue shortly, a sufficient condition for a canonical change of microcausal frame is that the flux term $\mathscr{F}^\upsilon$ vanishes, which happens when $Y_\upsilon$ is constant on $\mathcal{N}$, such that $\mathcal{C}^\upsilon = -X^\upsilon_-+X^\upsilon_+$. Moreover, $\mathcal{C}_\upsilon$ is in this case then promoted to the Hamiltonian generator of the change of frame, meaning the effect of this change on an observable $O$ is given by the Poisson bracket $\pb{\mathcal{C}_{\upsilon}}{O}$.

  To see why this is true, note that deforming the action via $S\to S-\lambda\mathcal{C}_{\upsilon}+\order{\lambda^2}$ results in a deformation of the on-shell field configuration $\phi\to\phi+\lambda\delta_\upsilon\phi+ \order{\lambda^2}$. When the flux $\mathscr{F}^\upsilon$ vanishes, the field variation $\delta_\upsilon\phi$ decomposes into two pieces:
  \begin{equation}
    \delta_\upsilon\phi = \delta^+_{X^-_\upsilon}\phi + \delta^-_{X^+_\upsilon}\phi,
  \end{equation}
  where $\pm\delta^\pm_{X^\mp_\upsilon}\phi$ is the field variation resulting from deforming the action by $X^\mp_\upsilon$. Since $X^\mp_\upsilon$ are located on the past/future Cauchy surfaces $\Sigma_\mp$ respectively, recalling the Peierls bracket machinery described in Section~\ref{Section: microcausality}, we can identify $\delta^+_{X^-_\upsilon}\phi$, $\delta^-_{X^+_\upsilon}\phi$ as precisely the retarded/advanced field variations generated by $X^-_\upsilon$, $X^+_\upsilon$ respectively. Moreover, the variation due to $\delta_\upsilon\phi$ of any observable $O$ located between $\Sigma_-$ and $\Sigma_+$ is given by
  \begin{align}
    \delta_\upsilon O &= \delta^+_{X^-_\upsilon}O + \delta^-_{X^+_\upsilon}O\\
    &= \delta_{X^-_\upsilon}O - \delta_{X^+_\upsilon}O,
  \end{align}
  where
  \begin{equation}
    \delta_{X^\pm_\upsilon}\phi = \delta^+_{X^\pm_\upsilon}\phi - \delta^-_{X^\pm_\upsilon}\phi
  \end{equation}
  is the full field variation generated by $X^\pm_\upsilon$ (c.f.\ \eqref{Equation: peierls difference}). Here we have used the fact that $O$ being between $\Sigma_\pm$ implies $\delta^\pm_{X^\pm_\upsilon}O=0$. Comparing to the definition of the Peierls/Poisson bracket~\eqref{Equation: Poisson Peierls bracket}, one finds that the field variation $\delta_\upsilon\phi$ due to the change of frames is canonical, and given by a bracket with $\mathcal{C}_\upsilon = X_+^\upsilon-X_-^\upsilon$, as required.

  Although it is likely that a generalization of this result exists for non-vanishing flux, for concreteness it will be convenient in what follows to set $\mathscr{F}^\upsilon=0$. Then the change of microcausal frames can be understood as resulting from acting with $X^\upsilon_-$ at $\Sigma_-$ to perform the change $\mathcal{F}\to\mathcal{F}'$, and later acting with $X^\upsilon_+$ at $\Sigma_+$ to change back $\mathcal{F}'\to\mathcal{F}$. More generally, when the flux vanishes, one can understand the observable
  \begin{equation}
    X^\upsilon = -\int_{K}(B_\upsilon+Y_\upsilon\alpha)\Hodge\dd{\alpha},
  \end{equation}
  where $K=\Sigma\cap\mathcal{N}$, as producing the change of microcausal frame at a given Cauchy surface $\Sigma$.

  The change of microcausal frames charge $X^\upsilon$ is not a large-gauge-invariant quantity. As a consequence, it has non-trivial brackets with the generators $Q[F],P[F]$ of large gauge transformations, which may be written
  \begin{equation}
    \pb{X^\upsilon}{Q[F]} = Y_\upsilon Q[F] + \int_K\epsilon B_\upsilon F, \qquad \pb{X^\upsilon}{P[F]} = Y_\upsilon P[F].
  \end{equation}
  We will not write down $\pb{X^\upsilon}{H}$ explicitly, but it may be checked that this bracket does not vanish. Thus, $X^\upsilon$ is not a conserved quantity, and it takes energy to change the microcausal frame.

  It is also notable that the change of frames does not act trivially on completely gauge-invariant observables $O$. Using the fact that such $O$ have a non-trivial bracket with $B_\upsilon$, which itself is completely gauge-invariant, and using the equation of motion $\dd{\Hodge\dd{\alpha}}=0$, one finds that the bracket of $\mathcal{C}_\upsilon$ with such an observable reduces to corner terms at $K_\pm$:
  \begin{align}
    \pb{\mathcal{C}_\upsilon}{O} &= -\int_{\mathcal{N}}\dd{\pb{B_\upsilon}{O}}\wedge\Hodge\dd{\alpha}
    = \int_{K_-}\pb{B_\upsilon}{O}\Hodge\dd{\alpha} - \int_{K_+}\pb{B_\upsilon}{O}\Hodge\dd{\alpha}.
    \label{Equation: C_upsilon O}
  \end{align}
  Thus, under the change of microcausal frame the observable is modified to
  \begin{equation}
    O' = O + \lambda\pb{C_\upsilon}{O} + \order{\lambda^2} = O - \lambda\int_{\partial\mathcal{N}}\pb{B_\upsilon}{O}\Hodge\dd{\alpha} + \order{\lambda^2}.
    \label{Equation: C_upsilon O exp}
  \end{equation}
  Note that the support of $O'$ is changed relative to that of $O$, by subsets of $K_\pm$. Moreover, $O'$ is no longer completely gauge-invariant, since the corner terms depend on $\alpha$, and therefore $O'$ transforms non-trivially under large gauge transformations.

  For completeness, let us also consider the algebraic relations between the charges $X^\upsilon$ for different choices of $\upsilon$. To simplify matters let us make some extra assumptions: that spacetime is stationary, and that $\gamma(y)$ and $\upsilon$ are tangent to some Cauchy surface $\Sigma$ normal to the timelike Killing vector field. Then $B_\upsilon(y)$ will only ever depend on the initial data on $\Sigma$, and assuming $K=\mathcal{N}\cap \Sigma$, we may consequently write $\pb{B_\upsilon(y)}{B_{\upsilon'}(y')}=0$ for any $y,y'\in K$, and any $\upsilon,\upsilon'$ obeying our assumptions. Since $B_\upsilon(y)$ is completely gauge-invariant, we also have that it commutes with $\alpha$ and $\Hodge\dd{\alpha}$. This makes it much simpler to compute the bracket of the charges $X^\upsilon,X^{\upsilon'}$. We find
  \begin{equation}
    \pb*{X^\upsilon}{X^{\upsilon'}}=\int_K(B_\upsilon Y_{\upsilon'}-B_{\upsilon'}Y_{\upsilon})\Hodge \dd{\alpha}.
  \end{equation}
  Thus, in general, distinct changes of microcausal frame do not commute with one another.

  \subsection{Multiple microcausal frames in the same theory?}
  \label{Section: multiple frames}

  As explained above, different frames $\mathcal{F},\mathcal{F}'$ correspond to different definitions of relational support, which we have denoted $\relsupp,\relsupp'$ respectively. They therefore yield different notions of microcausality/locality. Namely: when the action is given by~\eqref{Equation: action first frame}, we have already shown that two observables $O_1,O_2$ commute if $\relsupp(O_1)$ is spacelike separated from $\relsupp(O_2)$; similarly, when the action is given by~\eqref{Equation: action second frame}, we have that $O_1,O_2$ commute if $\relsupp'(O_1)$ is spacelike separated from $\relsupp'(O_2)$.
  It is clear that these are different statements of microcausality, because $\relsupp(O)$ does not coincide with $\relsupp'(O)$, for a general observable $O$. For example, we have (cf.\ \eqref{relsuppex})
  \begin{equation}
    \relsupp(\alpha(y)) = y \ne \gamma(y)\cup \gamma(y') = \relsupp'(\alpha(y)),
    \label{Equation: relsupp v relsupp'}
  \end{equation}
  where $f(y)=f'(y')$, and we are generically assuming $\gamma(y)$ and $\gamma'(y')$ are not tangent to each other.
  Indeed, since~\eqref{Equation: action first frame} and~\eqref{Equation: action second frame} are two different actions, they correspond a priori to entirely different theories with their own `preferred frames' and notions of microcausality.

  However, more generally, a `notion of microcausality' can be understood as a definition of spacetime support (such as $\relsupp$) for each observable, and a promise that if two observables' supports are spacelike-separated, then they will commute. We would like to point out now the more-or-less obvious fact that such spacelike separation is only a \emph{sufficient} criterion for commutativity, not a necessary one -- meaning two commuting observables need not have spacelike-separated support. This leaves open the possibility that more than one notion of microcausality applies to the same theory, despite the associated definitions of spacetime support being quite different, as in~\eqref{Equation: relsupp v relsupp'}.

  A prime example of this phenomenon may be found in holography, where boundary and bulk microcausality are both relevant. Indeed, operators will commute if they have spacelike separated \emph{boundary} support, by the microcausality of the boundary theory. But operators will commute also if they have spacelike separated \emph{bulk} support. An operator with sharply peaked support deep in the bulk will generically be delocalised over a large region on the boundary, as clarified by `subregion-subalgebra duality' \cite{Leutheusser:2021frk,Leutheusser:2021qhd,Leutheusser:2022bgi}, so these are clearly quite different notions of microcausality. Nevertheless, holographic duality permits them to simultaneously apply to the same theory.

  A similar situation applies to the present discussion, in the special case that the map $m$ between two theories defined with different microcausal frames $\mathcal{F}\to\mathcal{F}'$ is canonical (i.e.\ a symplectomorphism). Indeed, we have the original microcausality criterion:
  \begin{equation}
    \relsupp(O_1) \qq{spacelike separated from} \relsupp(O_2) \implies \pb{O_1}{O_2}=0.
  \end{equation}
  But we may also write down a \emph{second} microcausality criterion:
  \begin{equation}
    \relsupp'(m_*O_1) \qq{spacelike separated from} \relsupp'(m_*O_2) \implies \pb{O_1}{O_2}=0.
  \end{equation}
  In other words, we take each observable $O_{1,2}$, push it forward into the theory defined with $\mathcal{F}'$, and then test for spacelike separation of $m_*O_{1,2}$ with $\relsupp'$. This works because such spacelike separation implies
  \begin{equation}
    \pb{O_1}{O_2} = m^*\underbrace{\pb{m_*O_1}{m_*O_2}}_{=0} = 0,
  \end{equation}
  where the first equality follows from the fact that $m$ is canonical.

  In this way, we have two distinct notions of microcausality applying to the same theory (defined with a fixed microcausal frame $\mathcal{F}$), based on two distinct notions of support for observables $O$:
  \begin{equation}
    \relsupp(O),\qquad \relsupp'(m_*O).
  \end{equation}
  As in the holographic case, one of these supports can be sharply localised while the other is spread out, for the same observable. For example, consider a completely gauge-invariant observable $O$ with support in some small region $\mathcal{U}$. Then $\relsupp(O)=\mathcal{U}$, but from~\eqref{Equation: C_upsilon O exp} we see that $\relsupp'(m_*O)$ contains additional contributions on $K_\pm$. Nevertheless, both of these notions of microcausality are valid sufficient criteria for commutativity.

  Thus, two (or more) frames can be used to define different versions of microcausality in the same theory. One is the frame of the original theory (and so perhaps the most natural one), and others are `imported' from other theories via canonical maps $m$. In any case, all the corresponding relational supports (and associated microcausalities) are departures from the na\"ive notion of kinematical support.

  It is clear that imported frames must come from theories for which the change of microcausal frames map $m$ is canonical. We refer to such frames as `compatible'.

  \section{Quantisation}
  \label{Section: quantisation}

  We have so far considered a classical treatment of the theory described in previous sections. Let us now describe how to quantise this theory perturbatively. This will permit us to revisit certain challenges to microcausality in scalar QED, such as those discussed in \cite{Donnelly:2015hta,Donnelly:2016rvo,Donnelly:2017jcd,AHLT}, and explore whether our Goldstone boundary conditions admit any improvement of the situation. Here, we will focus exclusively on a canonical reduced quantisation, i.e.\ we first gauge fix the classical theory around a background solution and then quantise the fluctuations around it in canonical language.

  We will find the same result as in the classical setting: the quantum theory is strictly microcausal in the sense of the relational support of the microcausal frame used to define the Goldstone boundary conditions; two observables commute if their relational supports are spacelike separated, regardless of whether they are charged or not. Thus, as a proof of principle, the combination of suitable non-local boundary conditions, and the change of focus to the notion of \emph{relational} support, permits one to sidestep challenges to microcausality in QED. While non-local boundary conditions are a price to pay, this is nevertheless an interesting observation that opens up novel perspectives on microcausality in gauge theories. We emphasise that this is not in conflict with the findings of \cite{Donnelly:2015hta,Donnelly:2016rvo,Donnelly:2017jcd,AHLT}, which hold for certain classes of \emph{local} boundary conditions and the standard non-relational notion of support of an observable.

  In our exposition, we will also discuss implications of our results for the vacuum state and some of the axioms of algebraic QFT when including charged observables. These concepts will become frame-dependent and, again, the notion of relational support will sidestep challenges to including charged observables in local algebras. Our discussion will ignore questions about renormalisability, though this should follow from the renormalisability of the completely gauge-invariant sector, in view of the fact that the charged sector decouples.

  \subsection{Perturbative canonical quantisation}

  To simplify matters, let us assume that spacetime $\mathcal{M}$ is flat (i.e.\ that it is some subset of Minkowski spacetime), that $\Gamma,\mathcal{N}$ are both tangent to $\pdv{t}$, where $t$ is a Minkowski time coordinate, and let us only consider Cauchy surfaces $\Sigma$ normal to $\pdv{t}$. Let us also further refine the ``Dirichlet up to large gauge transformations'' boundary conditions~\eqref{Equation: Dirichlet up to gauge} so that $\abs{\psi}\big|_\Gamma=\abs*{\bar\psi}=1$. We will also assume the potential obeys $V'(1)=0$. We then have that $A=0,\psi=1$ obeys the boundary conditions, is a solution to the bulk equations of motion~\eqref{Equation: bulk equations of motion}, and, since it implies $\alpha=0$, it also satisfies the Goldstone boundary condition $\Box_{\mathcal{N}}\alpha=0$. We will perform a canonical quantisation of the theory, perturbatively around this background.

  It is useful to decompose the fields as
  \begin{equation}\label{eq61}
    A = \tilde{A} - \dd{\varphi}, \qquad \psi = \tilde\psi e^{-iq\varphi},
  \end{equation}
  where $\varphi$ is the global Goldstone mode of section~\ref{Section: boundary conditions} (suitably extended into the bulk), and $\tilde A|_\Gamma$ therefore vanishes. In other words $(A,\psi)$ is equal to $(\tilde A,\tilde \psi)$ with the field-dependent large gauge transformation $\varphi$ applied, so that both $(\tilde A,\tilde\psi)$ are large (but not yet small) gauge-invariant. We had already encountered $\tilde A$ in the presymplectic forms~\eqref{Equation: naive presymplectic form 2} and~\eqref{omegastar2}. We can make this decomposition unique by imposing a small gauge-fixing condition on $(\tilde A,\tilde\psi)$; we will pick\footnote{The two conditions on $\tilde\psi$ fix the phase of $\psi$ up to multiples of $2\pi$ which is all that is needed.}
  \begin{equation}
    \dd{\hodge \tilde{A}}=0,\qquad \Im(\tilde\psi)\Big|_{\Sigma^\pm}=0,\qquad \Re(\tilde\psi)\Big|_{\Sigma_\pm}>0.
    \label{Equation: small gauge fixed}
  \end{equation}
  This can always be done uniquely; finding the right decomposition amounts to solving $\Box \varphi=\nabla\cdot A$ for $\varphi$, subject to certain boundary conditions on $\Gamma,\Sigma^\pm$.

  Writing $\tilde{A} = \Delta\tilde{A}$, $\tilde\psi = 1 + \Delta\tilde\psi$, where $\Delta\tilde{A}$, $\Delta\tilde\psi$ are some small perturbations around the background, one finds that the linearised bulk equations of motion around $\tilde{A}=0,\tilde\psi=1$ are
  \begin{equation}
    \Box \Delta\tilde A_\mu = q^2 \Delta\tilde A_\mu, \qquad \Box \Re(\Delta\tilde\psi) = m^2 \Re(\Delta\tilde\psi), \qquad \pdv{t}\Im(\Delta\tilde\psi) = 0,
    \label{Equation: linearised eom}
  \end{equation}
  where $\Delta\tilde A_\mu$ denotes the components of $\Delta \tilde A$, and $m^2 = \frac12 V''(1)$. We can immediately solve the third equation in~\eqref{Equation: linearised eom} subject to the third condition in~\eqref{Equation: small gauge fixed}, to see that $\Im(\Delta\tilde\psi)=0$ everywhere, which also implies $\Delta\tilde\psi|_\Gamma=0$.

  The first two equations in~\eqref{Equation: linearised eom} then tell us that, perturbatively, we have a bulk gauge potential $\Delta\tilde A_\mu$ of mass $q$ in the Lorenz gauge,\footnote{The gauge potential has acquired a mass via the Higgs mechanism.} and a bulk real scalar $\Delta\tilde\psi$ of mass $m$. We also have the massless real scalar $\alpha$ on $\mathcal{N}$. We can perturbatively solve for $\varphi|_{\Gamma}$ in terms of these fields: for example, to linear order in the perturbations we have\footnote{This decomposition of the global Goldstone mode in terms of the regional one is similar to a decomposition in \cite{Araujo-Regado:2024dpr}.}
  \begin{equation}
    \varphi(f(y)) =  -\alpha(y) + \int_{\gamma(y)}\Delta\tilde{A} + \dots.
  \end{equation}
  The value of $\varphi$ away from $\Gamma$ is not a physical degree of freedom because of small gauge symmetry. Thus, the three fields $\tilde A, \tilde\psi,\alpha$ represent all the physical degrees of freedom, and $\alpha$ is the only large gauge charged field. This is sufficient information to perform the perturbative canonical quantisation of the theory.

  In particular, on a fixed Cauchy surface $\Sigma$, the spacelike components of the gauge potential $\tilde{A}_I$ and scalar $\tilde\psi$, and their conjugate momenta $-E_I=\partial_t\tilde{A}_I - \partial_I\tilde{A}_t$ and $\pi=\partial_t\Re(\tilde\psi)$ respectively, are promoted to operators obeying the canonical commutation relations
  \begin{align}
    \comm{\tilde{A}_I(x)}{E_J(x')} &= -i\hbar\qty(\delta_{IJ}-\frac{\partial_I\partial_J}{\delta_{MN}\partial_M\partial_N})\delta^{(\dim \Sigma)}(x-x'),\\
    \comm{\tilde\psi(x)}{\pi(x')} &= i\hbar\delta^{(\dim \Sigma)}(x-x')
  \end{align}
  where $x,x'\in\Sigma$. In addition, on $K=\Sigma\cap \mathcal{N}$, $\alpha$ and its conjugate momentum $\beta=\partial_t\alpha$ are also promoted to operators obeying the canonical commutation relation
  \begin{equation}
    \comm{\alpha(y)}{\beta(y')} = i\hbar\delta^{(\dim K)}(y-y'),
  \end{equation}
  where $y,y'\in K$. All other commutators between these operators vanish.

  Let us expand the operators as a sum over modes:
  \begin{align}
    \tilde{A}_I(x) &= \sum_{r,p} \frac1{2E^q_p} \qty(a_r(p)\, \overline{\epsilon^r_I(p)} \,\overline{F(p,x)} + a^\dagger_r(p) \epsilon^r_I(p) F(p,x)), \\
    E_I(x) &= i\sum_{r,p} \frac1{2} \qty(a_r(p)\, \overline{\epsilon^r_I(p)}\, \overline{F(p,x)} - a^\dagger_r(p) \epsilon^r_I(p) F(p,x)), \\
    \tilde\psi(x) &= 1+\sum_p \frac1{2E^m_p} \qty(b(p)\,\overline{F(p,x)} + b^\dagger(p) F(p,x)), \\
    \pi(x) &= -i\sum_p \frac1{2} \qty(b(p)\,\overline{F(p,x)} - b^\dagger(p) F(p,x)).
  \end{align}
  Here, $p$ is spatial momentum in $\Sigma$ with $E^q_p = \sqrt{\abs{p}^2+q^2}$, $E_p^m = \sqrt{\abs{p}^2+m^2}$, $r$ is polarisation, $\epsilon^r_I(p)$ are polarisation vectors, and $F(p,x)$ is the solution to $\Box F(p,x) = E_p^2F(p,x) = 0$ satisfying $F(p,x) = 0$ when $x\in\partial\Sigma$. We also can write
  \begin{align}
    \alpha(y) &= \frac{Q}{\operatorname{Area}[K]} + \sum_{k,j}\frac1{2e_k} \qty(c_j(k)\,\overline{G_j(k,y)} + c_j^\dagger(k)\, G_j(k,y)), \\
    \beta(y) &= \frac{P}{\operatorname{Area}[K]}-i\sum_{k,j} \frac1{2} \qty(c_j(k)\,\overline{G_j(k,y)} - c_j^\dagger(k)\, G_j(k,y)).
  \end{align}
  Here $k$ is momentum in $K$ with corresponding energy $e_k$, and $G_j(k,y)$ is an orthogonal basis of solutions to $\Delta_K G_j(k,y)=e_k^2 G_j(k,y)$, where $\Delta_K$ denotes the Laplacian of $K$ (for example, these could be spherical harmonics, if $K$ is a sphere). The first terms in $\alpha(y),\beta(y)$ are the zero mode and its conjugate momentum respectively (we are assuming that $K$ has only one connected component, so there is only one zero mode), and the factor involving $\operatorname{Area}[K]=\int_K\epsilon$ has been included to match with the generators of large gauge transformations given earlier in the paper (in particular we have $Q=Q[1]$, $P=P[1]$). The zero mode operators are Hermitian, $Q=Q^\dagger$, $P=P^\dagger$.

  The functions $F(p,x),G_j(k,y)$ may be normalised so that the operators defined above obey the following commutation relations:
  \begin{align}
    \comm{a_r(p)}{a^\dagger_{r'}(p')} &= 2\hbar E^q_p \delta_{rr'}\delta_{pp'}, \\
    \comm{b(p)}{b^\dagger(p')} &= 2\hbar E^m_p \delta_{p p'}, \\
    \comm{c_j(k)}{c^\dagger_{j'}(k')} &= 2\hbar e_k \delta_{jj'}\delta_{kk'}, \\
    \comm{Q}{P} &= i\hbar \operatorname{Area}[K],\label{Equation: P Q commutator}
  \end{align}
  with all other commutators vanishing. The operators $Q,P$ act on the zero modes of $\alpha$, while the other operators are the creation and annihilation operators for the fields $\tilde{A},\tilde\psi,\alpha$ respectively.

  It is simple to quantise the Hamiltonian~\eqref{Equation: Hamiltonian} by writing it in terms of these operators. Employing normal ordering, we have $\normord{H} = \normord{H_0} + \normord{H_I}$, where
  \begin{align}
    \normord{H_0} &= \sum_{r,p} a^\dagger_r(p)a_r(p) + \sum_p b^\dagger(p)b(p) + \sum_{k,j} c^\dagger_j(k) c_j(k) + \frac{Q^2}{2\operatorname{Area}[K]},\label{H0} \\
    H_I &= \frac12\int_\Sigma \operatorname{vol}_\Sigma\qty(q^2(\tilde\psi^2-1)\tilde{A}_\mu\tilde{A}^\mu + V(\tilde\psi^2) - \frac12(\tilde\psi^2-1)^2V''(1)).\label{HI}
  \end{align}
  Ignoring the interaction Hamiltonian $H_I$ momentarily, we see that $a_r^\dagger(p)$ creates photons of energy $E_p^q$, $b^\dagger(p)$ creates scalar quanta of energy $E_p^m$, and $c_j^\dagger(k)$ creates `regional Goldstone mode' quanta of energy $e_k$. The interaction Hamiltonian $H_I$ is of cubic order and higher in $\delta \tilde A, \delta\tilde\psi$, and thus produces perturbative corrections to this picture. We could, if we wanted, use it to compute scattering amplitudes for these fields, employing Feynman diagrams in the usual way. However, note that there are no interactions between the regional Goldstone mode and the bulk fields.

  It is also clear now what the Hilbert space $\mathcal{H}$ of the theory should be. It is the tensor product of the Fock spaces for $\tilde{A}$, $\tilde\psi$, and the non-zero modes of $\alpha$, and a Hilbert space $\mathcal{H}_0$ for the zero modes of $\alpha$:
  \begin{equation}
    \mathcal{H}=\mathcal{H}_{\tilde{A}}\otimes\mathcal{H}_{\tilde\psi}\otimes\mathcal{H}_{\alpha}\otimes\mathcal{H}_0.
  \end{equation}
  Noting that the zero mode behaves as a free particle in one dimension, and $\alpha\in [0,2\pi)$ is a phase, we have that the zero mode Hilbert space is $\mathcal{H}_0 = L^2(S^1)$. This is the space of square integrable wavefunctions of a single variable $Q\in[0,2\pi\operatorname{Area}[K])$; multiplication by $Q$ represents the operator $Q$, and $-i\hbar\operatorname{Area}[K]\dv{Q}$ represents the operator $P$.

  Note that $\mathcal{H}$ is a \emph{physical} Hilbert space -- meaning we do not have to impose any gauge constraints on states in $\mathcal{H}$. This is because we obtained $\mathcal{H}$ by directly quantising classically gauge-fixed fields, i.e.\ we pursued a reduced quantisation.

  \subsection{Quantum microcausality}

  Let us now argue that the QFT constructed above obeys the same form of microcausality as the classical theory.

  If we turn off the interaction Hamiltonian $H_I$, then we just have a collection of free fields. It is well-known that free fields obey microcausality. In this case, two of the fields ($\tilde{A}$, $\tilde\psi$) live in the full spacetime $\mathcal{M}$, whereas the other ($\alpha$) lives on the codimension 1 surface $\mathcal{N}$. Thus, whereas the large gauge-invariant $\tilde{A}$, $\tilde\psi$ respect microcausality according to the causal structure of the flat metric on $\mathcal{M}$, $\alpha$ respects microcausality according to the causal structure of the induced metric on $\mathcal{N}$. This is precisely the same behaviour as the classical fields.

  We can turn back on the interaction Hamiltonian using perturbative QFT. In particular, by using the formalism of causal perturbation theory~\cite{Epstein:1973gw,scharf2014finite}, one can do this in such a way that the microcausal properties of the free theory are preserved. The key to causal perturbation theory is the requirement that the renormalisation scheme one employs is compatible with the causal structure. There are different approaches to satisfying this requirement -- the Epstein-Glaser axioms (for example) provide a minimal list of properties to guarantee that it holds. We shall not get into the weeds of the formalism here.

  The only possible difference with ordinary causal perturbation theory is again the fact that some fields evolve in the bulk, and others evolve on the surface $\mathcal{N}$. But these two sets of fields are completely decoupled (even in the interacting theory). Indeed, the field on $\mathcal{N}$, i.e.\ $\alpha$, remains free in the interacting theory, since it does not appear in $H_{I}$. So we only need to apply causal perturbation theory to the bulk fields.

  Thus, we can conclude that the fields $\tilde A,\tilde\psi,\alpha$ all obey the correct form of microcausality, even in the interacting theory (so long as we use a physically reasonable renormalisation scheme). Moreover, other observables $\hat O$ built out of these fields (perhaps by substituting in these field operators to the classical observable $O$ and applying normal ordering) will also respect microcausality. The criterion for commutation is the same as in the classical theory: two field operators $\hat{O}_1,\hat{O}_2$ commute if the corresponding classical observables $O_1,O_2$ have spacelike separated relational support.

  \subsection{Different microcausal frames give different quantisations}

  The quantum theory constructed here depends on the choice of microcausal frame $\mathcal{F}=(\mathcal{N},f,\gamma)$, because the classical theory does through the boundary conditions. Choosing a different microcausal frame $\mathcal{F}'=(\mathcal{N}',f',\gamma')$ amounts to defining a different classical theory, which upon quantisation gives a different quantum theory.

  Let us use $\mathcal{H}^{\mathcal{F}}$ to denote the Hilbert space for the theory with a given $\mathcal{F}$. The bulk fields $\tilde{A},\tilde\psi$ do not depend on $\mathcal{F}$, so their Fock spaces $\mathcal{H}_{\tilde{A}},\mathcal{H}_{\tilde\psi}$ also do not depend on $\mathcal{F}$. However, the other tensor factors $\mathcal{H}_\alpha\otimes\mathcal{H}_0$ do depend on $\mathcal{F}$, through the geometry of $K$. Indeed, the Fock space $\mathcal{H}_\alpha$ depends on $\mathcal{F}$ through the spectrum of the Laplacian $\Delta_K$, and the zero mode space depends on $\mathcal{F}$ through $\operatorname{Area}[K]$ (recall~\eqref{Equation: P Q commutator}). So let us write
  \begin{equation}
    \mathcal{H}^{\mathcal{F}}=\mathcal{H}_{\tilde{A}}\otimes\mathcal{H}_{\tilde\psi}\otimes \mathcal{H}_\alpha^{\mathcal{F}}\otimes\mathcal{H}_0^{\mathcal{F}}.
  \end{equation}

  On the other hand, the Hilbert spaces $\mathcal{H}^{\mathcal{F}}$, $\mathcal{H}^{\mathcal{F}'}$, for different choices $\mathcal{F},\mathcal{F}'$ of microcausal frame, are unitarily equivalent. When the frames satisfy the requirements of Section~\ref{Section: canonical change}, such that the classical change of frames is a canonical map generated by $\mathcal{C}_\upsilon$, a natural unitary mapping between the Hilbert space is given by the quantisation of the canonical transformation $m$ between the corresponding classical theories, which is a unitary map that we shall denote
  \begin{equation}
    U_{\mathcal{F}\to\mathcal{F}'}: \mathcal{H}^{\mathcal{F}}\to\mathcal{H}^{\mathcal{F}'}.
  \end{equation}
  We can use this map to convert a general operator $\hat{O}'\in\mathcal{B}(\mathcal{H}^{\mathcal{F}'})$ acting in the second theory to one acting in the first theory:
  \begin{equation}
    \hat O = (U_{\mathcal{F}\to\mathcal{F}'})^\dagger \hat{O}' U_{\mathcal{F}\to\mathcal{F}'} \in\mathcal{B}(\mathcal{H}^{\mathcal{F}}).
  \end{equation}
  Since $U_{\mathcal{F}\to\mathcal{F}'}$ is unitary, this conversion furnishes an algebraic isomorphism, which implies that if two operators commute in the second theory, they also commute in the first theory:
  \begin{equation}
    \comm{\hat O_1}{\hat O_2} = \comm{ (U_{\mathcal{F}\to\mathcal{F}'})^\dagger \hat O'_1U_{\mathcal{F}\to\mathcal{F}'}}{ (U_{\mathcal{F}\to\mathcal{F}'})^\dagger \hat O'_2U_{\mathcal{F}\to\mathcal{F}'}} = (U_{\mathcal{F}\to\mathcal{F}'})^\dagger\underbrace{\comm{\hat O'_1}{\hat O'_2}}_{=0}U_{\mathcal{F}\to\mathcal{F}'} = 0.
  \end{equation}
  Therefore, just as in the classical case, multiple notions of microcausality, associated with different frames $\mathcal{F},\mathcal{F}'$, coexist within the same same quantum theory. Indeed, if $\hat{O}$ is the quantisation of some classical observable $O$, then $\hat{O}'$ is by definition the quantisation of $m_*O$, and the preceding argument is just the quantum translation of the equivalent classical story described in section~\ref{Section: multiple frames}.

  The map $U_{\mathcal{F}\to\mathcal{F}'}$ can be understood as implementing a change of boundary conditions of the kind depicted in Figure~\ref{Figure: change frame} in a path integral. When the change of frames is infinitesimal in the sense described in Section~\ref{Section: generators}, the change of boundary conditions can be understood as being implemented by deforming the action as $S\to S-\lambda\mathcal{C}_\upsilon$ in a path integral:
  \begin{equation}
    \int\Dd{\phi} e^{iS} \to \int\Dd{\phi} e^{i(S-\lambda\mathcal{C}_\upsilon)} = \int\Dd{\phi} e^{iS}e^{i\lambda X^\upsilon_+}e^{i\lambda \mathscr{F}^\upsilon}e^{-i\lambda X_\upsilon^-}.
    \label{Equation: path integral frame change}
  \end{equation}
  Path integrals compute time-ordered correlators of operators, so the deformation of the path integral above can be understood as the insertion of $e^{i\lambda X^\upsilon_+}$ on the left where it acts on the final state, and $e^{-i\lambda X^\upsilon_-}$ on the right where it acts on the initial state. When there is flux $\mathscr{F}_\upsilon$, this manifests as operators inserted in a time-ordered manner over the course of the path integral.

  \subsection{Frame dependence of the vacuum state}

  Each of the Fock space tensor factors in $\mathcal{H}^{\mathcal{F}}$ has a unique vacuum state. There is also a unique zero mode vacuum $\ket{0}_0\in\mathcal{H}_0$ (the state with constant $Q$ wavefunction). Thus, there would appear to be a unique vacuum state in the full theory:
  \begin{equation}
    \ket{\Omega} = \ket{0}_{\tilde{A}}\otimes\ket{0}_{\tilde\psi}\otimes\ket{0}_{\alpha}\otimes \ket{0}_0.
  \end{equation}
  Vacuum correlation functions may be computed with path integrals, using $\ket{\Omega}$ as the initial and final state.

  One may change microcausal frames in a vacuum correlation function by deforming the path integral as in~\eqref{Equation: path integral frame change}. For concreteness let us again assume that the flux vanishes $\mathscr{F}^\upsilon=0$, so that the total effect of this deformation is to act on the initial and final vacuum states with $e^{-i\lambda\hat X^\upsilon}$. Here $\hat X^\upsilon$ is the quantization of the classical frame change charge $X^\upsilon$, and we have dropped the ${}_\pm$ subscript. As we shall now demonstrate, the operator $\hat X^\upsilon$ does not preserve the vacuum.

  There are of course ordering ambiguities in quantising $X^\upsilon$. Suppose we make the unique normal-ordered Hermitian choice, which may be written
  \begin{equation}
    \hat{X}^\upsilon =\normord{X^\upsilon} = -\frac{Y_\upsilon(QP+PQ)}{2\operatorname{Area}[K]} - i\int_{y\in K}\epsilon \normord{B_\upsilon(y)} \sum_{k,j} c_j^\dagger(k) G_j(k,y) + \dots
  \end{equation}
  where the ellipsis contains terms that all have at least one annihilation operator on the right. Using $P\ket{0}_0=0$, the first term produces a constant scaling of the vacuum state
  \begin{equation}
    \frac{Y_\upsilon(QP+PQ)}{2\operatorname{Area}[K]}\ket{\Omega} = \frac{i\hbar Y_\upsilon}{2}\ket{\Omega},
  \end{equation}
  and the terms in the ellipsis annihilate the vacuum state, so we are just left with
  \begin{equation}
    \normord{X^\upsilon}\ket{\Omega} = \frac{i\hbar Y_\upsilon}2\ket{\Omega}-i\int_{y\in K}\epsilon \sum_{k,j}G_j(k,y)(\normord{B_\upsilon(y)}\ket{0}_{\tilde{A}}\otimes\ket{0}_{\tilde\psi}) \otimes c_j^\dagger(k)\ket{0}_\alpha \otimes \ket{0}_0
    \label{Equation: quantum frame change}
  \end{equation}
  The first term on the right is just a phase shift and not so interesting. But the second term corresponds to a genuine change of state.
  Indeed, depending on $\upsilon$, the operator $\hat{X}^\upsilon$ can create photons, scalar quanta, and regional Goldstone mode quanta of various energies (although the zero mode stays in its vacuum). To see that this effect can be non-trivial, consider the case where $\upsilon$ is tangent to $\pdv{t}$ and vanishes on $\mathcal{N}$. Then
  \begin{equation}
    \normord{B_\upsilon(y)} = \int_{\gamma(y)}\normord{\iota_\upsilon F} =-\int_{\gamma(y)}v_t E
  \end{equation}
  clearly creates many photons, polarised along the curves $\gamma(y),y\in\mathcal{N}$, in a way that can fairly arbitrarily depend on $y$. Thus, the right hand side of~\eqref{Equation: quantum frame change} will not be a pure phase shift.

  So, the vacuum of the theory depends on the choice of microcausal frame. It may at first appear that this frame-dependence of the vacuum goes in hand with the perhaps unsurprising theory-dependence of the vacuum. Indeed, when the theory and microcausal frame $\mathcal{F}$ are fixed via the action~\eqref{nonlocalaction}, also the Hamiltonian~\eqref{H0},~\eqref{HI} is fixed. At this point, the vacuum corresponding to the preferred microcausal frame $\mathcal{F}$ appears distinguished; it is the minimal energy eigenstate of the free contribution $\normord{H_0}$.

  However, this is not the complete story. The choice of split between photon $\tilde A$, matter $\tilde\psi$, and large gauge degrees of freedom $\alpha$ is clearly not unique even in a fixed Hamiltonian. For example, suppose we chose $\mathcal{F}$ to define the theory via~\eqref{nonlocalaction}, yielding the independent variables $(\tilde A,\tilde\psi,\alpha)$ that also appear in the corresponding Hamiltonian. Nothing stops us from changing variables to $(\tilde A,\tilde\psi,\alpha')$ in the same theory, where $\alpha'$ is associated with some other frame $\mathcal{F}'$. We can then write $\alpha=\alpha(\tilde A,\tilde\psi,\alpha')$ as a functional of the new variables and insert this into the original Hamiltonian. In terms of the new fields, the \emph{same} Hamiltonian will generically split differently into a free $\normord{H_O'}$ and interacting piece $\normord{H_I'}$.
  Accordingly, now the vacuum associated with frame $\mathcal{F}'$ will minimise the energy of $\normord{H_0'}$. Thus, even within the same theory, we have a frame dependence of the vacuum, which goes in hand with the choice of physical variables used to describe the theory.\footnote{In order for the argument to hold, it has to be ensured that the theory is perturbative around both choices of free sector.}

  This resonates with the quantum frame-dependence of the vacuum observed in parametrised field theory \cite{Hoehn:2023axh} and cavity QED \cite{cavity}, which  holds for a \emph{fixed} Hamiltonian. It is a consequence of the general observation that in gauge systems interactions are gauge frame-dependent \cite{Hoehn:2023ehz}. In particular, in QED light-matter splits are defined only relative to a gauge frame, or, equivalently, a gauge fixing (see \cite{cavity} for more discussion in cavity QED).

  In fact, there is a further source of such a frame dependence of the vacuum. So far we have only referred to different choices $\mathcal{F},\mathcal{F}'$, which amount to small gauge frames for dressing (or equivalently gauge fixing) on the surfaces $\mathcal{N},\mathcal{N}'$. However, a spacetime complete small gauge frame has to cover the entire bulk of our spacetime in order to build physical observables everywhere. Implicitly, we have made a specific choice of small gauge frame away from the surface $\mathcal{N}$. Indeed, we made specific small gauge choices to render the decomposition in \eqref{eq61} unique. But this decomposition also affects our physical light-matter split that enters the Hamiltonian. Fixing different small gauge-dependent degrees of freedom away from $\mathcal{N}$, corresponding to different choices of small gauge frame in the bulk, will also result in a distinct such split and thus in a distinct free contribution.

  It is also worth stressing that this frame dependence does not mean that a choice of split is gauge-dependent and thereby unphysical. Quite the contrary: the split only depends on the convention of which degrees of freedom are to be gauge fixed (these are the dynamical frames), but not on which precise configuration these degrees of freedom are gauge fixed to. In fact, the entire discussion can be made at the purely small gauge-invariant level in terms of different subalgebras of the total gauge-invariant observable algebra, and this goes under the name of subsystem relativity \cite{Hoehn:2023ehz,AliAhmad:2021adn,DeVuyst:2024pop,DeVuyst:2024uvd,Araujo-Regado:2025ejs,cavity}. In conclusion, the light-matter partition and with it the free contribution in the Hamiltonian depends in general on the choice of small gauge frame.

  This is similar to the general situation of QFT in curved space, where one has to pick a timelike Killing vector field (KVF) relative to which energies are measured, in order to define the vacuum. Different timelike KVFs give different vacuums, and this property is responsible for a variety of interesting phenomena, such as Hawking radiation and the Unruh effect. But note that the choice of timelike KVF is basically a choice of reference frame -- albeit a reference frame of a different kind to what we have been considering in this paper. Indeed, in the case of this paper, different choices of frame also lead to different definitions of energy and excitations. Thus, the non-preservation of the vacuum by the microcausal frame change operator $\hat X^\upsilon$ is a phenomenon bearing some similarity to those of Hawking and Unruh. A key difference, already alluded to in \cite{Hoehn:2023ehz}, is that different vacua in the case of Hawking and Unruh arise because different frames do not have access to the same degrees of freedom. Here, by contrast, the differences in vacua refer to global states in either case. This is qualitatively closer to the appearance of different vacua in cosmology. It would be very interesting to investigate this further, but we will leave this to future work.

  \subsection{Subregion algebras}

  Having constructed its canonical quantisation, let us now give some brief comments on the algebraic QFT approach to the theory, in light of the results of this paper.

  The main tool of algebraic QFT is a local net of observable algebras obeying the Haag-Kastler axioms~\cite{haag2012local,Fewster:2019ixc} (here, we shall only consider two of these). This is the specification for every subset $U$ of spacetime of a $*$- or even $C^*$-algebra $\mathcal{A}_U$ of observables,\footnote{When a representation on a Hilbert space $\mathcal{H}$ is given, the $\mathcal{A}_U$ are often taken to be von Neumann algebras.} such that $\mathcal{A}_{U_1}\subset\mathcal{A}_{U_2}$ if $D(U_1)\subset D(U_2)$ (`isotony'), and such that $\comm{\mathcal{A}_{U_1}}{\mathcal{A}_{U_2}}=0$ if $U_1,U_2$ are spacelike separated (`Einstein causality'). Recall, $D(U)$ denotes the domain of influence of $U$.

  Ordinarily, in a QFT without gauge symmetry, one defines the algebra $\mathcal{A}_U$ as containing observables whose support is contained within $D(U)$. In QFTs with gauge symmetry, a question arises due to the existence of nonlocal gauge-invariant observables -- in particular charged observables that extend all the way to the asymptotic boundary. How should they be incorporated while maintaining the algebraic QFT axioms, particularly isotony and Einstein causality? For uncharged observables, no fundamental obstruction arises, e.g.\ see \cite{Buchholz:2015epa,Buchholz:2021tpv} for the case of QED. Similarly, the algebra for a \emph{single} finite region $U$ in Maxwell theory -- including its boundary charges and edge modes -- can be built using techniques from algebraic QFT, as was recently demonstrated in \cite{Fewster:2025ijg}. However, the latter work did not consider the embedding of $U$ into a global spacetime and of $\mathcal{A}_U$ into a corresponding net of algebras. 

For ordinary completely local boundary conditions (such as those described in the examples in Section~\ref{Section: boundary conditions}), it is in fact perfectly consistent to proceed exactly as in a QFT without gauge symmetry. That is, one includes all gauge-invariant observables in $\mathcal{A}_U$ whose standard (non-relational) support is contained within $D(U)$. In this case, $\mathcal{A}_U$ contains a given boundary-dressed observable $O$, if $U$ itself extends to the asymptotic boundary such that it contains the standard support of $O$. On the other hand, within this standard setup, if $U$ is any compact (i.e.\ not extending to the asymptotic boundary) bulk subregion, then the corresponding algebra $\mathcal{A}_U$ does \emph{not} contain any boundary-dressed, and hence non-trivially globally charged, observables. It is reasonable to ask if one can change the situation so that there are boundary-dressed observables contained in such a $U$, and hence such that measurements of these observables can be consistently considered as a procedure local to the bulk.

 In the context of this paper, it is clear that one can do this by using instead non-local boundary conditions of the kind described in Section~\ref{Section: non-local}, and replacing the \emph{standard} support of an observable $O$ by its \emph{relational} support. Thus, $\mathcal{A}_U$ contains all observables whose relational support is contained within $D(U)$. By Section~\ref{Section: microcausality}, this assignment is consistent with Einstein causality. Furthermore, for any compact bulk region $U$ with $\mathcal{N}\cap U\ne \emptyset$, the algebra $\mathcal{A}_U$ will contain non-trivial boundary-dressed observables (such as $\alpha(y)$ for $y\in\mathcal{N}\cap U$). 

 There is a tradeoff in this approach. It is true that boundary-dressed observables that use the particular system of Wilson lines $\gamma(y)$ will be contained in the algebras of compact bulk regions -- because for such observables the relational support is smaller than the standard support. But for other kinds of boundary-dressed observables (perhaps using other Wilson lines), the relational support is larger than the standard support, as explained in Section~\ref{ssec_other}. Hence, they are still not contained in any compact bulk region algebras, and moreover are contained in \emph{fewer} non-compact region algebras, relative to the standard case. So, even though we have made some boundary-dressed observables \emph{more} local, we have made others \emph{less} local. Ultimately, we are simply trading the roles of the asymptotic $\Gamma$ and the finite-distance $\mathcal{N}$ also in terms of the support of charged observables inside the net of local algebras.

 Nevertheless, from the perspective of purely compact bulk region algebras, the situation has improved. In the standard case, no boundary-dressed observables are permitted into such algebras, whereas some are in the case with non-local boundary conditions. It is therefore worth taking this option seriously, and we shall do so now.

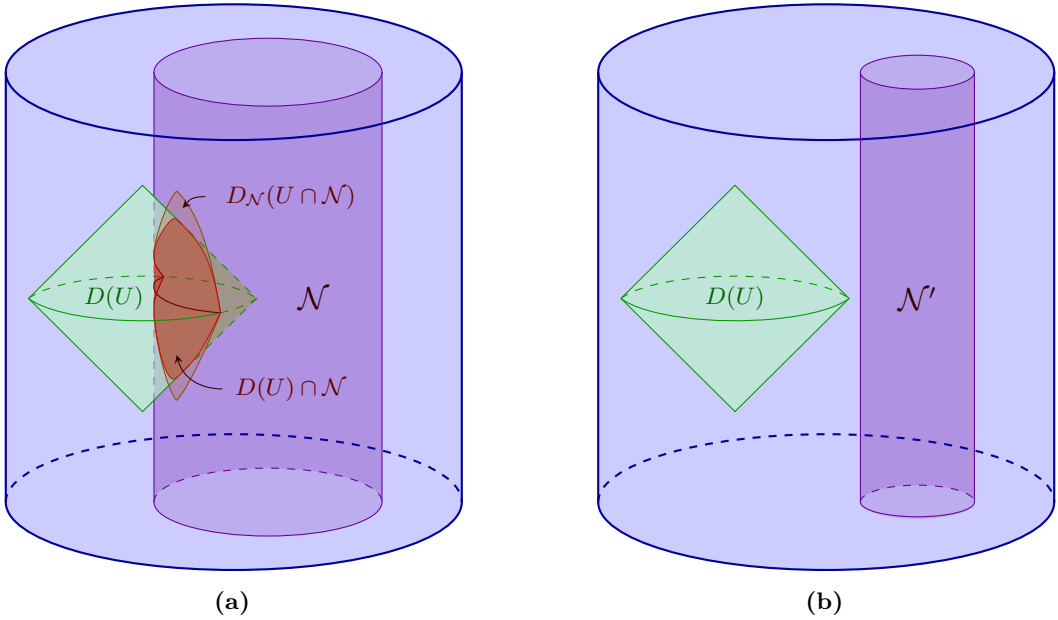
\begin{figure}
    \centering
    \begin{subfigure}{0.45\textwidth}
      \centering
      \begin{tikzpicture}[scale=1.5]
        \fill[blue!20] (0,2.2) -- (0,6) arc (180:0:2 and 0.6) -- (4,2.2) arc (360:180:2 and 0.6);

        \fill[red!40!blue,opacity=0.15] (1.3,2.2) -- (1.3,6) arc (180:0:1 and 0.3) -- (3.3,2.2) arc (0:180:1 and 0.3);
        \fill[red!40!blue,opacity=0.15] (1.3,2.2) -- (1.3,6) arc (180:360:1 and 0.3) -- (3.3,2.2) arc (360:180:1 and 0.3);

        \draw[thick,blue!60!black] (0,2.2) -- (0,6) arc (180:0:2 and 0.6) -- (4,2.2) arc (360:180:2 and 0.6);
        \draw[thick,blue!60!black] (0,6) arc (-180:0:2 and 0.6);
        \draw[thick,dashed,blue!60!black] (0,2.2) arc (180:0:2 and 0.6);

        \path (0.2,4) arc (160:80:{1/cos(20)} and 0.3) coordinate (b) arc (80:20:{1/cos(20)} and 0.3);
        \path (0.2,4) arc (200:310:{1/cos(20)} and 0.3) coordinate (a) arc (310:340:{1/cos(20)} and 0.3);
        \begin{scope}
          \clip (0.2,2) -- (1.2,5) -- (2.2,4) -- (1.2,3) -- cycle;
          \draw[dashed,red!40!blue] (3.3,6) -- (3.3,2.2) arc (360:180:1 and 0.3) -- (1.3,2.2) -- (1.3,6);
        \end{scope}
        \fill[green!30,opacity=0.5] (0.2,4) -- (1.2,5) -- (2.2,4) -- (1.2,3) -- cycle;

        \fill[orange!70!black,opacity=0.2] (1.3,4.4) .. controls (1.3,4.5) and (1.4,4.85) .. (1.5,4.95) .. controls (1.7,4.75) and (1.8,4.4) .. (a) .. controls (1.8,3.65) and (1.6,3.2) .. (1.5,3.1) .. controls (1.4,3.2) and (1.3,3.7) .. (1.3,3.9);
        \begin{scope}
          \clip (0,6.6) -- (1.2,5) -- (2.2,4) -- (1.2,3) -- (0,1.6) -- (4,1.6) -- (4,6.6) -- cycle;
          \draw[red!40!blue] (2.3,6) ellipse (1 and 0.3);
          \draw[red!40!blue] (3.3,6) -- (3.3,2.2) arc (360:180:1 and 0.3) -- (1.3,2.2) -- (1.3,6);
          \draw[dashed,red!40!blue] (1.3,2.2) arc (180:0:1 and 0.3);
        \end{scope}
        \draw[orange!60!black] (1.3,4.4) .. controls (1.3,4.5) and (1.4,4.85) .. (1.5,4.95);
        \draw[orange!60!black] (1.5,3.1) .. controls (1.4,3.2) and (1.3,3.7) .. (1.3,3.9);

        \draw[green!60!black,dashed] (1.5,4.7) -- (2.2,4) -- (1.5,3.3);
        \draw[green!60!black,dashed] (a) arc (310:340:{1/cos(20)} and 0.3);

        \draw[green!60!black,dashed] (0.2,4) arc (160:20:{1/cos(20)} and 0.3);
        \draw[orange!60!black] (1.5,4.95) .. controls (1.7,4.75) and (1.8,4.4) .. (a) .. controls (1.8,3.65) and (1.6,3.2) .. (1.5,3.1);

        \fill[green!30!purple,opacity=0.4] (1.3,3.9) .. controls (1.3,3.7) and (1.4,3.2) .. (1.5,3.3) -- (2.2,4) -- (1.5,4.7) .. controls (1.45,4.75) and (1.3,4.5) .. (1.3,4.4) -- cycle;

        \fill[red,opacity=0.2] (1.3,3.9) .. controls (1.3,4.0) .. (b) .. controls (1.3,4.3) .. (1.3,4.4);
        \fill[red,opacity=0.2] (1.3,3.9) .. controls (1.3,3.7) and (1.4,3.2) .. (1.5,3.3) .. controls (1.7,3.5) .. (a) .. controls (1.8,4.4) and (1.7,4.5) .. (1.5,4.7) .. controls (1.45,4.75) and (1.3,4.5) .. (1.3,4.4);

        \draw[red!60!black] (1.3,3.9) .. controls (1.3,4.0) .. (b) .. controls (1.3,4.3) .. (1.3,4.4);
        \draw[red!40!black] (a) .. controls (1.5,3.9) and (1.3,4) .. (1.3,4.1) .. controls (1.3,4.15) .. (b);
        \draw[red!80!black,opacity=0.8] (1.3,3.9) .. controls (1.3,3.7) and (1.4,3.2) .. (1.5,3.3) .. controls (1.7,3.5) .. (a) .. controls (1.8,4.4) and (1.7,4.5) .. (1.5,4.7) .. controls (1.45,4.75) and (1.3,4.5) .. (1.3,4.4);

        \draw[green!60!black] (1.5,3.3) -- (1.2,3) -- (0.2,4) -- (1.2,5) -- (1.5,4.7);
        \draw[green!60!black] (0.2,4) arc (200:310:{1/cos(20)} and 0.3);

        \node[red!20!black] at (2.7,4) {\large$\mathcal{N}$};
        \node[green!40!black] at (0.95,4) {\footnotesize$D(U)$};

        \node[red!40!black] at (2.5,3.2) {\footnotesize$D(U)\cap\mathcal{N}$};
        \draw[red!22!black,stealth-] (1.5,3.5) .. controls (1.55,3.3) and (1.7,3.2) .. (1.9,3.2);

        \node[red!40!black] at (2.5,4.9) {\footnotesize$D_{\mathcal{N}}(U\cap\mathcal{N})$};
        \draw[red!22!black,stealth-] (1.55,4.8) .. controls (1.6,4.85) and (1.6,4.9) .. (1.75,4.9);
      \end{tikzpicture}
      \caption{}
    \end{subfigure}
    \begin{subfigure}{0.45\textwidth}
      \centering
      \begin{tikzpicture}[scale=1.5]
        \fill[blue!20] (0,2.2) -- (0,6) arc (180:0:2 and 0.6) -- (4,2.2) arc (360:180:2 and 0.6);

        \begin{scope}[shift={(1,0)}]
          \fill[red!40!blue,opacity=0.15] (1.3,2.2) -- (1.3,6) arc (180:0:0.5 and 0.15) -- (2.3,2.2) arc (0:180:0.5 and 0.15);
          \fill[red!40!blue,opacity=0.15] (1.3,2.2) -- (1.3,6) arc (180:360:0.5 and 0.15) -- (2.3,2.2) arc (360:180:0.5 and 0.15);
        \end{scope}

        \draw[thick,blue!60!black] (0,2.2) -- (0,6) arc (180:0:2 and 0.6) -- (4,2.2) arc (360:180:2 and 0.6);
        \draw[thick,blue!60!black] (0,6) arc (-180:0:2 and 0.6);
        \draw[thick,dashed,blue!60!black] (0,2.2) arc (180:0:2 and 0.6);

        \fill[green!30,opacity=0.5] (0.2,4) -- (1.2,5) -- (2.2,4) -- (1.2,3) -- cycle;

        \draw[red!40!blue] (2.8,6) ellipse (0.5 and 0.15);
        \draw[red!40!blue] (3.3,6) -- (3.3,2.2) arc (360:180:0.5 and 0.13) -- (2.3,6);
        \draw[dashed,red!40!blue] (2.3,2.2) arc (180:0:0.5 and 0.15);
        \draw[green!60!black] (0.2,4) -- (1.2,5) -- (2.2,4) -- (1.2,3) -- cycle;
        \draw[green!60!black,dashed] (0.2,4) arc (160:20:{1/cos(20)} and 0.3);
        \draw[green!60!black] (0.2,4) arc (200:340:{1/cos(20)} and 0.3);

        \node[red!20!black] at (2.8,4) {\large$\mathcal{N}'$};
        \node[green!40!black] at (1.2,4) {\footnotesize$D(U)$};
      \end{tikzpicture}
      \caption{}
    \end{subfigure}
    \caption{The algebra $\mathcal{A}_U^{\mathcal{F}}$ of a spacetime subregion $U$ depends on the choice of frame $\mathcal{F}$. \mbox{\textbf{(a)}}\ For a given frame $\mathcal{F}=(\mathcal{N},f,\gamma)$, the algebra $\mathcal{A}_U^{\mathcal{F}}$ consists of completely gauge-invariant observables in $U$, plus large-gauge-dependent observables dressed to the boundary via Wilson lines $\gamma(y)$, for $y\in U\cap\mathcal{N}$, plus observables that depend on $\alpha|_{D_{\mathcal{N}}(U\cap\mathcal{N}})$, and any combinations of these three types of observables. \mbox{\textbf{(b)}}\ For another frame $\mathcal{F}'=(\mathcal{N}',f',\gamma')$, the algebra will be different if $U\cap\mathcal{N}\ne U\cap\mathcal{N}'$. In the case shown, we have $U\cap\mathcal{N}'=\emptyset$, so $\mathcal{A}_U^{\mathcal{F}'}$ only contains completely gauge-invariant observables in $U$.}
    \label{Figure: different algebras}
  \end{figure}

A subtlety now arises: the set of  observables with relational support in $D(U)$ does not always form an algebra. In particular, if $U$ intersects $\mathcal{N}$, then the set of observables contains $\alpha(y)$ for $y\in D(U)\cap\mathcal{N}$. But $\alpha$ behaves as a scalar field on $\mathcal{N}$, and so the local algebras it generates should be associated with domains of dependence \emph{according to the causal structure of $\mathcal{N}$}. We use $D_{\mathcal{N}}$ to denote such a domain of dependence. In general $D_{\mathcal{N}}(U\cap\mathcal{N})$ extends beyond $D(U)\cap\mathcal{N}$, see Figure~\ref{Figure: different algebras}. Thus, if we want the smallest algebra generated by observables with relational support in $U$, we really have to consider the following:
  \begin{equation}
    \mathcal{A}_U = \langle\tilde A(x),\tilde\psi(x),\alpha(y)\mid x\in D(U),y\in D_{\mathcal{N}}(U\cap\mathcal{N})\rangle
  \end{equation}
  (the notation $\langle\cdot\rangle$ here denotes the algebra generated by the operators in question).

  We have so far glossed over another subtlety. While $(\tilde A,\tilde\psi)$ correspond to completely gauge-invariant data and, accordingly, can be compactly smeared in any region $D(U)$ to turn them from operator-valued distributions into genuine operators, $\alpha$ is only defined on $\mathcal{N}$ (in terms of its relational support). Thus, we can only smear it on the codimension-one surface $\mathcal{N}$. Nevertheless, such codimension-one smearings turn $\alpha$ into operators on $\mathcal{H}_\alpha\otimes\mathcal{H}_0$ too. Bounded functions of such smeared operators may thus safely be included in the relevant algebras. This is consistent because $\alpha$ commutes with the other variables $(\tilde{A},\tilde\psi)$.\footnote{Turning such regional Goldstone modes  into well-defined operators by  smearing relative to the standard spacetime support and in the absence of non-local boundary conditions is significantly more challenging \cite{AHLT}.}

  So the algebra $\mathcal{A}_U$ contains all the observables with relational support within $D(U)$, but also $\alpha|_{D_{\mathcal{N}}(U\cap\mathcal{N})}$ (and any combinations thereof). Note that this doesn't mean $\mathcal{A}_U$ contains arbitrary dressed observables with relational support in $D_{\mathcal{N}}(U\cap\mathcal{N})$ -- it only does so in $D(U)$.

  This is of course quite different to the ordinary notion of $\mathcal{A}_U$, which, as described above, is only associated with observables in $D(U)$. We leave a full investigation of the consequences of these differences in structure to future work.
  However, here we already note that our new proposal in terms of the relational support obeys both the isotony and causality axioms, despite including charged observables. The causality axiom follows from our previous discussion of microcausality. Isotony, $\mathcal{A}_{U_1}\subset\mathcal{A}_{U_2}$ for $D(U_1)\subset D(U_2)$ is a simple consequence of the fact that if $D(U_1)\subset D(U_2)$ then also $D_\mathcal{N}(U_1\cap\mathcal{N})\subset D_\mathcal{N}(U_2\cap\mathcal{N})$. This is a rather promising feature of our proposal which thus warrants further exploration (especially whether our non-local boundary conditions could be relaxed).

  Let us here only further comment that this construction depends on which frame is employed. Indeed, the relational support is a frame dependent quantity -- for each frame $\mathcal{F}$ we have a different relational support $\relsupp_{\mathcal{F}}$, and as discussed in section~\ref{Section: multiple frames}, multiple choices of frame can be consistent with their own versions of microcausality in the same theory. For example, in principle, we could also build a net of algebras obeying isotony and causality for a given theory by mapping observables to another theory and using the relational support $\relsupp_{\mathcal{F}'}$ of a different frame there. In that case, a sufficient condition for $O\in\mathcal{A}_U^{\mathcal{F}'}$ would be if $\relsupp_{\mathcal{F}'}(m_*O)\subset D(U)$. The set of extra observables depending on $\alpha|_{D_{\mathcal{N}}(U\cap\mathcal{N})}$ also clearly varies with the frame. Thus, for each microcausal frame $\mathcal{F}$, one will get a different net of algebras $\mathcal{A}^{\mathcal{F}}_U$. So: \emph{the net of algebras}, a key ingredient of algebraic QFT, \emph{is frame-dependent}! We have not seen this possibility considered elsewhere in the literature. This observation will be further explored in \cite{AHLT}, together with other axioms and assumptions of algebraic QFT in Maxwell theory, but from a slightly different perspective not involving relational notions of support.

  Note that in order to be consistent with isotony and Einstein causality, the net of algebras must be strictly built using a single frame.  In particular, if $\mathcal{F}$ and $\mathcal{F}'$ define distinct microcausal frames, then there exist observables $O \in \mathcal{A}^{\mathcal{F}}_U$ and $O' \in \mathcal{A}^{\mathcal{F}'}_{U'}$ with $U$ spacelike to $U'$ such that $[O,O'] \neq 0$. Thus, Einstein causality cannot be imposed globally across frames, but only relative to a chosen frame. Similarly, we may for example have $\mathcal{A}_U^\mathcal{F}\not\subset\mathcal{A}_U^{\mathcal{F}'}$ if $U\cap\mathcal{N}\neq\emptyset$ but $U\cap\mathcal{N}'=\emptyset$, and so isotony too cannot be globally obeyed across frames.

  As a consequence, many quantities which we may think of as ``properties of the states and algebras in $U$'' are frame-dependent. This includes  all the various quantum information measures associated with subregions, such as entanglement entropies, relative entropies, and so on.\footnote{A version of this was already shown in the perturbative gravitational context~\cite{DeVuyst:2024pop,DeVuyst:2024uvd} and in lattice gauge theories \cite{Araujo-Regado:2025ejs}, where entanglement entropies of subregions depend on the chosen frame. Furthermore, in perturbative gravity, also the \emph{Type} of the corresponding von Neumann algebras may depend on the frame (e.g., Type $\rm{II}_1$ vs.\ $\rm{II}_\infty$) ~\cite{DeVuyst:2024pop,DeVuyst:2024uvd}, and this may happen more generally when applying the present discussion to gravity. In the present case of scalar QED, however, it should be emphasised that our regional algebras, defined invoking relational support, must be expected to be of Type $\rm{III}_1$, in line with general arguments in algebraic quantum field theory \cite{haag2012local,Fredenhagen:1984dc,Buchholz:1986bg}. Indeed, our regional algebras are given by tensor products of smeared versions of completely gauge-invariant observables $(\tilde A,\tilde\psi)$ and of $\mathcal{N}$-smearings of $\alpha$ (which is just a scalar field on $\mathcal{N}$), both of which should independently obey the assumptions leading to the Type $\rm{III}_1$ property. By contrast, in the gravitational case the Type $\rm{II}$ nature can come about due to certain dressings with modular flow, which are not present in our scalar QED setting. }
  It would be very interesting to understand this in more detail.

  \section{Conclusion}
  \label{Section: conclusion}

  In this paper, we have described in some detail a case study on the nature of microcausality in (scalar) QED. We have shown how \emph{non-local} Goldstone boundary conditions can be used to enhance the microcausality of boundary-dressed bulk observables, and we have explained how this can be understood in terms of a preferred `microcausal frame'. Moreover, we have described how these properties are maintained at the level of a perturbative quantisation. Let us conclude by first describing the general lessons we believe should be taken away from this case study.

  First, perhaps despite the prevailing intuition, it is entirely consistent in certain situations to think of boundary-dressed observables as local to the bulk, in a way that respects microcausality.

  Next, the question of which kinds of boundary-dressed observables have this status is determined by the boundary conditions one imposes.  Moreover, for a non-trivial setup, the boundary conditions generically ought to be non-local. This non-locality is confined entirely to the large gauge (i.e.\ Goldstone mode) sector of the theory, so the bulk gauge-independent physics is unaffected. Such boundary conditions have not been considered very much previously in the literature, but we believe the results of this paper mean they warrant further investigation. In particular, given a set of boundary conditions in a general theory, it is clearly of great interest to understand which frames are compatible (meaning observables dressed to them are consistent with microcausality).

A key feature of our construction is that microcausally local boundary-dressed observables are confined to a timelike codimension-1 region $\mathcal{N}$ in the bulk. This restriction appears intrinsic to boundary dressing: because the dressing must terminate on the boundary, locality can be restored only on a region that is in causal correspondence with it. This suggests a structural link with holographic ideas, in which boundary data naturally encode bulk physics.

  More broadly, our results illustrate that non-local boundary conditions do not necessarily degrade locality, but can instead be used to improve it for some observables. In this sense, we are fighting non-locality with non-locality.

  A central conceptual outcome is the frame dependence of microcausality. Each choice of microcausal frame admits a consistent, causally local net of algebras of observables, but different frames are generally incompatible. In the quantum theory this leads to further consequences, including a frame-dependent vacuum, the absence of a unique causally local net of observable algebras, and potentially frame-dependent notions of entropy. These features point to a much richer notion of locality in gauge theories than is usually appreciated.

  Let us now briefly comment on possible generalisations of our results.

  The ``up to large gauge'' component of the boundary conditions used here is conventional; while we have focused on Dirichlet conditions for concreteness, Neumann, Robin, or mixed boundary conditions could equally be employed without altering the conceptual picture.

  The non-local Goldstone boundary condition itself also admits straightforward generalisations. For example, one may endow $\alpha$ with a mass or introduce self-interactions, provided it remains decoupled from large-gauge-invariant bulk degrees of freedom. More generally, $\alpha$ could be replaced by any boundary field that transforms regularly under large gauge transformations and admits a well-defined boundary action. As long as the dynamics of $\alpha$ remains consistent with the causal structure of the surface $\mathcal{N}$, microcausality of the kind described in this paper should be maintained.

  The extension to non-Abelian gauge theories appears conceptually straightforward, with the primary modifications arising from the non-commutative structure of large gauge transformations. In particular, the analogue of the field $\alpha$ will not just be valued as a phase, as in QED, but rather as an element of the Lie algebra of the structure group of the gauge theory. 

  On the other hand, extending these ideas to gravity presents additional challenges. In a diffeomorphism-invariant theory, the analogue of $\mathcal{N}$ must itself be defined relationally and hence be field-dependent~\cite{ReconcilingBulkLocality,Carrozza:2021gju,Carrozza:2022xut,Freidel:2025ous}. Understanding how non-local boundary conditions, relational localisation, and background independence interplay in this context remains an important direction for future work.

  Finally, while we have focused on dressings that terminate on a boundary, the underlying mechanism is more general. One may equally consider dressings that attach observables to other extended structures, such as particle worldlines. In this case the dressing no longer serves as a reference frame for large gauge transformations, but instead implements a non-local representation of the gauge symmetry acting along the worldline. This again requires a non-local structure (now associated with the particle rather than the boundary) playing a role analogous to the non-local boundary conditions studied here. Note that the region on which dressed observables may be regarded as local should have the same dimension as the object to which the dressing attaches: codimension 1 for boundary dressings, and one-dimensional for worldline dressings. This suggests that the dimensionality of the `dressable' region is not accidental, but reflects the geometry of the structure that carries the gauge reference. Exploring such alternative dressings and their implications for locality and microcausality remains an interesting direction for future work.
A complementary discussion of microcausality and local nets of algebras in Maxwell theory will appear in \cite{AHLT}.

  \section*{Acknowledgements}
  We are grateful for helpful discussions with Gon\c{c}alo Ara\'ujo-Regado, Julian De Vuyst, Bianca Dittrich, Stefan Eccles, Laurent Friedel, Steve Giddings, Christophe Goeller, Alok Laddha, Don Marolf, Fabio Mele, and Bilyana Tomova at various stages of this project. This work was supported in part by funding from Okinawa Institute of Science and Technology Graduate University. Research at Perimeter Institute is supported in part by the Government of Canada through the Department of Innovation, Science and Economic Development and by the Province of Ontario through the Ministry of Colleges and Universities. This work was also made possible through the support of the ID\# 62312 grant from the John Templeton Foundation, as part of the project \href{https://www.templeton.org/grant/the-quantum-information-structure-of-spacetime-qiss-second-phase}{``The Quantum Information Structure of Spacetime'' (QISS)}. This work was further supported through the \href{https://withoutspacetime.org}{``WithOut SpaceTime project'' (WOST)}, led by the Center for Spacetime and the Quantum (CSTQ), and funded by Grant ID\# 63683 from the John Templeton Foundation. The opinions expressed in this work are those of the authors and do not necessarily reflect the views of the John Templeton Foundation.

  \printbibliography

  \end{document}